\documentclass[aip,pof,longbibliography,twocolumn,reprint]{revtex4-1}
\usepackage[utf8]{inputenc}
\usepackage[T1]{fontenc}
\usepackage{natbib}
\usepackage{amssymb,amsmath}
\usepackage{epsfig}
\usepackage{bm}
\usepackage{color}

\usepackage{graphicx}

\usepackage{epstopdf}

\usepackage{color}

\newcommand\beq{\begin{equation}}
\newcommand\eeq{\end{equation}}
\newcommand\beqa{\begin{eqnarray}}
\newcommand\eeqa{\end{eqnarray}}
\newcommand{\dd}{\text{d}}

\newcommand{\al}{\alpha}

\begin{document}

\title{Tracer diffusion coefficients in a moderately dense granular suspension. Stability analysis and thermal diffusion segregation}

\author{Rub\'en G\'omez Gonz\'alez}
\altaffiliation[Author to whom correspondence should be addressed: ]{ruben@unex.es}
\affiliation{Departamento de F\'{\i}sica,
Universidad de Extremadura, E-06006 Badajoz, Spain}

\author{Vicente Garz\'{o}}\email{vicenteg@unex.es} \homepage{https://fisteor.cms.unex.es/investigadores/vicente-garzo-puertos}
\affiliation{Departamento de F\'{\i}sica and Instituto de Computaci\'on Cient\'{\i}fica Avanzada (ICCAEx), Universidad de Extremadura, E-06006 Badajoz, Spain}

\begin{abstract}

The diffusion transport coefficients of a binary granular suspension where one of the components is present in tracer concentration are determined from the (inelastic) Enskog kinetic equation. The effect of the interstitial gas on the solid particles is accounted for in the kinetic equation through two different terms: (i) a viscous drag force proportional to the particle velocity and (ii) stochastic Langevin-like term defined in terms of the background temperature. The transport coefficients are obtained as the solutions of a set of coupled linear integral equations recently derived for binary granular suspensions with arbitrary concentration [G\'omez Gonz\'alez \emph{et al.}, Phys. Rev. E \textbf{101}, 012904 (2020)]. To achieve analytical expressions for the diffusion coefficients which can be sufficiently accurate for highly inelastic collisions and/or disparate values of the mass and diameter rations, the above integral equations are approximately solved by considering the so-called second Sonine approximation (two terms in the Sonine polynomial expansion of the distribution function). The theoretical results for the tracer diffusion coefficient $D_0$ (coefficient connecting the mass flux with the gradient of density of tracer particles) are compared with those obtained by numerically solving the Enskog equation by means of the direct simulation Monte Carlo method. Although the first-Sonine approximation to $D_0$ yields in general a good agreement with simulation results, we show that the second-Sonine approximation leads to an improvement over the first-Sonine correction, especially when the tracer particles are much lighter than the granular gas. The expressions derived here for the diffusion coefficients are also used for two different applications. First, the stability of the homogeneous steady state is discussed.  Second, segregation induced by a thermal gradient is studied. As expected, the results show that the corresponding phase diagrams for segregation clearly differ from those found in previous works when the effect of gas phase on grains is neglected.
\end{abstract}

\draft
\date{\today}
\maketitle

\section{Introduction
\label{sec1}}

The understanding of transport processes occurring in a granular suspension (granular particles immersed in a fluid, like air or water) is still a challenging open problem. This is essentially due to the fact that the corresponding granular flow is a multiphase process involving quite different spatial and temporal scales. The problem is much more difficult when one considers \emph{multicomponent} gas-solid flows (i.e ., when grains have different masses and sizes), specially when one intends to describe the granular suspension in terms of a set of coupled kinetic equations for each one of the one-particle velocity distributions of the different phases.        

Thus, because of the complexity embodied in the description of multicomponent granular suspensions, a coarse-grained approach is usually adopted where the influence of the background (interstitial) fluid on grains is accounted for through an effective fluid-solid force. \cite{K90,G94,J00,KH01} This assumption holds true under the condition that the concentration of granular particles is sufficiently low, ensuring that the interstitial gas remains unaffected by the presence of the solid particles. In simpler terms, the interstitial gas is treated as a \emph{thermostat} at a constant temperature. In some works, \cite{G94,ZW21,CBC23} the gas-solid force is simply proportional to the relative mean flow velocity between the solid and gas phases.  A more sophisticated model \cite{GTSH12} accounts for the thermal fluctuations arising from the fluid and incorporates a stochastic Langevin-like term defined in terms of the (known) temperature of the interstitial gas. The stochastic term takes into account the energy transferred to  grains due to their ``interactions'' with the background particles. In this paper, we adopt the latter model, which is incorporated in the corresponding kinetic equation via a Fokker--Plack term. Consistent with other studies, \cite{TK95,SMTK96,WZLH09,PS12,H13,WGZS14,SA17,ASG19,SA20} our analysis is limited to the regime of low Reynolds numbers (Stokes flow). In this regime, the inertia of the fluid is negligible compared to its viscous forces.  Consequently, the virtual mass force resulting from the difference in accelerations of particle and fluid is not considered. This force could be important when the solid/gas particle mass ratio is close to unity. However, as recently showed, \cite{GG22} the transport coefficients derived by explicitly considering the (elastic) collisions between grains and the molecular gas particles reduce to that using the Fokker-Planck effective model in the Brownian limit. This occurs when the solid particles are much heavier than that of the molecular gas. Therefore, the results reported in this paper are expected to apply for to situations where the solid/gas particle mass ratio is large.

The Navier--Stokes transport coefficients of a binary granular suspension have been recently obtained \cite{GKG20} in the context of the (inelastic) Enskog kinetic equation.  As in Refs.\ \onlinecite{GTSH12,GGG19a}, a coarse-grained description has been also adopted and hence, the influence of the interstitial gas on solid particles is via a viscous drag force plus a stochastic Langevin-like term. The Enskog equation has been solved by means of the Chapman--Enskog method \cite{CC70} where, in contrast to other perturbation schemes, \cite{SG98,SGNT06,ZW21} the Knudsen number (which measures the strength of the spatial variations of the hydrodynamic fields on the scale of the mean free path) and not the degree of inelasticity has been used as the only relevant perturbation parameter. As a consequence, although the results reported in Ref.\ \onlinecite{GKG20} are limited to the Navier--Stokes order (first order in spatial gradients), they go beyond the weak dissipation limit and hence can be in principle applicable to a wide range of coefficients of restitution. In addition, they take into account the nonequipartition of granular energy and can be easily extended to any number of components. In this context, the kinetic theory exposed in Ref.\ \onlinecite{GKG20} extends the theory developed by Garz\'o \emph{et al.} \cite{GDH07,GHD07} for \emph{dry} (no gas phase) granular mixtures to gas-solid flows. It is important to note that the theoretical results for the shear viscosity coefficient of a binary granular mixture obtained in Refs.\ \onlinecite{GDH07,GHD07} from the Chapman--Enskog solution has been recently shown \cite{G21,ChG23} to compare much better with computer simulations than those obtained by assuming Maxwellian distributions for each species. \cite{SM21}

As occurs for elastic \cite{CC70} and inelastic \cite{GDH07,GHD07} collisions, the explicit forms of the kinetic contributions to the Navier--Stokes transport coefficients displayed in Ref.\ \onlinecite{GKG20} requires to solve a set of coupled linear integral equations. The usual way of obtaining these kinetic contributions consists of approximating the solutions to these integral equations by Maxwellian distributions (defined at different partial temperatures) corrected by truncated Sonine polynomial expansions. For the sake of simplicity, the leading Sonine polynomial (the so-called first Sonine approximation) is retained in most of the works. However, although the results obtained from this approximation for \emph{dry} (no gas phase) granular mixtures compare in general quite well with simulations for not too large values of the coefficients of restitution, \cite{GM04,GM07,ChG23} they exhibit significant discrepancies with computer simulations for strong inelasticity and/or disparate values of the mass and diameter ratios. \cite{GV09,GV12} These differences can be mitigated in part if one retains second-order terms in the Sonine polynomial expansion (second Sonine approximation). Thus, it is important to assess the degree of accuracy of at least the two first Sonine solutions for dense binary granular suspensions via a comparison of the corresponding theoretical results with appropriate computer simulations. 

Needless to say, the determination of the Navier--Stokes transport coefficients for gas-solid flows beyond the first Sonine approximation is a quite cumbersome problem. These technical difficulties increase considerably when one studies multicomponent systems due specially to the coupling among the different integral equations obeying the transport coefficients. Thus, in order to gain some insight, one considers simple situations where explicit results can be in principle attainable. Here, we consider the so-called \emph{tracer} limit, namely, a granular binary suspension where the concentration of one of the species (of mass $m_0$ and diameter $\sigma_0$) is much smaller than the other one (of mass $m$ and diameter $\sigma$). As usual, in the tracer limit, one can assume that (i) the state of the excess component is not perturbed by the presence of the tracer particles and (ii) one can also neglect the collisions among tracer particles in its kinetic equation. In these conditions, while the distribution function $f$ of the granular fluid obeys a nonlinear (closed) Enskog equation, the distribution $f_0$ of the tracer species verifies the (linear) Enskog--Lorentz equation. The influence of the interstitial gas in both kinetic equations on the distributions $f$ and $f_0$ is accounted for by means of Fokker--Planck terms with different coefficients of friction $\gamma$ and $\gamma_0$, respectively.   

In the tracer limit, the stress tensor and the heat flux of the binary granular suspension (granular gas plus tracers) are the same as those for the monodisperse granular suspension. The expressions of these fluxes were already determined in Ref.\ \onlinecite{GGG19a}. Consequently, the mass transport of tracer particles $\mathbf{j}_0$ is the relevant flux of the problem. The constitutive equation of $\mathbf{j}_0$ to Navier--Stokes order is given by\cite{GKG20}   
\beq
\label{1.1}
\mathbf{j}_0^{(1)}=-\frac{m_0^2}{\rho}D_{0}\nabla n_0-\frac{m m_0}{\rho}D\nabla n-\frac{\rho}{T} D^T\nabla T-D^U \Delta\mathbf{U},
\eeq
where $\rho=m n$ is the total mass density, $n$ is the number density of the gas particles, $n_0$ is the number density of tracer particles, $T$ is the granular temperature, and $\Delta\mathbf{U}=\mathbf{U}-\mathbf{U}_g$; $\mathbf{U}$ and $\mathbf{U}_g$ being the mean flow velocities of the granular and the interstitial gas, respectively. In Eq.\ \eqref{1.1}, $D_0$ is the tracer diffusion coefficient, $D$ is the mutual diffusion coefficient, $D^T$ is the thermal diffusion coefficient, and $D^U$ is the velocity diffusion coefficient. One of the main goals of the present paper is to obtain the diffusion transport coefficients $D_0$, $D$, $D^T$, and $D^U$ up to the second Sonine approximation in terms of the coefficients of restitution for the tracer-grain ($\al_0$) and grain-grain ($\al$) collisions, the masses and diameters, the solid volume fraction $\phi$ occupied by the granular gas, and the background temperature $T_\text{ex}$. The expression of the coefficient $D_0$ has been obtained in a previous work \cite{GGABYG23} where a complete study of the mean square displacement of an intruder in a granular suspension has been carried out. 

To check the accuracy of both Sonine approximations, the theoretical predictions for the tracer diffusion coefficient $D_0$ are compared with numerical solutions of the Enskog equation obtained from the direct simulation Monte Carlo (DSMC) method. \cite{B94} As in previous papers, \cite{GM04,GV09,GV12,GGABYG23} the coefficient $D_0$ is computed from the Einstein formula connecting the tracer diffusion coefficient with the mean square displacement of intruders in a granular suspension in a homogeneous steady state. Here, to complement the simulations performed in Ref.\ \onlinecite{GGABYG23} (which were carried out for mixtures with $\al=\al_0$) we consider situations where both coefficients of restitution are different ($\al\neq \al_0$). Comparison between theory and simulations shows an excellent agreement, especially in the case of the second Sonine approximation for systems with high degree of inelasticity and/or very disparate mass and diameter ratios.      
    
The knowledge of the complete set of diffusion coefficients opens up the possibility of applying our results to specific problems. As a first application, we perform a linear stability analysis of the homogeneous steady state of the mixture. This study is important by itself and also because this state plays the role of the reference state in the Chapman--Enskog expansion. Since the linearized hydrodynamic equations associated with the hydrodynamic fields of the granular gas ($n$, $\mathbf{U}$, and $T$) are decoupled from the tracer density $n_0$, the stability analysis (involving $d+3$ differential equations) is quite simple since a previous study \cite{GGG19a} has shown that the $d+2$ equations corresponding to the fields $n$, $\mathbf{U}$, and $T$ are linearly stable. The fact that the coefficient $D_0$ is positive implies automatically the stability of the homogeneous steady state of the system (tracer plus granular gas). As a second application, thermal diffusion segregation of an intruder immersed in a granular suspension is analyzed. A segregation criterion showing the transition between the well-known Brazil-nut effect (BNE) and the reverse Brazil-nut effect (RBNE) by varying the parameters of the system is derived. The present study complements a previous analysis recently carried in the low-density regime. \cite{GG22a} As expected, our results show that the form of the phase-diagrams for the BNE/RBNE transition depends sensitively on the presence of the interstitial gas.

The plan of the paper is as follows. In Sec.\ \ref{sec2}, the Enskog kinetic equation for the granular suspension is introduced and the homogeneous steady state is analyzed. Section \ref{sec3} deals with the Enskog--Lorentz kinetic equation for the tracer particles immersed in a granular suspension. We study first the corresponding homogeneous steady state for intruders where it is shown that their partial temperature is different from that of the granular gas. The set of integral equations obeying the diffusion transport coefficients is displayed in Sec.\ \ref{sec4} while the the first and second Sonine approximations to these coefficients are obtained in Sec.\ \ref{sec5}. These coefficients are explicitly determined in terms of both the granular and the background temperatures, the volume fraction, the coefficients of restitution, and the masses and diameters of the mixture. Some technical details of the calculations are offered in four appendices. The dependence of the diffusion coefficients on the parameter space is illustrated in Sec.\ \ref{sec6} where the theoretical predictions of the coefficient $D_0$ are also compared with Monte Carlo simulations. The stability analysis of the homogeneous steady state is carried out in Sec.\ \ref{sec6} while the thermal diffusion segregation is studied in Sec.\ \ref{sec8}. We close the paper in Sec.\ \ref{sec7} with a brief discussion of the results reported in this work.

\section{Granular suspension}
\label{sec2}


Let us consider a set of solid particles of mass $m$ and diameter $\sigma$ with collision rules according to the smooth hard sphere model. In this case, collisions among solid particles are \emph{inelastic} and are characterized by a positive constant coefficient of normal restitution $\al \leq 1$, where $\al=1$ corresponds to elastic collisions (ordinary gases). We assume that grains are immersed in a molecular gas of viscosity $\eta_g$. The system is subjected to the action of the gravitational field and hence, each particle feels the action of the force $m \mathbf{g}$ where $\mathbf{g}$ is the gravity acceleration.

As said in Sec.\ \ref{sec1}, we adopt here a coarse-grained description and the influence of the interstitial gas on solid particles is via an instantaneous fluid force. For low Reynolds numbers, this force is composed by two independent terms. The first term accounts for the friction of grains on the surrounding molecular gas and hence, it is proportional to the particle velocity $\mathbf{v}$ (viscous drag force). The second term tries to model the energy gained by the solid particles due to their ``collisions'' with the more energetic particles of the gas phase. \cite{WM96} Both terms are represented in the Enskog equation by a Fokker--Planck collision operator. \cite{NE98} 

According to this way of modeling a granular suspension,  at moderate densities, the one-particle velocity distribution function of solid particles  $f(\mathbf{r}, \mathbf{v}; t)$ obeys the Enskog kinetic equation \cite{GGG19a}
\beqa
\label{2.1}
& &\frac{\partial f}{\partial t}+\mathbf{v}\cdot \nabla f-\gamma\Delta \mathbf{U}\cdot \frac{\partial f}{\partial \mathbf{v}}+\mathbf{g}\cdot \frac{\partial f}{\partial \mathbf{v}}-\gamma\frac{\partial}{\partial \mathbf{v}}\cdot \mathbf{V} f
\nonumber\\
& &
-\gamma \frac{T_{\text{ex}}}{m}\frac{\partial^2 f}{\partial v^2}=J[\mathbf{r},\mathbf{v}|f(t),f(t)],
\eeqa
where $T_{\text{ex}}$ is the background temperature, $\Delta \mathbf{U}=\mathbf{U}-\mathbf{U}_g$, $\mathbf{U}_g$ being the mean fluid velocity of the gas phase and
\beq
\label{2.5}
\mathbf{U}(\mathbf{r},t)=\frac{1}{n(\mathbf{r},t)}\int d\mathbf{v}\; \mathbf{v} f(\mathbf{r},\mathbf{v},t)
\eeq
is the mean particle velocity. In addition, the Enskog collision operator $J[\mathbf{r}, \mathbf{v}|f,f]$ is
\beqa
\label{2.2}
& & J\left[ \mathbf{r}_{1},\mathbf{v}_{1}\mid f(t), f(t)\right] \nonumber\\
&=& \sigma^{d-1}\int
d\mathbf{v}_{2}\int d\widehat{{\boldsymbol\sigma}}\Theta ( \widehat{{\boldsymbol\sigma}}\cdot
\mathbf{g}_{12})(\widehat{{\boldsymbol\sigma}}
\cdot \mathbf{g}_{12})\nonumber\\
& &  \times \Big[ \alpha^{-2}\chi\left( \mathbf{r}_{1},\mathbf{r}
_{1}-{\boldsymbol\sigma}\right) f(\mathbf{r} _{1},\mathbf{v}_{1}^{\prime \prime
};t)f(\mathbf{r}_{1}-{\boldsymbol\sigma},\mathbf{v}_{2}^{\prime \prime };t)\nonumber\\
& & -\chi\left( \mathbf{r}_{1},\mathbf{r}_{1}+{\boldsymbol\sigma}\right)
f(\mathbf{r}_{1},\mathbf{v} _{1};t)f(\mathbf{r}_{1}+{\boldsymbol\sigma},\mathbf{v}_{2};t)\Big],
\eeqa
where $d$ is the dimensionality of
the system ($d=2$ for disks and $d=3$ for spheres), $\boldsymbol
{\sigma}=\sigma \widehat{\boldsymbol {\sigma}}$, $\widehat{\boldsymbol
{\sigma}}$ being a unit vector, $\Theta $ is the Heaviside step function, and ${\bf g}_{12}={\bf v}_{1}-{\bf v}_{2}$. The double primes on the velocities in the Enskog operator \eqref{2.2} denote the initial values $\{\mathbf{v}_1'', \mathbf{v}_2''\}$ that lead to $\{\mathbf{v}_1, \mathbf{v}_2\}$ following a binary collision:
\beq
\label{2.3}
{\bf v}_{1}''={\bf v}_{1}-\frac{1}{2}\left( 1+\alpha^{-1}\right)(\widehat{{\boldsymbol {\sigma }}}\cdot {\bf g}_{12})\widehat{\boldsymbol {\sigma}},
\eeq
\beq
\label{3.3.1}
{\bf v}_{2}''={\bf v}_{2}+\frac{1}{2}\left( 1+\alpha^{-1}\right)
(\widehat{{\boldsymbol {\sigma }}}\cdot {\bf g}_{12})\widehat{
\boldsymbol {\sigma}}.
\eeq
In addition, $\chi[{\bf r},{\bf r}\pm\boldsymbol{\sigma}|\{n(t)] $ is the equilibrium pair correlation function at contact as a functional of the nonequilibrium density field $n({\bf r}, t)$ defined by
\begin{equation}
\label{2.4}
n({\bf r}, t)=\int\; d{\bf v} f({\bf r},{\bf v},t).
\end{equation}
Apart from $n$ and $\mathbf{U}$, the other relevant hydrodynamic field is the \emph{granular} temperature $T(\mathbf{r},t)$ defined as
\beq
\label{2.6}
T(\mathbf{r},t)=\frac{m}{d n(\mathbf{r},t)} \int d\mathbf{v}\; V^2 f(\mathbf{r},\mathbf{v},t),
\eeq
where $\mathbf{V}=\mathbf{v}-\mathbf{U}$ is the peculiar velocity.

It is important to note that the Enskog equation \eqref{2.1} still assumes the validity of the 
molecular chaos assumption. This means that we neglect correlations between the pre-collisional velocities of colliding particles. However, spatial correlations between those particles are taken into consideration through the pair correlation function $\chi$. Therefore, it is imperative to evaluate the reliability of the Enskog equation through computer simulations. The theoretical results have shown good agreement with molecular dynamics simulations for all $\alpha$ at densities $n\sigma^3 \lesssim 0.25$ and for moderately high densities at $\alpha\gtrsim 0.9$. \cite{LBD02,DHGD02,MDCPH11,MGH14}  Within the scope of this paper,  moderately dense systems refer to those that meet the above requirements. \cite{G19}

As in previous works, \cite{GKG20,GGG19a} the drift (or friction) coefficient $\gamma$ is assumed to be a scalar quantity for the sake of simplicity. It is proportional to the gas viscosity $\eta_g$.  Given that in the dilute regime every particle is only subjected to its respective Stokes drag force, then for hard spheres ($d=3$) $\gamma$ is defined as
\beq
\label{2.7}
\gamma\equiv \gamma_\text{St}=\frac{3 \pi \sigma \eta_g}{m}.
\eeq
For moderate densities and very low Reynolds numbers, $\gamma$ can be written as
\beq
\label{2.8}
\gamma=\gamma_\text{St}R(\phi),
\eeq
where $R(\phi)$ is a function of the solid volume fraction
\beq
\label{2.9}
\phi=\frac{\pi^{d/2}}{2^{d-1}d\Gamma \left(\frac{d}{2}\right)}n\sigma^d.
\eeq
The calculations performed along this paper are extended to systems with a general dimensionality $d$. Nonetheless, in order to obtain explicit results, it is necessary to know the form of the function $R(\phi)$. The density dependence of the dimensionless function $R$ can be inferred from computer simulations. It should be noted that 
the available literature primarily offers simulations exclusively for three-dimensional systems ($d=3$). We are not aware of any expression in the granular literature of $R$ for hard disks (two-dimensional systems $d=2$) or hard rods (one-dimensional systems $d=1$). 
Based on the current state of knowledge, a good approximation for $R$ is given by \cite{HBK05,BHK07,YS09b}
\beq
\label{2.10}
R(\phi)=\frac{10 \phi}{(1-\phi)}+\left(1-\phi\right)^3\left(1+1.5 \sqrt{\phi}\right).
\eeq
A more sophisticated model incorporates a parameter in $R$ that mimics lubrication forces between two approaching particles. \cite{K90,KS99,GGG19a} However, this model only applies in principle for monocomponent granular gases. Given our intention to study granular mixtures in which one species is present in tracer concentration, the selection of $R(\phi)$ is made to ensure the recovery of the same drift coefficient $\gamma$ for intruders and grains when the particles are mechanically identical. We expect our findings to exhibit weak dependence on the choice of $R$.

The macroscopic balance equations for the granular suspension are obtained when one
multiplies the Enskog equation \eqref{2.9}  by $\{1, m{\bf v}, m v^2\}$ and integrates over
velocity. After some algebra, one achieves the balance equations
\begin{equation}
D_{t}n+n\nabla \cdot {\bf U}=0\;, \label{2.11}
\end{equation}
\begin{equation}
D_{t}{\bf U}+\rho ^{-1}\nabla \cdot \mathsf{P}=\mathbf{g}-\gamma \Delta \mathbf{U}\;,
\label{2.12}
\end{equation}
\begin{equation}
D_{t}T+\frac{2}{dn} \left( \nabla \cdot {\bf q}+\mathsf{P}:\nabla {\bf U}\right) =2\gamma \left(T_\text{ex}-T\right)-\zeta \,T.
\label{2.13}
\end{equation}
In the above equations, $D_{t}=\partial_{t}+{\bf U}\cdot \nabla$ is the material derivative. The
cooling rate $\zeta$ is proportional to $1-\alpha^2$ and is due to dissipative collisions. The pressure tensor ${\sf P}({\bf r},t)$ and
the heat flux ${\bf q}({\bf r},t)$ have both {\em kinetic} and {\em collisional transfer} contributions, i.e., ${\sf P}={\sf P}^\text{k}+{\sf P}^\text{c}$ and ${\bf q}={\bf q}^\text{k}+{\bf q}^\text{c}$. The microscopic expressions of $\zeta$, $\mathsf{P}$, and $\mathbf{q}$ in terms of $f$ can be found in Ref.\ \onlinecite{GKG20}. They will be omitted here for brevity.

\subsection{Homogeneous steady state}

Let us assume that the granular suspension is in a homogeneous steady state. In this state, $\mathbf{U}=\mathbf{U}_g$, $\nabla n=\nabla T=0$, $\partial_t n=\partial_t T=0$ and hence, there is an exact balance between the cooling effects (coming from the viscous friction and the inelasticity of collisions) and the energy gained by grains due to their interaction with particles of the bath. The balance equation \eqref{2.13} for the granular temperature $T$ reads
\beq
\label{2.14}
2\gamma\left(T_\text{ex}-T\right)=T\zeta.
\eeq
For elastic collisions ($\al=1$), $\zeta=0$, and $T=T_\text{ex}$, as expected. Moreover, it is quite simple to show that the Maxwellian distribution at the temperature $T_\text{ex}$ is a solution of the Enskog equation \eqref{2.1}. This is a consequence of the fluctuation-dissipation theorem. \cite{K07} For inelastic collisions ($\al \neq 1$), it follows from dimensional analysis that $f(\mathbf{v})$ has the scaling
form
\beq
\label{2.15}
f(\mathbf{v})=n \pi^{-d/2} v_\text{th}^{-d} \varphi(\mathbf{c}),
\eeq
where $v_\text{th}=\sqrt{2T/m}$ is the thermal velocity of the granular gas and $\mathbf{c}\equiv \mathbf{v}/v_\text{th}$. The exact form of the scaling distribution $\varphi$ is not known to date. On the other hand, previous calculations \cite{GGG19a} has shown that the fourth-cumulant (measuring the deviation of $f$ from its Maxwellian form) is in general very small. Thus, a good estimate for the temperature $T$ can be made by approximating $f$ by the Maxwellian distribution 
\beq
\label{2.16}
f_{\text{M}}(\mathbf{v})=n \left(\frac{m}{2\pi T}\right)^{d/2} \exp \left(-\frac{m v^2}{2T}\right).
\eeq
In this case, the cooling rate $\zeta$ is
\beq
\label{2.17cooling}
\zeta=\frac{\sqrt{2}\pi^{(d-1)/2}}{d\Gamma\left(\frac{d}{2}\right)}\chi (1-\al^2)\nu,
\eeq
where
\beq
\label{4.10}
\nu=n\sigma^{d-1}\sqrt{\frac{2T}{m}}
\eeq
is an effective collision frequency.  Equation \eqref{2.14} can be more explicitly written when one takes into account Eq.\ \eqref{2.17cooling} for $\zeta$. In dimensionless form, Eq.\ \eqref{2.14} can be rewritten as
\beq
\label{2.18}
2\lambda \theta^{-1/2}\left(\theta^{-1}-1\right)=\zeta^*,
\eeq
where $\theta\equiv T/T_\text{ex}$, $\zeta^*=\zeta/\nu$, and
\beq
\label{2.20}
\lambda=\frac{\sqrt{2}\pi^{d/2}}{2^d d \Gamma\left(\frac{d}{2}\right)\phi}\frac{R(\phi)}{\sqrt{T_\text{ex}^*}}.
\eeq
In Eq.\ \eqref{2.20}, the (reduced) bath temperature $T_\text{ex}^*\equiv T_\text{ex}/(m\sigma^2 \gamma_\text{St}^2)$. Equation \eqref{2.18} is a cubic equation for the (reduced) temperature $\theta$. It physical solution gives $\theta$ in terms of the density $\phi$, the (reduced) background temperature $T_\text{ex}^*$, and the coefficient of restitution $\al$. It must satisfy the requirement $\theta=1$ for $\al=1$ and any value of $\phi$ and $T_\text{ex}^*$. In contrast, for inelastic collisions, $\theta<1$ since the granular temperature is smaller than that of the interstitial gas. More explicitly, the physical solution to Eq.\ \eqref{2.18} is given by \cite{GGABYG23}
\beq
\label{2.20.1}
\theta=\frac{\left(\Xi^{1/3}+\Xi^{-1/3}-1\right)^2}{9\varepsilon^2},
\eeq
where $\varepsilon=\zeta^*/2\lambda$ and 
\beq
\label{2.20.2}
\Xi=\frac{3\sqrt{3}\sqrt{27\varepsilon^4-4\varepsilon^2}+27\varepsilon^2-2}{2}.
\eeq
For elastic collisions ($\al=1$), $\varepsilon\to 0$, and Eq.\ \eqref{2.20} yields $\theta\to 1$ as expected.

\section{Tracer particles immersed in a granular suspension}
\label{sec3}

\begin{figure}
\includegraphics[width=0.8\columnwidth,angle=0]{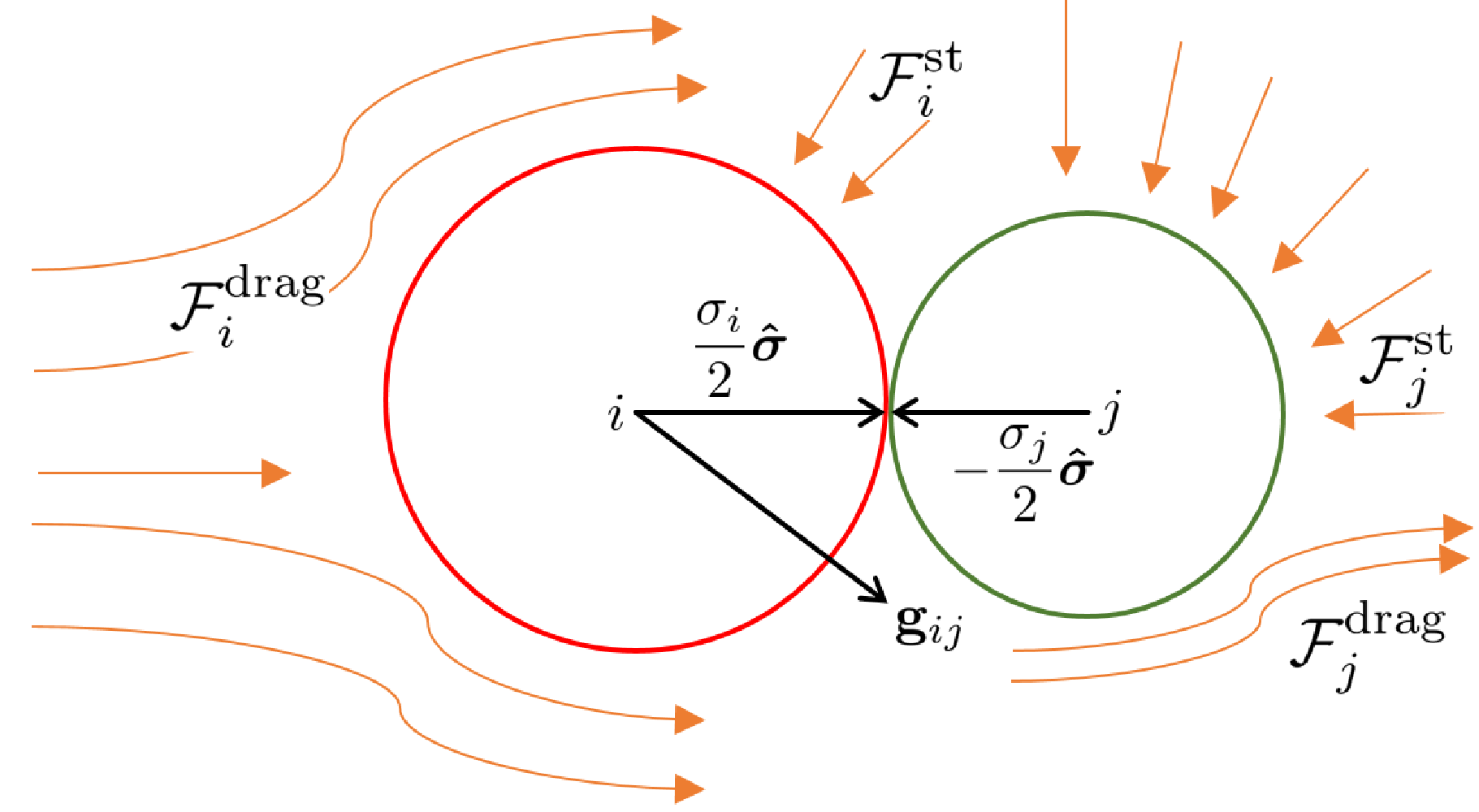}
\caption{(color online) A sketch representation of a collision between a particle of species $i$ and a particle of species $j$ together with the drag and stochastic gas-solid forces acting on these particles. \label{collision}}
\end{figure}

Let us suppose now that a few tracer or intruder particles of mass $m_0$ and diameter $\sigma_0$
are added to the granular suspension. Since the concentration of the tracer particles is negligible, their presence does not have any effect on the state of the granular suspension. This implies that the velocity distribution function $f(\mathbf{v})$ of the granular particles still obeys the (nonlinear) Enskog equation \eqref{2.1}. In addition, the balance equations for the macroscopic flow velocity $\mathbf{U}$ and temperature $T$ for the multicomponent granular suspension composed by the (monocomponent) granular suspension plus the tracer particles are the same as those for the granular suspension, namely they are given by Eqs.\ (\ref{2.12}) and (\ref{2.13}), respectively.  Under these conditions, the velocity distribution function $f_0({\bf r}, {\bf v},t)$ of the tracer species satisfies the linear Enskog-Lorentz kinetic equation
\beqa
\label{2.14b}
& &\frac{\partial f_0}{\partial t}+\mathbf{v}\cdot \nabla f_0-\gamma_0\Delta \mathbf{U}\cdot \frac{\partial f_0}{\partial \mathbf{v}}+\mathbf{g}\cdot \frac{\partial f_0}{\partial \mathbf{v}}-\gamma_0\frac{\partial}{\partial \mathbf{v}}\cdot \mathbf{V} f_0
\nonumber\\
& &
-\gamma_0 \frac{T_{\text{ex}}}{m_0}\frac{\partial^2 f_0}{\partial v^2}=J_0[\mathbf{r},\mathbf{v}|f_0(t),f(t)],
\eeqa
where the Enskog--Lorentz collision operator $J_{0}[{\bf v}|f_0(t),f(t)]$ is
\begin{eqnarray}
\label{2.14c}
& & J_{0}[{\bf r}_1, {\bf v}_1|f_0(t),f(t)]\nonumber\\
&=&
\overline{\sigma}^{d-1} \int d\mathbf{v}_{2}\int
d\widehat{{\boldsymbol\sigma}}\Theta ( \widehat{{\boldsymbol\sigma}}\cdot
\mathbf{g}_{12})(\widehat{{\boldsymbol\sigma}}
\cdot \mathbf{g}_{12})\nonumber\\
& & \times \Big[ \alpha_0^{-2}\chi_0\left(\mathbf{r}_{1},\mathbf{r} _{1}-\overline{{\boldsymbol\sigma}}\right)
f_0(\mathbf{r}_{1},\mathbf{v}_{1}^{\prime \prime
};t)f(\mathbf{r}_{1}-\overline{{\boldsymbol\sigma}}
,\mathbf{v}_{2}^{\prime \prime };t)\nonumber\\
& & -\chi_0\left( \mathbf{r}_{1},\mathbf{r}_{1}+\overline{{\boldsymbol\sigma}}
\right) f_0\left(\mathbf{r}_{1},\mathbf{v}
_{1};t\right)f\left(\mathbf{r}_{1}+\overline{{\boldsymbol\sigma}},\mathbf{v}_{2};t\right)\Big].\nonumber\\
\end{eqnarray}
Here, $\overline{{\boldsymbol\sigma}}=\overline{\sigma}
\widehat{{\boldsymbol\sigma}}$,
$\overline{\sigma}=(\sigma_0+\sigma)/2$, $\alpha_0 \leq 1$ is the (positive) 
coefficient of restitution for tracer-gas collisions, and $\chi_0$ is the pair
correlation function for tracer-gas pairs at contact. The relationship between the  precollisional velocities $\left\{\mathbf{v}_1'', \mathbf{v}_2''\right\}$ and the postcollisional velocities $\left\{\mathbf{v}_1, \mathbf{v}_2\right\}$ is
\begin{equation}
\label{2.15.a}
{\bf v}_{1}^{\prime \prime}={\bf v}_{1}-M\left( 1+\alpha _{0}^{-1}\right)
(\widehat{{\boldsymbol {\sigma }}}\cdot {\bf
g}_{12})\widehat{{\boldsymbol {\sigma }}} ,
\eeq
\beq
\label{2.15.1}
{\bf v}_{2}^{\prime \prime}={\bf v}_{2}+M_0\left( 1+\alpha
_{0}^{-1}\right) (\widehat{{\boldsymbol {\sigma }}}\cdot {\bf g}_{12})\widehat{ \boldsymbol
{\sigma}},
\end{equation}
where $M=m/(m+m_0)$ and $M_0=m_0/(m+m_0)$.  To summarize and clarify, a collision diagram along with the forces acting on the grains can be seen in Figure\ \ref{collision}.

The number density for the tracer particles is defined as
\begin{equation}
\label{2.16a}
n_0({\bf r},t)=\int\; d{\bf v}f_0({\bf r},{\bf v},t).
\end{equation}
The tracer or intruder may freely lose or gain momentum and energy in its interactions with the
particles of the granular gas. This means that these quantities are not invariants of the collision operator
$J_{0}[{\bf v}|f_0,f]$. Only the number density $n_0$ is conserved; its continuity
equation is directly obtained from the Enskog--Lorentz equation \eqref{2.13}:
\begin{equation}
\label{2.17}
D_{t}n_0+n_0\nabla \cdot {\bf U}+\frac{\nabla \cdot {\bf j}_0}{m_0}=0\;,
\end{equation}
where
\begin{equation}
{\bf j}_{0}(\mathbf{r}, t)=m_{0}\int d{\bf v}\,{\bf V}\,f_0({\bf r},{\bf v},t)
\label{2.18b}
\end{equation}
is the mass flux for the tracer particles relative to the mean local flow ${\bf U}$ of the granular gas. Apart from $n_0$, an interesting quantity at a kinetic level is the local temperature of the intruder $T_0$. It is defined as
\begin{equation}
\label{2.19}
T_0({\bf r}, t)=\frac{m_0}{d n_0({\bf r}, t)}\int \; d{\bf v}\, V^2
f_0({\bf r},{\bf v},t).
\end{equation}
The partial temperature $T_0$ measures the mean kinetic energy of the intruders; it is in general different from that of the granular gas ($T\neq T_0$).

The friction coefficient $\gamma_0$ takes into account the ``interaction'' between the tracer particles and the background molecular gas. As the coefficient $\gamma$, it can be written as $\gamma_0=\gamma_{0,\text{St}} R_0(\phi)$ where for hard spheres ($d=3$)
\beq
\label{2.20.1a}
\gamma_{0,\text{St}}=\frac{3 \pi \sigma_0 \eta_g}{m_0}=\frac{\sigma_0 m}{\sigma m_0}\gamma_{\text{St}}.
\eeq
As in the case of the function $R$, one must chose an specific form of $R_0$ to get explicit results. Computer simulations \cite{HBK05,BHK07,YS09b} indicate that a reliable choice for the function $R_0$ is 
\beq
\label{2.21}
R_0=1+\left(R-1\right)\left[a \frac{\sigma_0}{\sigma}+(1-a)\frac{\sigma_0^2}{\sigma^2}\right],
\eeq
where
\beq
\label{2.22}
a(\phi)=1-2.660\phi+9.096\phi^2-11.338\phi^3.
\eeq
Note that for mechanically equivalent particles, $R_0=R$ as expected. It must remarked that our results apply for any specific choice of the functions $R$ and $R_0$.

\subsection{Homogeneous steady state}

As in the case of the granular gas, the study of the homogeneous steady state for the tracer particles is a crucial step since this state plays the role of the reference state in the Chapman--Enskog perturbative method. \cite{CC70} In the absence of spatial gradients, the Enskog--Lorentz equation \eqref{2.14b} becomes
\beq
\label{2.23}
-\gamma_0\frac{\partial}{\partial \mathbf{v}}\cdot \mathbf{V} f_0-\gamma_0 \frac{T_{\text{ex}}}{m_0}\frac{\partial^2 f_0}{\partial v^2}=J_0[f_0(t),f(t)],
\eeq
where use has been made of the identity $\mathbf{U}=\mathbf{U}_g$ for homogeneous states. The equation for the (steady) partial temperature $T_0$ can be easily obtained from Eq.\ \eqref{2.23} as
\beq
\label{2.24}
2\gamma_0 \left(T_\text{ex}-T_0\right)=T_0 \zeta_0,
\eeq
where
\beq
\label{2.25}
\zeta_0=-\frac{1}{d n_0 T_0}\int d\mathbf{v}\; m_0 v^2\; J_0[f_0,f]
\eeq
is the partial cooling rate characterizing the rate of energy dissipated by tracer-grain collisions. As in the case of the granular suspension, for elastic collisions ($\al_0=\al=1$), $\zeta_0=0$, $T_\text{ex}=T_0$, and the Maxwellian distribution is an exact solution of Eq.\ \eqref{2.23}. Again, for inelastic collisions ($\al \neq 1$) the solution is not known to date. However, since the corresponding cumulant associated with the distribution $f_0(\mathbf{v})$ is in general very small, \cite{GGG21} an accurate estimate of $\zeta_0$ is obtained from the Maxwellian distribution defined at the partial temperature $T_0$, i.e.,
\beq
\label{2.26}
f_{0,\text{M}}(\mathbf{v})=n_0 \left(\frac{m_0}{2\pi T_0}\right)^{d/2} \exp \left(-\frac{m v^2}{2T_0}\right).
\eeq
In this approximation,
\beqa
\label{2.27}
\zeta_0&=&\frac{4\pi^{(d-1)/2}}{d\Gamma\left(\frac{d}{2}\right)} \chi_0^{(0)} M \left(\frac{\overline{\sigma}}{\sigma}\right)^{d-1} \left(\frac{1+\beta}{\beta}\right)^{1/2}(1+\al_{0})\nonumber\\
& & \times\left[1-\frac{1}{2}M (1+\beta)(1+\al_{0})\right] \nu,
\eeqa
where $\chi_0^{(0)}$ is the tracer-gas pair correlation at zeroth-order and 
\beq
\label{2.28}
\beta=\frac{m_0 T}{m T_0}
\eeq
is the ratio between the mean square velocities of intruders and grains.

In dimensionless form, Eq.\ \eqref{2.24} reads
\beq
\label{2.29}
2\lambda_0 \theta^{-1/2}\left(\theta^{-1}-\tau_0\right)=\zeta_0^* \tau_0,
\eeq
where $\tau_0=T_0/T$ is the temperature ratio, $\zeta_0^*=\zeta_0/\nu$, and
\beq
\label{3.10b}
\lambda_0=\frac{\sigma_0 m}{\sigma m_0} \frac{R_0}{R} \lambda.
\eeq
We recall that $\lambda$ is defined by Eq.\ \eqref{2.20}. When intruder and granular gas particles are mechanically equivalent ($m=m_0$, $\sigma=\sigma_0$, and $\al=\al_0$), $\lambda=\lambda_0$, $\zeta^*=\zeta_0^*$, $T=T_0$, and hence energy equipartition applies. However, in the general case (namely, when collisions are inelastic and intruder and grains are mechanically different), the solution to the cubic equation \eqref{2.29} provides  the temperature ratio $\tau_0$ in terms of the parameter space of the system. As in the free cooling case, there is a breakdown of the energy equipartition ($\tau_0\neq 1$) as expected.

\section{Diffusion transport coefficients}
\label{sec4}

The diffusion transport coefficients associated with the mass flux $\mathbf{j}_0$ of intruders can be obtained by solving the Enskog--Lorentz equation \eqref{2.14b} by means of the Chapman--Enskog method \cite{CC70} conveniently adapted to account for the inelastic character of collisions. \cite{G19} This long and hard task has been recently carried out in Ref.\ \onlinecite{GKG20} up to first-order in spatial gradients and arbitrary concentration. Using those results, we consider here the tracer limit ($x_0\equiv n_0/n\to 0$) of the linear integral equations obeying the corresponding diffusion transport coefficients.

The first-order contribution $\mathbf{j}_0^{(1)}$ to the mass flux is given by Eq.\ \eqref{1.1} where the diffusion transport coefficients $D^T$, $D_0$, $D$, and $D^U$ are defined, respectively, as
\begin{equation}
D^{T}=-\frac{m_0}{\rho d}\int d\mathbf{v}\mathbf{V}\cdot
\boldsymbol{\mathcal{A}}_{0}\left( \mathbf{V}\right),  \label{3.2}
\end{equation}
\begin{equation}
D_{0}=-\frac{\rho}{m_{0}n_{0}d}\int d\mathbf{v}\mathbf{V}\cdot
\boldsymbol{\mathcal{B}}_{0}\left( \mathbf{V}\right), \label{3.3}
\end{equation}
\begin{equation}
D=-\frac{1}{d}\int d\mathbf{v}\mathbf{V}\cdot \boldsymbol{\mathcal{C}}_{0}\left(
\mathbf{V}\right), \label{3.4}
\end{equation}
and 
\beq
\label{3.5}
D^U=-\frac{m_0}{d}\int d\mathbf{v}\mathbf{V}\cdot \boldsymbol{\mathcal{E}}_{0}\left(
\mathbf{V}\right).
\eeq
For arbitrary concentration $x_0$, the quantities $\boldsymbol{\mathcal{A}}_{0}(\mathbf{V})$,
$\boldsymbol{\mathcal{B}}_{0}\left( \mathbf{V}\right)$,
$\boldsymbol{\mathcal{C}}_{0}\left( \mathbf{V}\right)$, and $\boldsymbol{\mathcal{E}}_{0}(\mathbf{V})$ are the solutions of the set of coupled linear integral equations given by Eqs.\ (73)--(74) and (77), respectively, of Ref.\ \onlinecite{GKG20}.  In order to write this set of integral equations in the limit $x_0\to 0$, one has to take into account that while in the tracer limit $D_0$ is independent of $x_0$, the coefficients $D$, $D^T$, and $D^U$ are proportional to $x_0$. This dependence must be then self-consistently confirmed. This means that $\boldsymbol{\mathcal{A}}_{0}\propto x_0$, $\boldsymbol{\mathcal{C}}_{0}\propto x_0$, and $\boldsymbol{\mathcal{E}}_{0}\propto x_0$. Thus, in the tracer limit ($x_0\to 0$), the set of coupled linear integral equations for the unknowns $\boldsymbol{\mathcal{A}}_{0}(\mathbf{V})$,
$\boldsymbol{\mathcal{B}}_{0}\left( \mathbf{V}\right)$,
$\boldsymbol{\mathcal{C}}_{0}\left( \mathbf{V}\right)$, and $\boldsymbol{\mathcal{E}}_{0}(\mathbf{V})$ are:
\begin{widetext}
\beq
\label{3.6}
-\Bigg(2\gamma \theta^{-1}+\frac{1}{2}\zeta^{(0)}\Bigg)\boldsymbol{\mathcal{A}}_{0}
-\gamma_0\frac{\partial}{\partial {\bf V}}\cdot \left( {\bf V}\boldsymbol{\mathcal{A}}_{0}\right)
-\gamma_0 \frac{T_{\text{ex}}}{m_0}\frac{\partial^2 \boldsymbol{\mathcal{A}}_{0}}{\partial v^2}-J_0^{(0)}[\boldsymbol{\mathcal{A}}_{0},f^{(0)}]=\mathbf{A}_{0}+J_{0}^{(0)}[f_0^{(0)},\boldsymbol{\mathcal{A}}],
\eeq
\beq
\label{3.7}
-\gamma_0\frac{\partial}{\partial {\bf V}}\cdot \left( {\bf V}\boldsymbol{\mathcal{B}}_{0}\right)
-\gamma_0 \frac{T_{\text{ex}}}{m_0}\frac{\partial^2 \boldsymbol{\mathcal{B}}_{0}}{\partial v^2}-J_0^{(0)}[\boldsymbol{\mathcal{B}}_{0},f^{(0)}]=\mathbf{B}_{0},
\eeq
\beq
\label{3.8}
-\gamma_0\frac{\partial}{\partial {\bf V}}\cdot \left( {\bf V}\boldsymbol{\mathcal{C}}_{0}\right)
-\gamma_0 \frac{T_{\text{ex}}}{m_0}\frac{\partial^2 \boldsymbol{\mathcal{C}}_{0}}{\partial v^2}-J_0^{(0)}[\boldsymbol{\mathcal{C}}_{0},f^{(0)}]=
\mathbf{C}_0+J_{0}^{(0)}[f_0^{(0)},\boldsymbol{\mathcal{C}}]+\Bigg[n\frac{\partial \zeta^{(0)}}{\partial n}-2n \gamma\left(\theta^{-1}-1\right)\frac{\partial \ln \gamma}{\partial n}\Bigg]\boldsymbol{\mathcal{A}}_{0},
\eeq
\beq
\label{3.9}
-\gamma_0\frac{\partial}{\partial {\bf V}}\cdot \left( {\bf V}\boldsymbol{\mathcal{E}}_{0}\right)
-\gamma_0 \frac{T_{\text{ex}}}{m_0}\frac{\partial^2 \boldsymbol{\mathcal{E}}_{0}}{\partial v^2}-J_0^{(0)}[{\boldsymbol{\mathcal{E}}}_{0},f^{(0)}]
=\mathbf{E}_0.
\eeq
\end{widetext}
In the integral equations \eqref{3.6}--\eqref{3.9}, $\zeta^{(0)}$ is the zeroth-order approximation to the cooling rate $\zeta$, and $J_0^{(0)}[X,Y]$ is the operator
\beqa
\label{3.10}
J_{0}^{(0)}\left[{\bf v}_{1}|X,Y\right]&=&\chi_0^{(0)}\overline{\sigma}^{d-1}
\int d\mathbf{v}_{2}\int d\widehat{{\boldsymbol\sigma}}\Theta (
\widehat{{\boldsymbol\sigma}}\cdot \mathbf{g}_{12})(\widehat{{\boldsymbol\sigma}} \cdot
\mathbf{g}_{12})\nonumber\\
& & \times \left[ \alpha_0^{-2} X(\mathbf{V}_{1}^{\prime \prime
})Y(\mathbf{V}_{2}^{\prime \prime })- X(\mathbf{V}_{1})Y(\mathbf{V}_{2})\right],\nonumber\\
\eeqa
where $\chi_0^{(0)}$ is the intruder-gas pair correlation function at zeroth order. The inhomogeneous terms appearing in Eqs.\ \eqref{3.6}--\eqref{3.9} are given by
\begin{equation}
\mathbf{A}_{0}\left( \mathbf{V}\right)=-\mathbf{V}T\frac{\partial f_{0}^{(0)}}{\partial T}-\frac{p}{\rho}\frac{\partial f_{0}^{(0)}}{\partial \mathbf{V}}
-\boldsymbol{\mathcal{K}}_0\Bigg[T\frac{\partial f^{(0)}}{\partial T}\Bigg],
\label{3.11}
\end{equation}
\begin{equation}
{\bf B}_{0}\left( \mathbf{V}\right) = -{\bf V} n_0 \frac{\partial f_{0}^{(0)}}{\partial n_0},  \label{3.12}
\end{equation}
\beqa
\label{3.13}
\mathbf{C}_{0}\left(\mathbf{V}\right) &=&-\mathbf{V}  n \frac{\partial f_{0}^{(0)}}{\partial n}-m^{-1}\frac{\partial p}{\partial n}\frac{\partial f_{0}^{(0)}}{\partial \mathbf{V}}-\frac{(1+\omega)^{-d}}{\chi_0^{(0)}T}\nonumber\\
& & \times
\left(\frac{\partial \mu_0}{\partial \phi}\right)_{T,n_0}
\boldsymbol{\mathcal{K}}_0\left[f^{(0)}\right],
\eeqa
\beq
\label{3.14}
\mathbf{E}_0(\mathbf{V})=\left(\gamma_0-\gamma\right)\frac{\partial f_{0}^{(0)}}{\partial \mathbf{V}}.
\eeq
In Eqs.\ \eqref{3.11}--\eqref{3.14},
\begin{equation}
p=nT\left[1+2^{d-2}\chi^{(0)}\phi(1+\alpha)\right]
\label{3.15}
\end{equation}
is the pressure, $\omega\equiv\sigma_0/\sigma$ is the size ratio, $\mu_0$ is the chemical potential of the intruder and the operator $\boldsymbol{\mathcal{K}}_0[X]$ is defined as
\beqa
\label{3.16}
\boldsymbol{\mathcal{K}}_0[X]&=&\overline{\sigma}^{d}\chi _{0}^{(0)}\int d
\mathbf{v}_{2}\int d\widehat{\boldsymbol {\sigma }}\Theta (\widehat{\boldsymbol
{\sigma}} \cdot \mathbf{g}_{12})(\widehat{\boldsymbol {\sigma}}\cdot \mathbf{g}_{12})
\boldsymbol{\widehat{\sigma}} \nonumber\\
& & \times \left[ \alpha _{0}^{-2}f_{0}^{(0)}(\mathbf{V}_{1}^{\prime
\prime})X(\mathbf{V}_{2}^{\prime \prime
})+f_{0}^{(0)}(\mathbf{V}_{1})X(\mathbf{V}_{2})\right].\nonumber\\
\eeqa
In addition, upon writing Eq.\ \eqref{3.13}, we have neglected the non-Gaussian corrections to $f^{(0)}$ and $f_0^{(0)}$, have taken into account that $I_{121}\propto x_0\to 0$ and have used the explicit form of $I_{122}$ given by \cite{GDH07,G19}
\beq
\label{3.17}
I_{122}=\frac{d\Gamma\left(\frac{d}{2}\right)}{\pi^{d/2}T\overline{\sigma}^d \chi_0^{(0)}}\left(\frac{\partial \mu_0}{\partial n}\right)_{T,n_0}-2-n\frac{\partial \ln \chi_0^{(0)}}{\partial n}.
\eeq
The explicit expressions of the quantities $I_{ij\ell}$ for a binary mixture can be found in the Appendix A of chapter 5 of Ref.\ \onlinecite{G19}.

Note that in Eqs.\ \eqref{3.6}, \eqref{3.8}, and \eqref{3.9}, the quantities $\boldsymbol{\mathcal{A}}_{0}(\mathbf{V})$ and $\boldsymbol{\mathcal{C}}_{0}(\mathbf{V})$ associated with the tracer particles are coupled with their corresponding counterparts $\boldsymbol{\mathcal{A}}(\mathbf{V})$ and $\boldsymbol{\mathcal{C}}(\mathbf{V})$ of the host granular gas. A direct consequence of this coupling is that the mass flux $\mathbf{j}_0^{(1)}$ inherits terms coming from the autonomous host integral equations (those proportional to the spatial gradients $\nabla n$ and $\nabla T$). The expressions of $\boldsymbol{\mathcal{A}}$ and $\boldsymbol{\mathcal{C}}$ up to the second Sonine approximation are displayed in the Appendix \ref{appA}.

\section{Second Sonine polynomial approximation}
\label{sec5}

As for ordinary (elastic) mixtures, \cite{CC70} the integral equations \eqref{3.6}--\eqref{3.9} are solved by expanding the unknowns $\boldsymbol{\mathcal{A}}_{0}$, $\boldsymbol{\mathcal{B}}_{0}$, $\boldsymbol{\mathcal{C}}_{0}$, and $\boldsymbol{\mathcal{E}}_{0}$ in a complete set of orthogonal polynomials with a Gaussian measure. In kinetic theory, one employs the Sonine (or generalized Laguerre) polynomials. On the other hand, for practical purposes, to achieve explicit results one has to resort to some kind of truncation in the series expansion. Usually, the leading term in the above series expansion is retained. In this paper, as said in Sec.\ \ref{sec1}, to provide accurate results for the diffusion transport coefficients for quite inelastic systems and/or mixtures with quite disparate masses and diameters, we will consider two terms in the series expansion. This approximation is referred to as the second Sonine approximation. In this case, the quantities $\boldsymbol{\mathcal{A}}_{0}$, $\boldsymbol{\mathcal{B}}_{0}$, $\boldsymbol{\mathcal{C}}_{0}$, and $\boldsymbol{\mathcal{E}}_{0}$ are given by
\begin{equation}
\label{4.1}
\boldsymbol{\mathcal{A}}_{0}({\bf V})\to -f_{0,M}({\bf V})\left[\frac{\rho}{n_0T_0}{\bf
V}D^T+a_0{\bf S}_0({\bf V}) \right],
\end{equation}
\begin{equation}
\label{4.2} \boldsymbol{\mathcal{B}}_{0}({\bf V})\to -f_{0,M}({\bf V})\left[\frac{m_0^2}{\rho T_0}{\bf
V}D_{0}+b_0{\bf S}_0({\bf V}) \right],
\end{equation}
\begin{equation}
\label{4.3} \boldsymbol{\mathcal{C}}_{0}({\bf V})\to -f_{0,M}({\bf V})\left[\frac{m_0}{ n_0T_0}{\bf
V}D+c_0{\bf S}_0({\bf V}) \right],
\end{equation}
\beq
\label{4.3.1}
\boldsymbol{\mathcal{E}}_0\to -f_{0,M}({\bf V})\left[\frac{1}{n_0T_0}{\bf V}D^U+e_0{\bf S}_0({\bf V}) \right],
\eeq
where
\begin{equation}
\label{4.4} {\bf S}_0({\bf V})=\left(\frac{1}{2}m_0V^2-\frac{d+2}{2}T_0\right){\bf V},
\end{equation}
and $f_{0,M}({\bf V})$ is the Maxwellian distribution \eqref{2.26} with the change $\mathbf{v}\to \mathbf{V}$.
While the diffusion coefficients are defined by Eqs.\ \eqref{3.2}--\eqref{3.5}, the second Sonine coefficients $a_0$, $b_0$, $c_0$, and $e_0$ are defined as
\begin{equation}
\label{4.5} a_0=-\frac{2}{d(d+2)}\frac{m_0}{n_0T_0^3}\int\;d{\bf v}\; {\bf S}_0({\bf
V})\cdot \boldsymbol{\mathcal{A}}_{0}({\bf V}),
\end{equation}
\begin{equation}
\label{4.6} b_0=-\frac{2}{d(d+2)}\frac{m_0}{n_0T_0^3}\int\;d{\bf v}\; {\bf S}_0({\bf
V})\cdot \boldsymbol{\mathcal{B}}_{0}({\bf V}),
\end{equation}
\begin{equation}
\label{4.7} c_{0}=-\frac{2}{d(d+2)}\frac{m_0}{n_0T_0^3}\int\;d{\bf v}\; {\bf S}_0({\bf
V})\cdot \boldsymbol{\mathcal{C}}_{0}({\bf V}),
\end{equation}
\beq
\label{4.7.1}
e_0=-\frac{2}{d(d+2)}\frac{m_0}{n_0T_0^3}\int\;d{\bf v}\; {\bf S}_0({\bf
V})\cdot \boldsymbol{\mathcal{E}}_{0}({\bf V}).
\eeq

The transport coefficients $D_{0}$, $D$, $D^T$ and $D^U$ as well as the second Sonine coefficients $a_0$, $b_0$, $c_0$ and $e_0$ are determined by substitution of Eqs.\ (\ref{4.1})--(\ref{4.3.1}) into the integral equations (\ref{3.6})--(\ref{3.9}), multiplication of these equations by $m_0{\bf V}$ and by ${\bf S}_0({\bf V})$, and integration over velocity. Some technical details on these calculations are given in the Appendix \ref{appB}.

To provide the final expressions, it is convenient to write them in dimensionless form. Let us introduce the reduced transport coefficients
\beq
\label{4.8}
D^{T*}=\frac{\rho\nu}{n_0 T}D^T, \quad D_0^*= \frac{m_0^2\nu}{\rho T}D_0,
\eeq
\beq
\label{4.9}
D^*=\frac{m_0\nu}{n_0 T}D, \quad D_U^*= \rho_0^{-1}D^U,
\eeq
where $\rho_0=m_0 n_0$ is the mass density of the intruders and $\nu$ is defined by Eq.\ \eqref{4.10}. 

According to the results exposed in the Appendix \ref{appB}, the first Sonine approximations $D^{T*}[1]$, $D_0^*[1]$, $D^*[1]$, and $D_U^*[1]$ can be written as
\beq
\label{4.11}
D^{T*}[1]=\frac{X_1^*}{\nu_1^*-2\gamma^* \theta^{-1}-\frac{1}{2}\zeta^*+\gamma_0^*},
\eeq
\beq
\label{4.12}
D_0^*[1]=\frac{\tau_0}{\nu_1^*+\gamma_0^*},
\eeq
\beq
\label{4.13}
D^{*}[1]=\frac{Y_1^*}{\nu_1^*+\gamma_0^*},
\eeq
\beq
\label{4.14}
D_U^*[1]=\frac{\gamma_0^*-\gamma^*}{\nu_1^*+\gamma_0^*}.
\eeq
In Eqs.\ \eqref{4.11}--\eqref{4.14}, $X_1^*$, $a^*=\nu T^2 a$, $Y_1^*$, and $c^*=\nu n T c$ are given by Eqs.\ \eqref{b16}, \eqref{a5}, \eqref{b23}, and \eqref{a6}, respectively. The dimensionless cooling rate $\zeta^*=\zeta^{(0)}/\nu$ is defined by Eq.\  \eqref{2.17cooling} while the 
expressions of the (reduced) collision frequencies $\nu_i^*=\nu_i/\nu$ are displayed in the Appendix \ref{appC}. The forms of the second Sonine approximations $D^{T*}[2]$, $D_0^*[2]$, $D^*[2]$, and $D_U^*[2]$ are given by Eqs.\  \eqref{b18}, \eqref{b20}, \eqref{b22}, and \eqref{b25}, respectively.

In the low-density limit ($\phi\to 0$), the first Sonine approximations to the diffusion coefficients reduce to
\beq
\label{4.14.0}
D^{T*}[1]=\frac{\tau_0+\theta \Delta_{\theta,0}-m_0/m}{\nu_1^*-2\gamma^* \theta^{-1}-\frac{1}{2}\zeta^*+\gamma_0^*},
\eeq
\beq
\label{4.14.1}
D_0^*[1]=\frac{\tau_0}{\nu_1^*+\gamma_0^*}, 
\eeq
\beq
\label{4.14.2}
D^{*}[1]=\frac{\zeta^* D^{T^*}[1]-m_0/m}{\nu_1^*+\gamma_0^*},
\eeq
where $\nu_1^*$ is given by Eq.\ \eqref{c1} with $\chi_0^{(0)}=1$ while Eq.\ \eqref{d4} gives the derivative $\Delta_{\theta,0}\equiv \partial\tau_0/\partial_\theta$. This derivative arises from the fact that the zeroth-order distribution $f_0^{(0)}$ depends not only on hydrodynamic quantities but, in addition, on the partial temperature $\tau_0$. Therefore, when performing the Chapman--Enskog expansion around  $f_0^{(0)}$, there are contributions to the transport coefficients coming from the derivatives of the kinetic quantities in the vicinity of the steady state. These contributions can be seen as a measure of the departure of the perturbed time-dependent state from the steady reference state. The expressions \eqref{4.14.0}--\eqref{4.14.2} are consistent with those obtained in a previous work \cite{GG22a} from the Boltzmann equation. \footnote{Note that there is a mistake in Eq.\ (35) for the coefficient $D^*$. Equation \eqref{4.14.2} corrects this mistake.}

It is quite apparent that, in general, the first and second Sonine approximations to the diffusion transport coefficients have a complex dependence on the coefficients of restitution, the mass and diameter ratios, the solid volume fraction, and the background temperature. Thus, before studying this dependence, it is quite instructive to consider some special limits. 

In the limiting case of mechanically equivalent particles ($m_0=m$, $\sigma_0=\sigma$, $\al_0=\al$, $\gamma=\gamma_0$), as expected one gets $D^{T*}[2]=D_U^*[2]=0$, $D_0^*[2]=-D^*[2]$ and so,
\beq
\label{4.15}
\mathbf{j}_0^{(1)}=-\frac{n T}{\nu}D_0^*[2] \nabla x_0.
\eeq
The expression of $D_0^*[2]$ for arbitrary $d$ and finite $\phi$ is still a very long expression. In the particular case of a dilute ($\phi=0$) suspension of hard spheres ($d=3$), $D_0^*[2]$ is  
\begin{widetext}
\beq
\label{4.16}
D_0^*[2]=\frac{98+\al\left[47+3\al_0(4\al-13)\right]+\frac{90\sqrt{2}}{\sqrt{\pi}}\gamma^*}{\frac{4\sqrt{2\pi}}{3}(1+\al)^2\left[50+\al(4\al-25)\right]+\gamma^*\left\{(1+\al)\left[218+3\al(4\al-17)\right]+\frac{90\sqrt{2}}{\sqrt{\pi}} \gamma^*\right\}}.
\eeq
\end{widetext}
Apart from $D_0^*[2]$, the first Sonine approximation $D_0^*[1]$ reaches the simple form
\beq
\label{4.17}
D_0^*[1]=\frac{1}{\frac{2\sqrt{2\pi}}{3}(1+\al)+\gamma^*}.
\eeq

In the case of elastic collisions ($\al=\al_0=1$) but mechanically different particles, $\tau_0=\theta=1$ and the tracer diffusion coefficient for a three-dimensional dilute system reads
\begin{widetext}
\beq
\label{4.18}
D^*_0[2]=\frac{D^*_0[1]}{1-\left[\left(1+\frac{3}{2}\mu M_0^{-1/2}(1+\omega)^{-2}\gamma_0^*\right)\left(30\mu^2+(16+\frac{45}{\sqrt{\pi}}M^{-2}M_0^{-1/2}(1+\omega)^{-2}\gamma_0^*)\mu+13\right)\right]^{-1}},
\eeq
\end{widetext}
where $\mu\equiv m_0/m$ is the mass ratio and 
\beq
\label{4.19}
D^*_0[1]=\left[\frac{2\sqrt{\pi}}{3}\mu^{-1}M_0^{1/2}(1+\omega)^2+\gamma_0^*\right]^{-1}.
\eeq
Here, $\omega\equiv \sigma_0/\sigma$ is the diameter ratio. When $\gamma^*=\gamma_0^*=0$, the expression \eqref{4.18} agrees with the one derived many years ago for a gas mixture of elastic hard spheres. \cite{M54,LC84} All these results show the self-consistency of the present results with those reported in previous works in some limiting cases.


\section{Some illustrative systems and comparison with computer simulations}
\label{sec6}

\begin{figure}
\includegraphics[width=0.7 \columnwidth,angle=0]{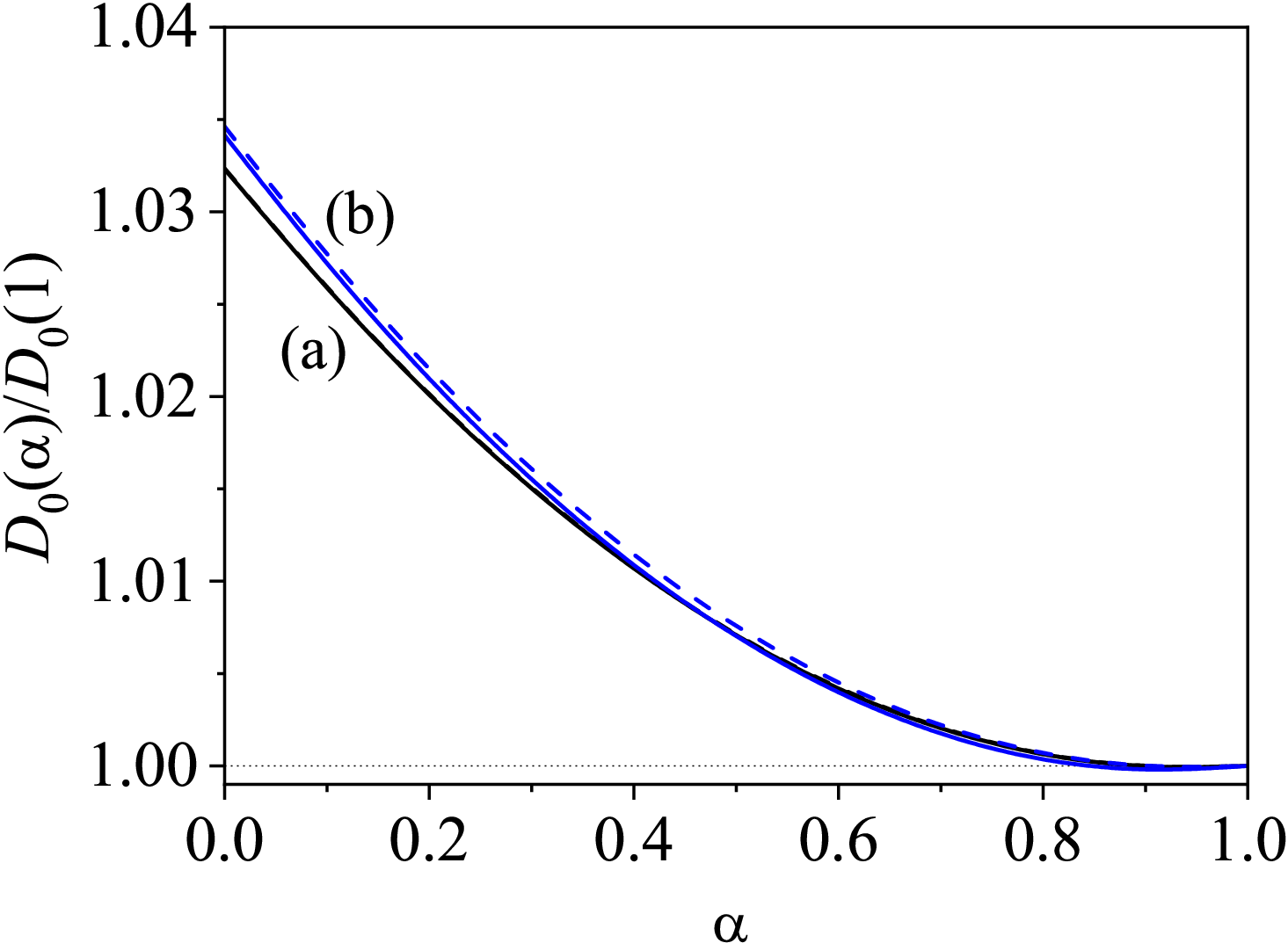}
\caption{(color online) Plot of reduced tracer diffusion coefficient $D_{0}(\alpha)/D_{0}(1)$ as a function of the (common)
coefficient of restitution $\alpha$ for the systems $(m_0/m=4,\sigma_0/\sigma=2)$ (a) and
$(m_0/m=0.5,\sigma_0/\sigma=0.8)$ (b) in the case of a three-dimensional gas with $\phi=0.1$ and $T_\text{ex}^*=0.1$. The solid lines correspond to
the second Sonine approximation while the dashed lines refer to the first Sonine approximation. Here,
$D_{0}(1)$ is the elastic value of the tracer diffusion coefficient consistently obtained in each approximation. \label{fig1}}
\end{figure}
\begin{figure}
\includegraphics[width=0.7 \columnwidth,angle=0]{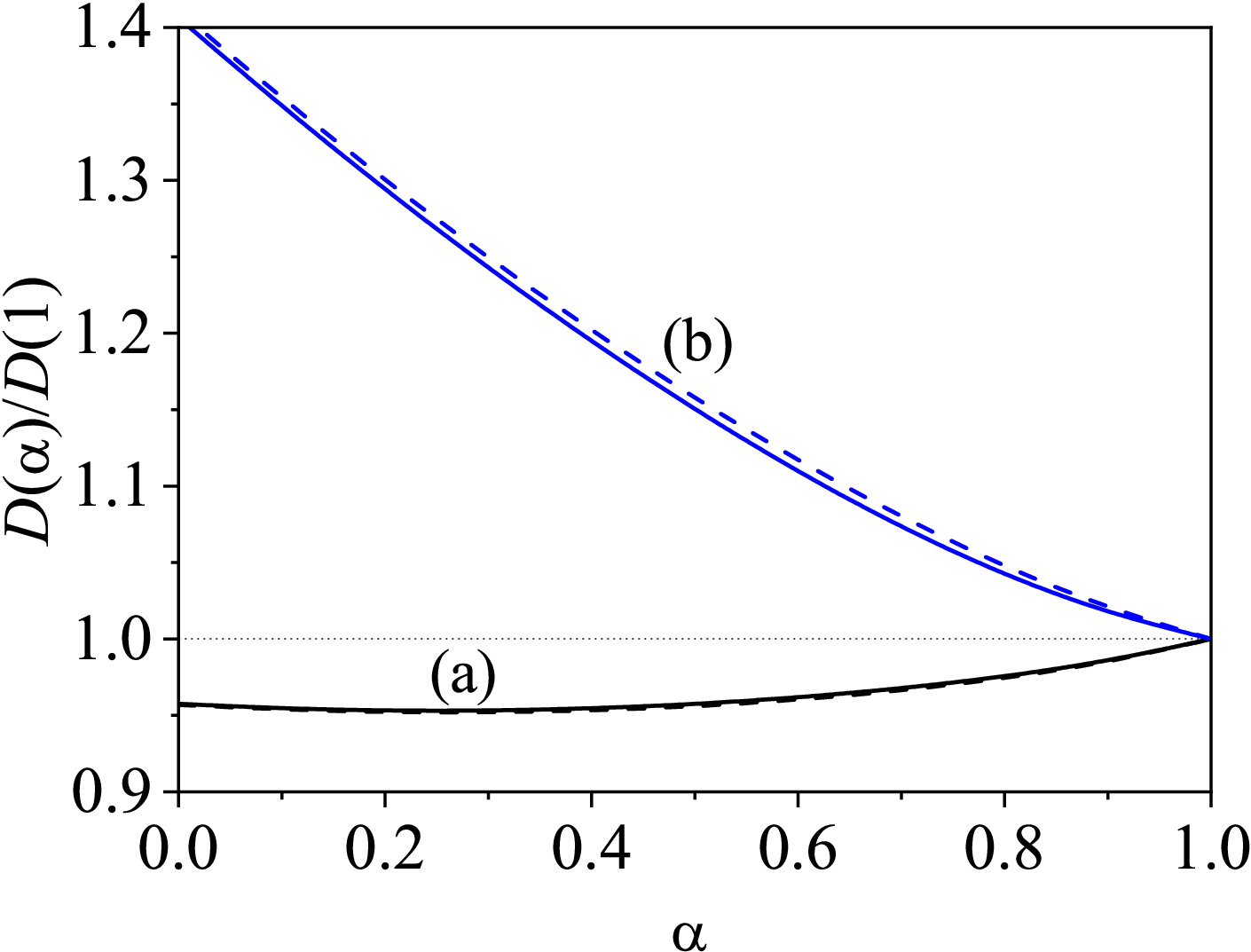}
\caption{(color online) Plot of reduced mutual diffusion coefficient $D(\alpha)/D(1)$ as a function of the (common)
coefficient of restitution $\alpha$ for the systems $(m_0/m=4,\sigma_0/\sigma=2)$ (a) and
$(m_0/m=0.5,\sigma_0/\sigma=0.8)$ (b) in the case of a three-dimensional gas with $\phi=0.1$ and $T_\text{ex}^*=0.1$. The solid lines correspond to
the second Sonine approximation while the dashed lines refer to the first Sonine approximation. Here,
$D(1)$ is the elastic value of the tracer diffusion coefficient consistently obtained in each approximation.\label{fig2}}
\end{figure}
\begin{figure}
\includegraphics[width=0.7 \columnwidth,angle=0]{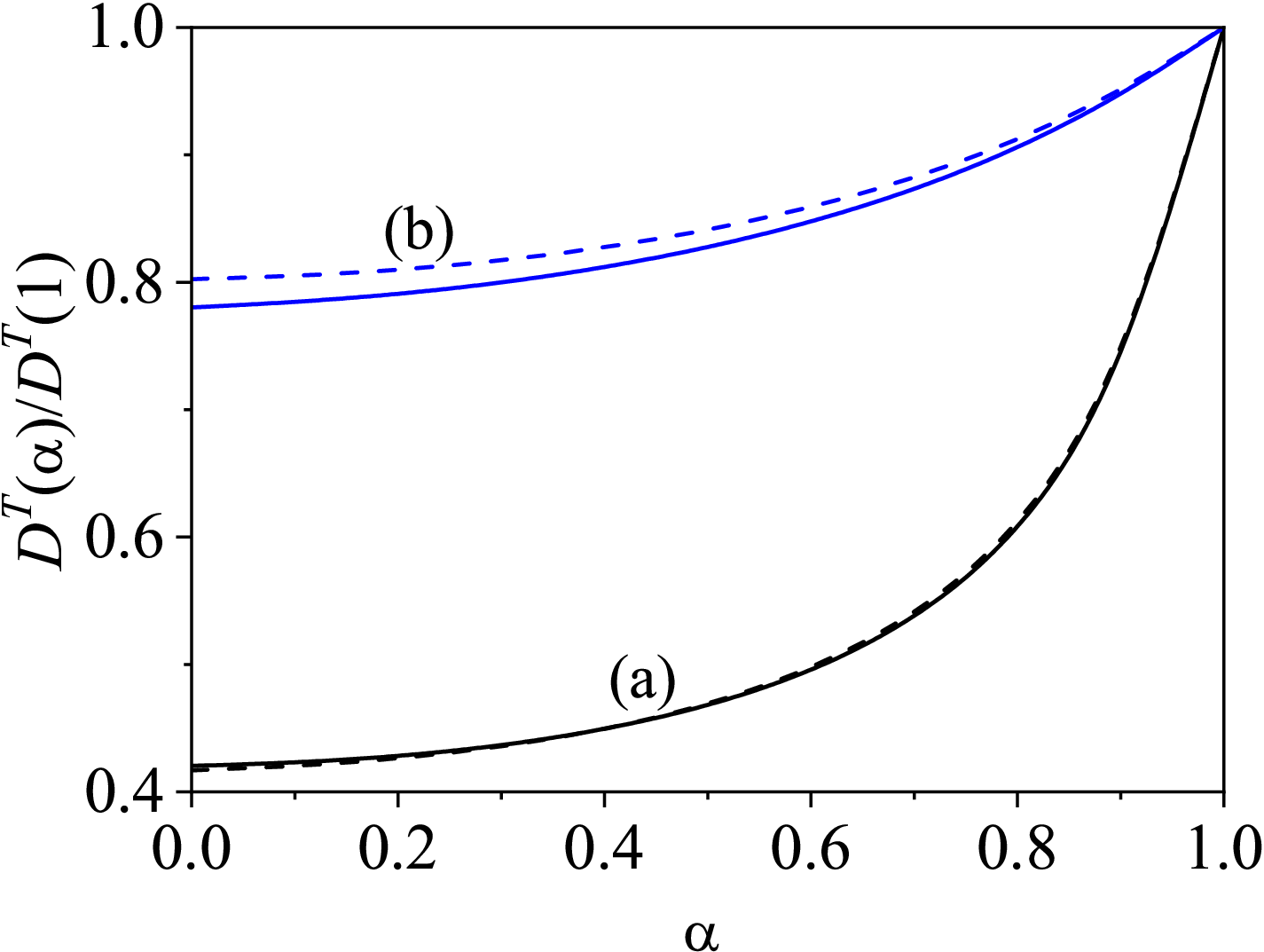}
\caption{(color online) Plot of reduced thermal diffusion coefficient $D^T(\alpha)/D^T(1)$ as a function of the
(common) coefficient of restitution $\alpha$ for the systems $(\mu=4,\omega=2)$ (a)
and $(\mu=0.5,\omega=0.8)$ (b) in the case of a three-dimensional gas with $\phi=0.1$ and $T_\text{ex}^*=0.1$. The solid lines
correspond to the second Sonine approximation while the dashed lines refer to the first Sonine approximation. Here, $D^T(1)$ is the elastic value of the thermal diffusion coefficient consistently obtained in each
approximation. \label{fig3}}
\end{figure}
\begin{figure}
\includegraphics[width=0.7 \columnwidth,angle=0]{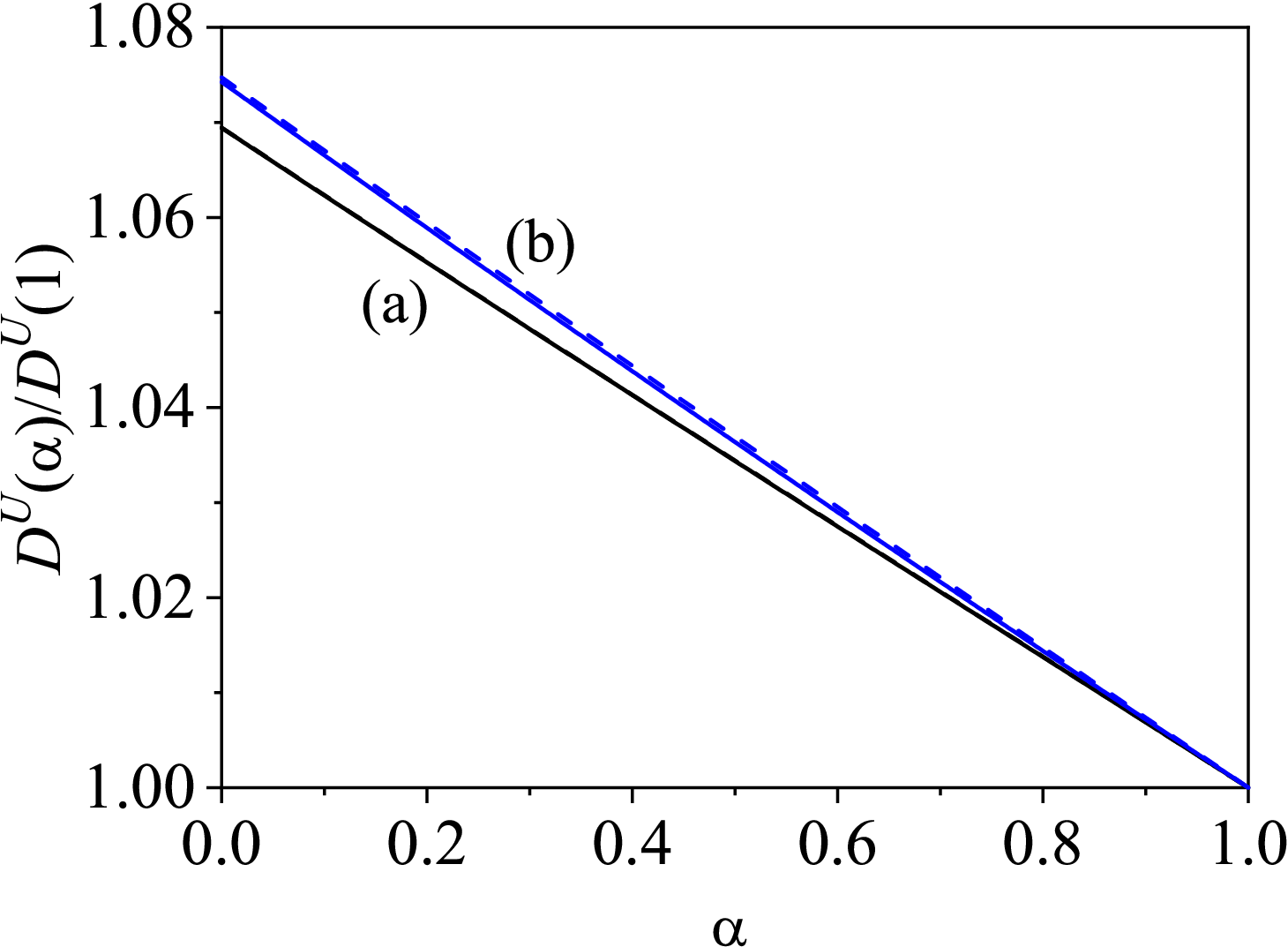}
\caption{(color online) Plot of reduced velocity diffusion coefficient $D^U(\alpha)/D^U(1)$ as a function of the
(common) coefficient of restitution $\alpha$ for the systems $(\mu=4,\omega=2)$ (a)
and $(\mu=0.5,\omega=0.8)$ (b) in the case of a three-dimensional gas with $\phi=0.1$ and $T_\text{ex}^*=0.1$. The solid lines
correspond to the second Sonine approximation while the dashed lines refer to the first Sonine approximation. Here, $D^U(1)$ is the elastic value of the velocity diffusion coefficient consistently obtained in each
approximation. \label{fig4}}
\end{figure}

The expressions derived in Sec.\ \ref{sec4} for the (reduced) diffusion coefficients depend on many parameters: $\left\{d, m_0/m, \sigma_0/\sigma, \al, \al_0, \phi, T_\text{ex}^*\right\}$. This complexity appears in the elastic limit as well, except for the dependence on the coefficients of restitution $\al$ and $\al_0$. Thus, as we have done in previous papers, \cite{GV09,GV12} we normalize the transport coefficients with respect to their values for elastic collisions ($\al=\al_0=1$). This will show us more clearly the impact of inelasticity in collisions on the diffusion transport coefficients. Also, for the sake of simplicity, we consider the physical case of hard spheres ($d=3$) and in some plots we take a common coefficient of restitution $\al=\al_0$. This reduces the parameter space of the system to five independent quantities: $\left\{m_0/m, \sigma_0/\sigma, \al, \phi, T_\text{ex}^*\right\}$.

In order to get the explicit dependence of the transport coefficients on the above parameters one has to give the forms of the pair correlation functions $\chi^{(0)}$ and $\chi_0^{(0)}$ for a three-dimensional system. In this case, a good approximation for $\chi^{(0)}$ is provided by the Carnahan-Starling form \cite{CS69}
\begin{equation}
\label{5.1} 
\chi^{(0)}=\frac{1-\frac{1}{2}\phi}{(1-\phi)^3},
\end{equation}
while the intruder-gas pair correlation function is given by \cite{B70}
\begin{equation}
\label{5.2}
\chi_0^{(0)}=\frac{1}{1-\phi}+3\frac{\omega}{1+\omega}\frac{\phi}{(1-\phi)^2}+2
\frac{\omega^2}{(1+\omega)^2}\frac{\phi^2}{(1-\phi)^3}.
\end{equation}
The expression for the chemical potential of the intruder consistent with the approximation (\ref{5.2}) is \cite{RG73}
\begin{eqnarray}
\label{5.3} 
\frac{\mu_0}{T}&=&C_3+\ln n_0-\ln (1-\phi)+3\omega \frac{\phi}{1-\phi}\nonumber\\
& & +3\omega^2\left[\ln (1-\phi)+\frac{\phi(2-\phi)}{(1-\phi)^2}\right]\nonumber\\
& & -\omega^3\left[2\ln (1-\phi)+\frac{\phi(1-6\phi+3\phi^2)}{(1-\phi)^3}\right],
\end{eqnarray}
where $C_3$ is a constant.

Figures \ref{fig1}--\ref{fig4} show the dependence of the diffusion transport coefficients on inelasticity for two different systems. As said before, each transport coefficient has been reduced with respect to its elastic value consistently obtained in each approximation. The dashed and solid lines refer to the first and second Sonine approximations, respectively. We observe first that in general both Sonine approximations yield quite similar results even for strong inelasticity and quite disparate values of the mass and/or diameter ratios. These results clearly differ from those reported in the case of \emph{dry} (no gas phase) granular mixtures [see, for instance, Figs.\ 1-3 of Ref.\ \onlinecite{GV09} for the same mixture parameters] where both Sonine approximations lead to  quite different results in the range of large inelasticity when the tracer particles are lighter than the particles of the granular gas. This means that the Sonine polynomial expansion exhibits a better convergence for granular suspensions than in the dry granular case. 

With respect to the influence of inelasticity on diffusion coefficients, we observe that the impact of $\al$ on the diffusion transport coefficients is in general relevant (since their forms differ clearly from their elastic counterparts), although less important than in the dry granular case. The comparison of Figs.\ \ref{fig1}--\ref{fig3} with Figs.\ 1--3 of Ref.\ \onlinecite{GV09} reveals important differences for diffusion between the results obtained with and without the gas-phase. Thus, while the (scaled) tracer diffusion coefficient $D_0(\al)/D_0(1)$ 
exhibits in some cases a non-monotonic dependence on inelasticity for gas-solid flows, this reduced coefficient increases with decreasing $\al$ in the dry granular limit. In addition, Fig.\ \ref{fig1} also shows the tiny dependence of  $D_0(\al)/D_0(1)$ on both the mass and diameter ratios. 
In the case of the (scaled) coefficients $D(\al)/D(1)$ and $D^T(\al)/D^T(1)$, qualitative differences between the predictions with and without the gas phase are observed for large mass and/or diameter ratios. At a more quantitative level, we find that the effect of $\al$ on those coefficients is more significant in dry granular mixtures. The reduced velocity diffusion coefficient $D^U(\alpha)/D^U(1)$ is not present in the dry case. Figure \ref{fig4} shows in general a weak influence of inelasticity on this coefficient.  

There are several competing effects involved in the mechanisms in the diffusion of tracer particles in a granular suspension. On the one hand, the collisional effects are accounted for in the cooling rate $\zeta^*$ and the (reduced) collision frequencies $\nu_i^*$. On the other hand, the influence of the interstitial gas on mass transport appears through the (reduced) friction coefficients $\gamma^*$ and $\gamma_0^*$, which forms also depend on the volume fraction $\phi$, the mass and diameter ratios, and the (reduced) background temperature $T_\text{ex}^*$. Given the size of the parameter space of the problem (six independent parameters), it is generally difficult to predict which effect (collisions between grains or interstitial gas) dominates.  

In summary, as already noted in previous papers on granular mixtures, \cite{GKG20} the diffusion transport coefficients for a binary granular suspension where one of the species is present in tracer concentration differ in general significantly from those obtained for dry granular mixtures. \cite{GDH07,GHD07,GV09} These differences between both systems increase with increasing inelasticity in some cases, and depending on the cases, the influence of both mass and diameter ratios can be important.

\subsection{Tracer diffusion coefficient. DSMC results}

It is quite apparent that the accuracy of the Sonine approximations must be assessed via a comparison with computer simulation results. Here, as in a previous work, \cite{GGABYG23} we have numerically solved the Enskog equation by means of the DSMC method \cite{B94} for the tracer particles immersed in a granular suspension in the homogeneous steady state. In this situation, $\Delta \mathbf{U}=\mathbf{0}$, $\nabla n=\nabla T=0$, and hence Eq.\ \eqref{2.17} yields
\beq
\label{5.4}
\frac{\partial x_0}{\partial t}=2d \frac{m_0 D_0}{\rho}\nabla^2 x_0,
\eeq
where use has been made of the constitutive equation \eqref{1.1}. Equation \eqref{5.4} is a standard diffusion equation with the time-independent diffusion coefficient $m_0D_0/\rho$. It follows that the mean square displacement of the tracer particles after a time interval $t$ is \cite{M89} 
\beq
\label{5.5}
\frac{\partial}{\partial t}\langle |\mathbf{r}(t)-\mathbf{r}(0)|^2\rangle=2 d \frac{m_0 D_0}{\rho}.
\eeq
Here, $|\mathbf{r}(t)-\mathbf{r}(0)|$ is the distance traveled by the tracer particles from $t=0$ until time $t$. Equation \eqref{5.5} is the Einstein formula relating the tracer diffusion coefficient $D_0$ with the mean square displacement. The relation \eqref{5.5} is also employed in Monte Carlo simulations to obtain the coefficient $D_0$.   

Since we want to numerically solve the Enskog–Lorentz kinetic equation \eqref{2.14b} under steady conditions, we slightly adapt the method proposed by Montanero and Garz\'o \cite{MG02} oriented to model freely cooling granular mixtures. The main changes to be taken into account are three: (i) the intruders are in tracer concentration, and so we do not need to account for collisions among intruder particles themselves; (ii) if a granular-intruder collision is accepted, only the velocities of the intruders are updated; and (iii) to account for the action of the interstitial fluid on grains we update the velocity $\mathbf{v}_k$ of every single grain of species $i$ (intruder or granular particles) after each time step $\delta t$ according to the rule: \cite{KG14}
\beq
\label{DSMC}
\mathbf{v}_k\to e^{-\gamma_i\delta t}\mathbf{v}_k+\left(\frac{6\gamma_i T_\text{ex} \delta t}{m_i}\right)^{1/2}\boldsymbol{\varpi}_k.
\eeq
Here, $\boldsymbol{\varpi}_k$ is a random vector of zero mean and unit variance. The Fokker-Planck equation is recovered when the time step $\delta t$ is much shorter than the mean free time between collisions. \cite{KG14}

\begin{figure}
\includegraphics[width=0.7 \columnwidth,angle=0]{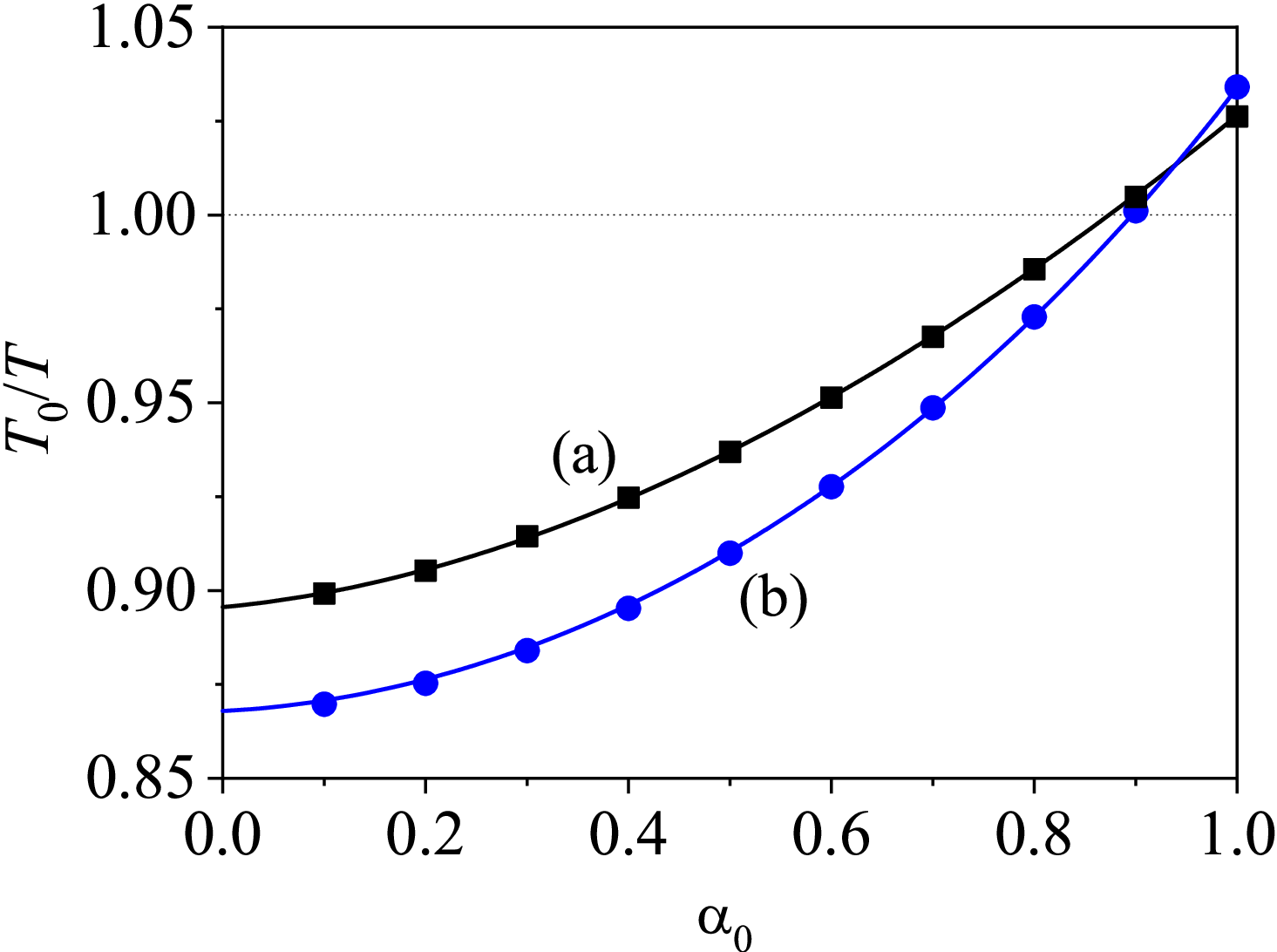}
\caption{(color online) Plot of the temperature ratio $T_0/T$ as a function of the intruder-grain coefficient of restitution $\alpha_0$ for the systems ($\mu=8$, $\omega=2$) (a) and  ($\mu=0.5$, $\omega=0.5^{1/3}$) (b) in the case of a three-dimensional gas with $\alpha=0.9$, $\phi=0.2$, and $T^*_\text{ex}=1$. The solid lines correspond to the theoretical results  whereas the symbols refer to DSMC results.\label{fig_DSMC1}}
\end{figure}

\begin{figure}
\includegraphics[width=0.7 \columnwidth,angle=0]{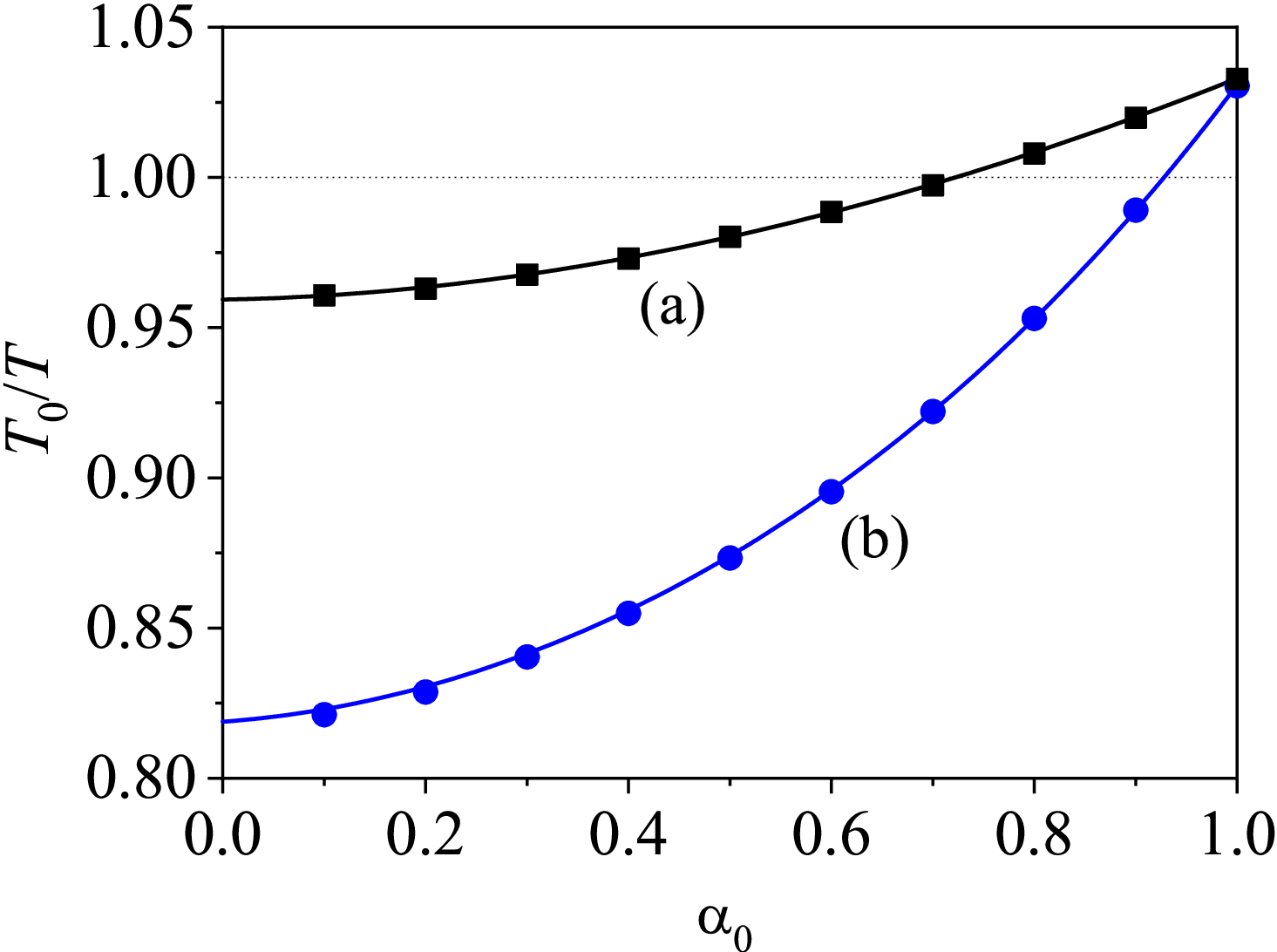}
\caption{(color online) Plot of the temperature ratio $T_0/T$ as a function of the intruder-grain coefficient of restitution $\alpha_0$ for the systems ($\mu=10$, $\omega=5$) (a) and  ($\mu=0.5$, $\omega=0.5$) (b) in the case of a three-dimensional gas with $\alpha=0.9$, $\phi=0.2$, and $T^*_\text{ex}=1$. The solid lines correspond to the theoretical results  whereas the symbols refer to DSMC results.\label{fig_DSMC2}}
\end{figure}

To complement the simulation results displayed in Ref.\ \onlinecite{GGABYG23} that were obtained considering a common coefficient of restitution $\alpha=\alpha_0$, we simulate here systems where $\alpha\neq\alpha_0$. Moreover, to assess the influence of $\alpha_0$ in the tracer diffusion coefficient $D_0$, the same mixtures as in Sec. V.C of Ref.\ \onlinecite{GGABYG23} are studied. However, we consider denser suspensions ($\phi=0.2$) to enlarge the difference between the first and second Sonine approximations to $D_0$. 

Figure \ref{fig_DSMC1} depicts the temperature ratio $T_0/T$ versus the coefficient of restitution $\alpha_0$ for $d=3$, $\alpha=0.9$, $T^*_\text{ex}=1$, and $\phi=0.2$. Two different mixtures with the same mass density ($m/\sigma^3 = m_0/\sigma_0^3$) are considered. As in the case of a common coefficient of restitution, we find a small influence of the mechanical properties (masses and sizes) on $T_0/T$ when the mass densities are equal. However, we observe an enhance breakdown of energy equipartition when $\alpha\neq\alpha_0$, as expected. This occurs because $T_0\sim T$ when $\alpha_0\sim\alpha$ since the loss of energy is comparable in both species. Using similar arguments to those presented in Sec. VI.B of Ref.\ \onlinecite{GGABYG23}, we conclude that for moderately dense suspensions, the action of the external bath is accentuated when $\omega$ increases and/or $\mu$ decreases. Although a combined effect of both masses and sizes is observed, we can conclude that the influence of the diameter ratio $\omega$ on the breakdown of energy equipartition is more relevant than the one caused by the mass ratio $\mu$. This is due to the action of the interstitial gas resulting from gas-solid collisions, whose dependence on particle size ($\gamma_{0,\text{St}}\propto \sigma_0/\sigma$) can be inferred from the change of an effective surface, which overcomes the inertial forces determined by particle masses. For this reason, the bath thermalizes the larger species more effectively causing its temperature to rise and therefore, the curve (a) of Fig.\ \ref{fig_DSMC1} ($\omega>1$) appears above the curve (b) 
of Fig.\ \ref{fig_DSMC1} ($\omega<1$). Furthermore, we note that a departure from energy equipartition occurs when $\alpha_0=1$. In contrast to Fig. 4 of Ref.\ \onlinecite{GGABYG23}, where $T_\text{ex} = T_0 = T$ for $\alpha_0=1$, here we still observe energy dissipation since $\alpha\neq1$. Specifically, the loss of kinetic energy when  $\alpha_0=1$ ($\alpha=0.9$) is greater for grains than for intruders, resulting in $T_0 > T$. Figure \ref{fig_DSMC1} also highlights an excellent agreement between theory and DSMC results, ensuring the reliability of the Maxwellian approximations \eqref{2.16} and \eqref{2.26} taken for the distribution functions.

We now consider two different mixtures in Fig.\ \ref{fig_DSMC2}. In contrast to the prior case depicted in Fig.\ \ref{fig_DSMC1}, the systems under consideration possess different mass densities, accentuating the disparity between the temperature ratios. Similar to the observations made in Fig.\ \ref{fig_DSMC1}, we note that for larger intruders, the thermalization of the temperature by the bath occurs more efficiently. Thus, the action of the bath facilitates the classification of diverse mixtures based on their mechanical properties and deviation from energy equipartition. For moderate mass densities ($\mu\sim\omega$), the curves exhibited in both Figs.\ \ref{fig_DSMC1} and \ref{fig_DSMC2} display a bottom-to-top ordering as the value of $\omega$ increases. We can also analyze the temperature ratio $T_0/T$ in the specific case in which $\alpha=\alpha_0=0.9$. While energy equipartition practically holds for the system considered in Fig.\ \ref{fig_DSMC1}, a more noticeable breakdown appears in Fig.\ \ref{fig_DSMC2}. This result has been previously discussed in Ref.\ \onlinecite{GGABYG23} and is due in part to the mechanical differences between the mixtures. We can conclude again that the theoretical results perfectly reproduce the DSMC simulations, as demonstrated in Fig.\ \ref{fig_DSMC2}.

\begin{figure}
\includegraphics[width=0.7 \columnwidth,angle=0]{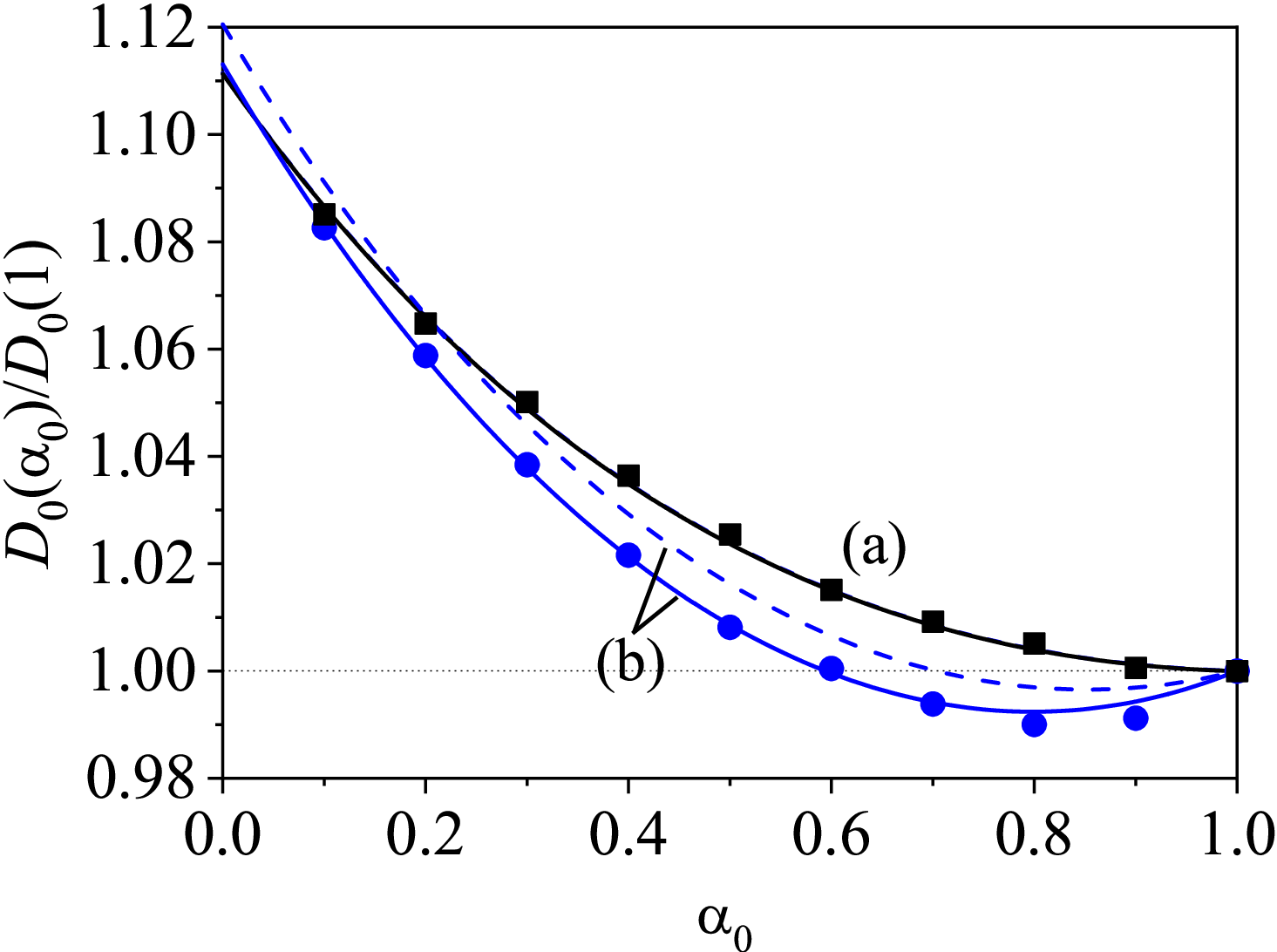}
\caption{(color online) Plot of the reduced tracer diffusion coefficient $D_0(\alpha_0)/D_0(1)$ as a function of the intruder-grain coefficient of restitution $\alpha_0$ for the systems ($\mu=8$, $\omega=2$) (a) and  ($\mu=0.5$, $\omega=0.5^{1/3}$) (b) in the case of a three-dimensional gas with $\alpha=0.9$, $\phi=0.2$, and $T^*_\text{ex}=1$.  Solid and dashed lines are for the second and first Sonine approximations, respectively, whereas the symbols refer to DSMC results. Here, $D_0(1)$ is the
value of the tracer diffusion coefficient consistently obtained
in each approximation  when $\alpha_0=1$.\label{fig_DSMC3}}
\end{figure}

\begin{figure}
\includegraphics[width=0.7 \columnwidth,angle=0]{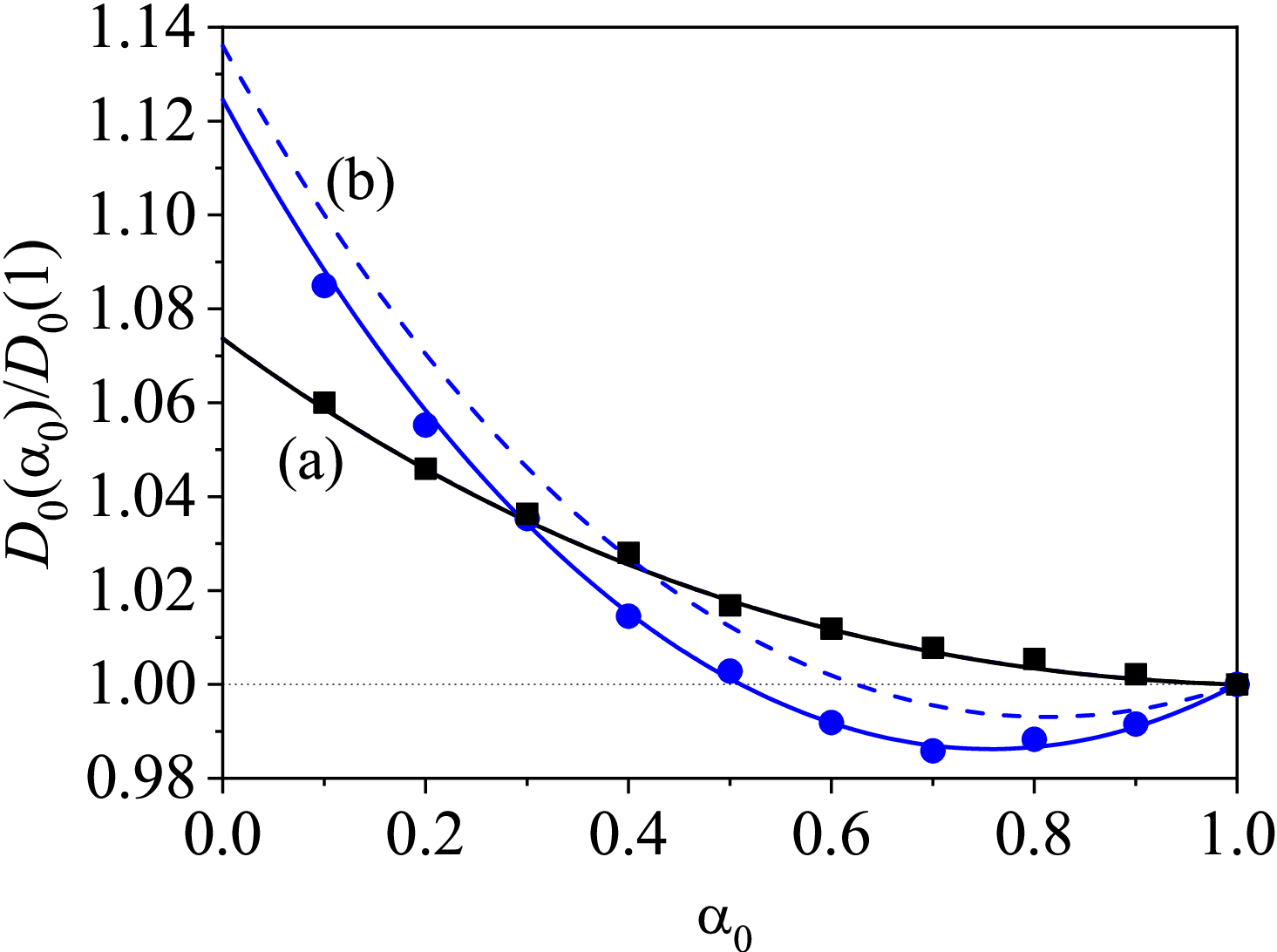}
\caption{(color online) Plot of the reduced tracer diffusion coefficient $D_0(\alpha_0)/D_0(1)$ as a function of the intruder-grain coefficient of restitution $\alpha_0$ for the systems ($\mu=10$, $\omega=5$) (a) and  ($\mu=0.5$, $\omega=0.5$) (b) in the case of a three-dimensional gas with $\alpha=0.9$, $\phi=0.2$, and $T^*_\text{ex}=1$. Solid and dashed lines are for the second and first Sonine approximations, respectively, whereas the symbols refer to DSMC results. Here, $D_0(1)$ is the
value of the tracer diffusion coefficient consistently obtained
in each approximation  when $\alpha_0=1$.\label{fig_DSMC4}}
\end{figure}

Finally, the accuracy of the first and second Sonine approximations to the tracer diffusion coefficient $D_0$ is evaluated by performing DSMC simulations. The ratio $D_0(\alpha_0)/D_0(1)$ is plotted in Figs.\ \ref{fig_DSMC3} and \ref{fig_DSMC4} as a function of the intruder-grain coefficient of restitution $\alpha_0$ ($\al\neq \al_0$). Here, $D_0(1)$ denotes the tracer diffusion coefficient value when $\alpha_0=1$. Initially, it is observed that the tracer diffusion coefficient $D_0$ exhibits minimal sensitivity to the mechanical properties of the particles when the mass densities are equal. However, in Fig.\ \ref{fig_DSMC4}, significant qualitative differences in the behavior of $D_0$ for the two mixtures are evident. It is also quite apparent that a non-monotonic trend in the $\al_0$-dependence of $D_0$ is consistently observed in both cases. Furthermore, a deliberate selection of a density value of $\phi=0.2$ and a reduced background temperature of $T_\text{ex}^*=1$ allows for a larger discrepancy between both Sonine approximations, enabling a clearer assessment of their accuracy through simulations. As seen in the dry granular case when $\alpha=\alpha_0$, the convergence between both Sonine approximations is practically achieved when the intruders possess both greater mass and size compared to the grains. Although noticeable deviations are observed as $\omega$ and $\mu$ decrease, these discrepancies are smaller than those observed in the dry granular case. \cite{G19} The distinctions between the Sonine approximations are used to evaluate their accuracy in comparison to simulations. Thus, Figs.\ \ref{fig_DSMC3} and \ref{fig_DSMC4} provide clear evidence that the second Sonine approximation for the ratio $D_0(\alpha_0)/D_0(1)$ improves the theoretical predictions of the first Sonine approximation, yielding excellent agreement with the simulation results.

\section{Stability analysis of the homogeneous steady state}
\label{sec7}

The knowledge of the diffusion transport coefficients along with the Navier--Stokes transport coefficients of the granular suspension opens up the possibility of solving the hydrodynamic equations for the densities $n_0$ and $n$, the flow velocity $\mathbf{U}$, and the granular temperature $T$ for states near to the homogeneous steady state. This solution will give us information on the linear stability of the above homogeneous state. The linearized hydrodynamic equations are obtained when one substitutes the constitutive equations for the mass flux, pressure tensor, and heat flux into Eqs.\ \eqref{2.11}--\eqref{2.13} and \eqref{2.17}. These equations are given by
\begin{widetext}
\beq
\label{6.1} 
D_t n_0+ n_0\nabla\cdot \mathbf{U}=\nabla \cdot \Bigg(\frac{m_0}{\rho}D_0\nabla  n_0+\frac{m_0}{\rho}D\nabla n+\frac{\rho}{m_0 T}D^T\nabla T\Bigg),
\eeq
\begin{equation}
\label{6.2}
D_t n+n\nabla\cdot\mathbf{U}=0,
\end{equation}
\begin{equation}
\label{6.3}
D_t U_i+\rho^{-1}\partial_i p=\rho^{-1}\partial_j\left[\eta\left(\partial_iU_j+\partial_jU_i-\frac{2}{d}\delta_{ij}
\nabla\cdot\mathbf{U}\right)+\eta_\text{b}\delta_{ij}\nabla\cdot\mathbf{U}\right]-\gamma \Delta \mathbf{U},
\end{equation}
\beqa
\label{6.4}
\Big(D_t+2\gamma\left(1-\theta^{-1}\right)+\zeta^{(0)}\Big)T&=&\frac{2}{dn}\nabla\cdot\left(\kappa\nabla T+\mu\nabla n\right)+\frac{2}{dn}\bigg[\eta\left(\partial_iU_j+\partial_jU_i-\frac{2}{d}\delta_{ij}\nabla\cdot\mathbf{U}\right)\nonumber\\
& & +
\eta_\text{b}\delta_{ij}\nabla\cdot\mathbf{U}\bigg]\partial_iU_j-T\zeta_U\nabla\cdot\mathbf{U}-\frac{2}{dn}p\nabla\cdot\mathbf{U}.
\eeqa
\end{widetext}
Here, $\eta$ and $\eta_\text{b}$ are the shear and bulk viscosities, respectively, $\kappa$ is the thermal conductivity, $\mu$ is the diffusive heat conductivity, and $\zeta_U$ is the first-order contribution to the cooling rate. All these quantities have been determined in Ref.\ \onlinecite{GGG19a} as functions of the coefficient of restitution $\al$, the volume fraction $\phi$, and the background temperature $T_\text{ex}$. 

In the homogeneous steady state, the hydrodynamic fields take the homogeneous steady values $n_{0,\text{s}}\equiv\text{const.}$, $n_\text{s}\equiv\text{const.}$, $T_\text{s}\equiv\text{const.}$, and $\mathbf{U}_g=\mathbf{U}\equiv \mathbf{0}$. The subscript $s$ means that the quantities are evaluated in the steady state. For small spatial gradients, we assume that the deviations $\delta y_{\beta}(\mathbf{r},t)=y_{\beta}(\mathbf{r},t)-y_{\beta,\text{s}}$ are small, where $\delta y_{\beta}(\mathbf{r},t)$ denotes the deviations of the hydrodynamic fields $\left\{y_\beta; \beta=1,\cdots, d+3\right\}=\left\{n_0,n, \mathbf{U}, T\right\}$ from their values in the homogeneous steady state. Moreover, as usual we also suppose that the interstitial fluid is not perturbed and hence, $\mathbf{U}_g=\mathbf{U}=\mathbf{0}$.

It is quite apparent that the hydrodynamic equations \eqref{6.2}--\eqref{6.4} for $n$, $\mathbf{U}$, and $T$ are decoupled from Eq.\ \eqref{6.1} for $n_0$. This is a consequence of the tracer limit. Equations \eqref{6.2}--\eqref{6.4} were already solved in the Fourier space in Ref.\ \onlinecite{GGG19a} showing that this set of equations is linearly stable. This means that the corresponding perturbations ($\delta n$, $\delta \mathbf{U}$, $\delta T$) tend to zero  for sufficiently long times. In this case, Eq.\ \eqref{6.1} becomes an autonomous differential equation for $\delta n_0$. To solve it, as in Ref.\ \onlinecite{GGG19a}, we introduce 
the following space and time variables:
\beq
\label{6.5}
t'=\frac{1}{2}n_\text{s}\sigma^{d-1}\sqrt{\frac{T_\text{s}}{m}}t,\quad \mathbf{r}'=\frac{1}{2}n_\text{s}\sigma^{d-1}\mathbf{r}.
\eeq
The dimensionless time scale $t'$ measures the average number of collisions per particle in the time interval between 0 and $t$. The unit length $\mathbf{r}'$ is proportional to the mean free path of solid particles. Moreover, 
\begin{equation}
\label{6.6}
\rho_{0,\mathbf{k}}(\tau)=\frac{\delta n_{0,\mathbf{k}}(\tau)}{n_{0,\text{s}}}
\end{equation}
denotes the Fourier transformed dimensionless tracer density
where 
\beq
\label{6.6b}
\delta n_{0,\mathbf{k}}(\tau)=\int\dd\mathbf{r}'\text{e}^{-i\mathbf{k}\cdot\mathbf{r}'}\delta n_0(\mathbf{r}',\tau).
\eeq
Note that here the wave vector $\mathbf{k}$ is dimensionless. In terms of the above dimensionless variables, $\rho_{0,\mathbf{k}}(\tau)$ verifies the equation 
\beq
\label{6.7}
\frac{\partial}{\partial t'}\rho_{0,\mathbf{k}}=-\frac{1}{2\sqrt{2}}\frac{m}{m_0}D_0^* k^2 \rho_{0,\mathbf{k}},
\eeq
whose solution is 
\beq
\label{6.8}
\rho_{0,\mathbf{k}}(t')=\rho_{0,\mathbf{k}}(0) \exp\left(-\frac{1}{2\sqrt{2}}\frac{m}{m_0}D_0^* k^2 t'\right).
\eeq
The first Sonine approximation to $D_0^*$ given by Eq.\ \eqref{4.12} clearly shows that $D_0^*[1]$ is always positive. As expected, a careful analysis of the dependence of the second Sonine approximation $D_0^*[2]$ on the parameter space of the system also shows that $D_0^*[2]>0$. As a consequence, $\rho_{0,\mathbf{k}}$ tends to zero for $t'\to \infty$ and hence, the homogeneous steady state for the system (tracer plus granular suspension) is \emph{linearly} stable.

\section{Thermal diffusion segregation}
\label{sec8}

Another nice and interesting application of the results derived in sections \ref{sec4} and \ref{sec5} is the study of thermal diffusion segregation of an intruder in a granular suspension. The understanding of the physical mechanisms involved in segregation and mixing of dissimilar grains is likely one of the most challenging problems in granular mixtures. In addition, the problem is relevant not only from a fundamental point of view but also from a more practical perspective. Needless to say, segregation has been widely studied in the past few years for dry granular mixtures. Among the different segregation problems, the so-called Brazil-nut effect (BNE) is perhaps one of the most famous examples of (size) segregation in vertically vibrated mixtures. In the Brazil-nut effect, a relatively large particle (intruder) tends to climb to the top of the sample against gravity. \cite{RSPS87, DRC93,CWHB96} On the other hand, another different experimental works \cite{SM98,HQL01} have observed the reverse buoyancy effect: the so-called reverse Brazil-nut effect (RBNE) where the intruder may also sink to the bottom of the sample under certain conditions. Several theoretical  \cite{AW98,JY02,TAH03,ATH06,YJ06,G06,G08a,GV09,G09} and computational \cite{GDH05,BRM05,BRM06,BEGS08,BS09} works have attempted to explain the segregation induced by the presence of a thermal gradient (thermal diffusion segregation) and/or the gravitational force.  All these results refer to situations where the effect of the interstitial gas (bath) on the dynamics of tracer particles has been neglected. Our objective here is to assess the influence of the background gas on the segregation criterion.

\begin{figure}[h]
\includegraphics[width=\columnwidth,angle=0]{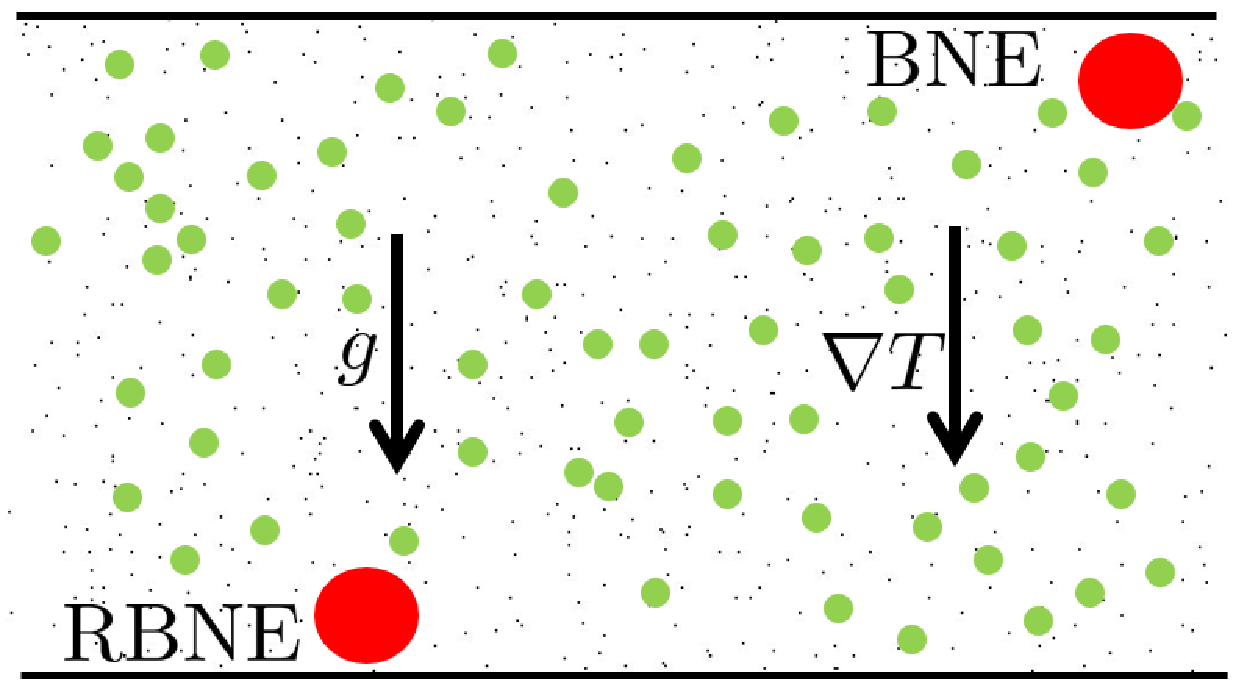}
\caption{(color online)  Illustration of the behavior of the gas-solid system. The small circles in the diagram represent granular particles, while the large circles depict intruders. All of them are immersed in a sea of molecular particles (very small particles). Specifically, the BNE (RBNE) effect occurs when the intruder rises (falls) to the top (bottom) plate of the system. \label{BNE}
}
\end{figure}

Thermal diffusion is the transport of matter caused by the presence of a thermal gradient. As a result of the motion of the species of the mixture, a steady state is finally achieved where the remixing effect of ordinary diffusion is balanced by the segregation effect arising from thermal diffusion. \cite{KCL87} The amount of segregation parallel to the thermal gradient may be characterized   by the so-called thermal diffusion factor $\Lambda$. This quantity is defined in an inhomogeneous non-convecting ($\mathbf{U}=\mathbf{U}_g=\mathbf{0}$) steady state with zero mass flux ($\mathbf{j}_0^{(1)}=\mathbf{0}$) through the relation
\beq
\label{8.1}
-\Lambda \frac{\partial \ln T}{\partial z}=\frac{\partial}{\partial z}\ln \Big(\frac{n_0}{n}\Big),
\eeq
where only gradients along the $z$ axis have been assumed for the sake of simplicity. Let us assume that the tracer is larger than the particles of the granular gas ($\sigma_0>\sigma$) and also that gravity and
the thermal gradient point in parallel directions (i.e., the bottom plate is hotter than the top plate, $\partial_z\ln T<0$, see Fig.\ \ref{BNE}). If the temperature gradient is sufficiently small, one can assume that $\Lambda$ is practically constant over the relevant ranges of composition and temperature of the system. Thus, according to Eq.\ \eqref{8.1}, when $\Lambda>0$ the intruders (or tracer particles) tend to rise with respect to the particles of the granular gas (i.e, $\partial_z \ln (n_0/n)>0$) and hence, the intruder particles tend to accumulate near the cold plate (BNE). On the other hand, when $\Lambda<0$ the intruders tend to fall with respect to the particles of the granular gas (i.e, $\partial_z \ln (n_0/n)<0$) and so, the intruder particles tend to accumulate near the hot plate (RBNE).

Let us determine the thermal diffusion factor. The mass flux $j_{0,z}^{(1)}$ of tracer particles is given by Eq.\ \eqref{1.1} with $\Delta \mathbf{U}=\mathbf{0}$. Since $j_{0,z}^{(1)}=0$ in the steady state and $\mathbf{U}=\mathbf{U}_g=\mathbf{0}$, then the momentum balance equation \eqref{2.12} yields $\partial_z p=-\rho g$. In dimensionless form, this relation can be written as
\beq
\label{8.2}
p^*+\left(p^*+\phi \frac{\partial p^*}{\partial \phi}\right)\frac{\partial_z \ln n}{\partial_z \ln T}
=-g^*,
\eeq
where $p^*=p/(nT)$ and $g^*=\rho g/n \partial_z T<0$ is a dimensionless parameter measuring the gravity relative to the thermal gradient. The condition $j_{0,z}^{(1)}=0$ along with Eq.\ \eqref{8.2} allows one to express the factor $\Lambda$ in terms of the (dimensionless) diffusion coefficients $D_0^*$, $D^*$, and $D^{T*}$ as  
\beq
\label{8.3}
\Lambda=\frac{\xi D^{T^*}-\left(D_{0}^*+D^*\right)\left(g^*+p^*\right)}{\xi D_{0}^*},
\eeq
where $\xi=p^*+\phi \partial_\phi p^*$. The dependence of $\Lambda$ on the parameters of the system can be obtained when one substitutes the expressions of the (reduced) pressure $p^*$ [Eq.\ \eqref{3.15}] and the (reduced) diffusion transports coefficients into Eq.\ \eqref{8.3}. Since we have seen that the first Sonine approximation to these coefficients is quite accurate in the complete range of values of the coefficients of restitution, we will consider henceforth those expressions for the sake of simplicity. In this approximation, the coefficients $D^{T*}$, $D_0^*$, and $D^*$ are given by Eqs.\ \eqref{b8}, \eqref{b19}, and \eqref{b21}, respectively.      

As expected, it is quite apparent from Eq.\ \eqref{8.3} that the influence of the parameter space of the system on the sign of $\Lambda$ is quite complex, given the large number of parameters involved. On the other hand, since Eq.\ \eqref{b19} clearly shows that $D_0^*>0$, then the condition $\Lambda=0$ yields the relation 
\beq
\label{8.4}
\xi D^{T^*}=\left(D_{0}^*+D^*\right)\left(g^*+p^*\right).
\eeq
Equation \eqref{8.4} gives the curves delineating the regimes between the segregation towards the cold and the hot wall (BNE/RBNE transition). However, the explicit dependence of the segregation criterion \eqref{8.4} on the parameter space is still cumbersome even if one employs the first Sonine solution to the diffusion transport coefficients. For this reason, to disentangle the different competing mechanisms appearing in the segregation problem, it is first convenient to consider some simple situations where a more simple criterion may be offered.   

\subsection{Mechanically equivalent particles}

In this limit case ($m_0=m$, $\sigma_0=\sigma$, $\al=\al_0$, $\gamma=\gamma_0$), $\tau_0=1$ and according to Eqs.\ \eqref{b8}, \eqref{b19}, and \eqref{b21}, $D^{T^*}=0$ and $D_0^*=-D^*=(\nu_1^*+\gamma^*)^{-1}$. Thus, Eq.\ \eqref{8.4} holds for any value of the coefficients of restitution, masses, diameters, and volume fraction. In this case, as expected, no segregation is possible.

\subsection{Low density limit}

Let us consider a binary granular suspension in the low-density regime ($\phi\to 0$) and in the absence of gravity ($g^*=0$). In this regime, $p^*=1$, $\xi=1$, and the diffusion coefficients are given by Eqs.\ \eqref{4.14.0}--\eqref{4.14.2}. According to these expressions, the segregation criterion \eqref{8.4} yields  
\beqa
\label{8.5}
& & \left(\tau_0-\mu\right)\left(\nu_1^*-2\gamma^* \theta^{-1}-\frac{1}{2}\zeta^*+\gamma_0^*\right)
=\left(\nu_1^*+\gamma_0^*-\zeta^*\right)\nonumber\\
& & \times\left(\tau_0+\theta \Delta_{\theta,0}-\mu\right).
\eeqa
The segregation criterion \eqref{8.5} is still quite a complex relation in comparison with the one obtained in the absence of gas phase (dry limit) where $\Lambda=0$ when $\mu=\tau_0$. \cite{BRM05,G06} For elastic collisions, $\zeta^*=0$, $\theta=\tau_0=1$ and hence, Eq.\ \eqref{8.5} becomes
\beq
\label{8.6}
\left(1-\mu\right)\left(\nu_1^*-2\gamma^*+\gamma_0^*\right)
=\left(\nu_1^*+\gamma_0^*\right)\Delta_{\theta,0}.
\eeq
As for inelastic collisions, Equation \eqref{8.6} clearly differs from the one derived many years ago for molecular mixtures of hard spheres \cite{KCL87} where the line separating the segregation towards the cold or hot wall is simply $\mu=1$ in the first Sonine approximation. 

In the case that the inhomogeneities
in both the temperature and the mixture volume fraction
are neglected ($\partial_z T\to 0$) but gravity is different from zero, then $|g^*|\to \infty$ (thermalized systems). In this case, the segregation dynamics of the intruders is
essentially driven by the gravitational force. This situation (gravity dominates the temperature gradient) can be achieved in the shaken or sheared systems employed in numerical simulations and physical experiments. \cite{HQL01,WHP01,BEKR03,SBKR05} Under these conditions ($|g^*|\to \infty$),  the criterion \eqref{8.5} leads to $D_0^*+D^*=0$. In the low-density limit, one gets
\beq
\label{8.7}
\tau_0-\mu=\frac{\tau_0+\theta \Delta_{\theta,0}-\mu}{\nu_1^*-2\gamma^* \theta^{-1}-\frac{1}{2}\zeta^*+\gamma_0^*}\zeta^*.
\eeq
For elastic collisions, $\zeta^*=0$ and Eq.\ \eqref{8.7} reduces to $1-\mu=0$, as expected.  

\subsection{Dense granular suspensions}

\subsubsection{Absence of gravity ($|g^*|\to 0$)}

Let us now study the segregation dynamics for moderately dense systems. We analyze first situations where the influence of gravity can be neglected 
($|g^*|\to 0$). In this particular scenario, the determination of the segregation criterion is exclusively governed by the temperature gradient.

\begin{figure}
\includegraphics[width=0.7 \columnwidth,angle=0]{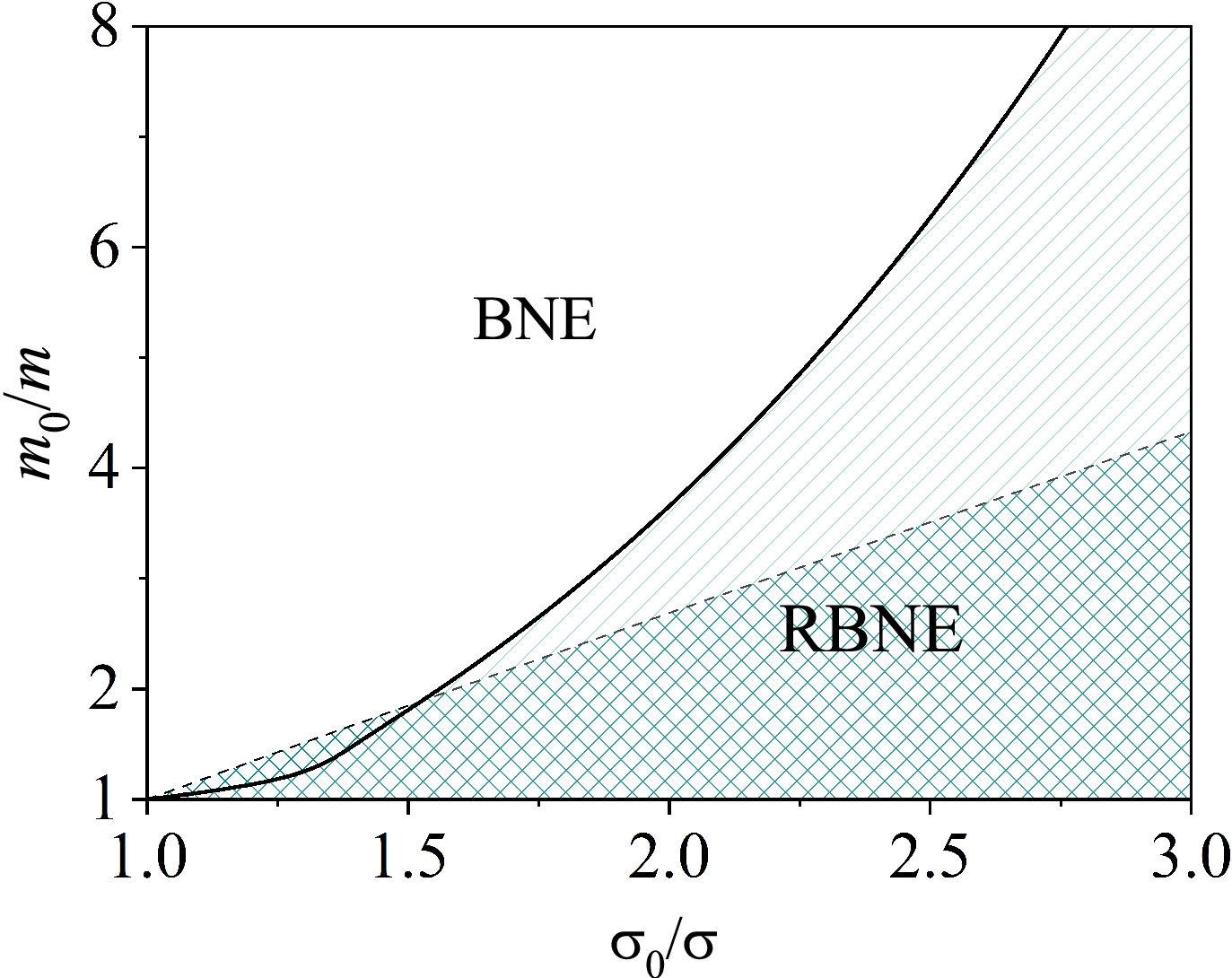}
\caption{(color online) Plot of the marginal segregation curve ($\Lambda=0$) for $d = 3$, $\alpha=\alpha_0=1$, $\phi=0.2$, $T^*_\text{ex}=0.1$, and $|g^*|\to 0$. The solid line corresponds to the segregation
criterion for molecular suspensions while the dashed line refers to the one derived for molecular
mixtures (no gas phase). The points below the curve correspond to $\Lambda<0$ (RBNE), while the points above the curve correspond to $\Lambda>0$ (BNE).\label{fig_seg1}}
\end{figure}

Let us consider first a three dimensional dry \emph{molecular} mixture ($\alpha=\alpha_0=1$) at moderate densities ($\phi=0.2$); this is illustrated by the dashed line of Fig.\ \ref{fig_seg1}. In this case, the particles of the molecular fluid (``grains'') that are close to the hot plate move faster. When tracer particles are introduced, they undergo substantial velocity changes upon colliding with the grains as they approach to the hot plate. As the size of the intruder particles increases (keeping the mass ratio $m_0/m$ fixed), collisions between grains and intruders become more frequent due to the larger effective surface area of the intruders. Consequently, these collisions cause the intruders to rebound and migrate back to the colder side of the system (RBNE). On the other hand, if we increase the mass ratio $m_0/m$ (keeping the size ratio $\sigma_0 /\sigma=$ fixed), the inertia of the massive particle leads it to retain its initial trajectory upon collision. Consequently, a greater number of collisions is required to modify the trajectory of the intruder. The application of a thermal gradient implicitly induces a gradient in particle density ($\partial_z n>0$) because the grains experience a higher frequency of collisions in the hot plate, resulting in a lower density compared to the cold plate. As a consequence of this density gradient, collisions between the intruder and grains increase in the cold plate, impeding the progress of the massive particle and causing it to migrate toward the hotter zone (BNE). 

When the mixture is immersed in an interstitial gas, we observe that the RBNE zone is expanded, as depicted in the solid line of Fig.\ \ref{fig_seg1}. This phenomenon can be understood by considering the arguments previously discussed in Sec. \ref{sec6}. First, the presence of the interstitial gas facilitates the thermalization of the temperatures $T$ and $T_0$, leading to an equipartition of the kinetic energy ($T=T_0=T_\text{ex}$). Thus, when grains are located in the hotter (colder) region, the interstitial gas tends to decrease (increase) its mean kinetic energy to establish thermal equilibrium. This effect is more pronounced when larger particles are present, as their increased surface amplifies the action of the interstitial gas ($\gamma_{0,\text{St}}\propto \sigma_0/\sigma$). When larger intruders are located in the hotter region, the interstitial gas undergoes a faster temperature decrease compared to that of the gas particles. This leads to an increased disparity in the mean velocities between grains and intruders. This discrepancy enables the grains to more effectively modify the trajectory of the intruders, causing them to move back towards the colder zone. As a result, the cooling effect caused by the presence of the bath combines with the size effect observed in a dry gas and the region where the RBNE effect occurs expands. Conversely, when the intruders are heavier than the grains, the influence of the interstitial gas on intruders is weakened ($\gamma_{0,\text{St}}\propto (m_0/m)^{-1}$), thereby accentuating the inertial effects observed in the dry case. This explains the reduction of the BNE region for suspensions when smaller size ratios are considered.

\begin{figure}
\includegraphics[width=0.7 \columnwidth,angle=0]{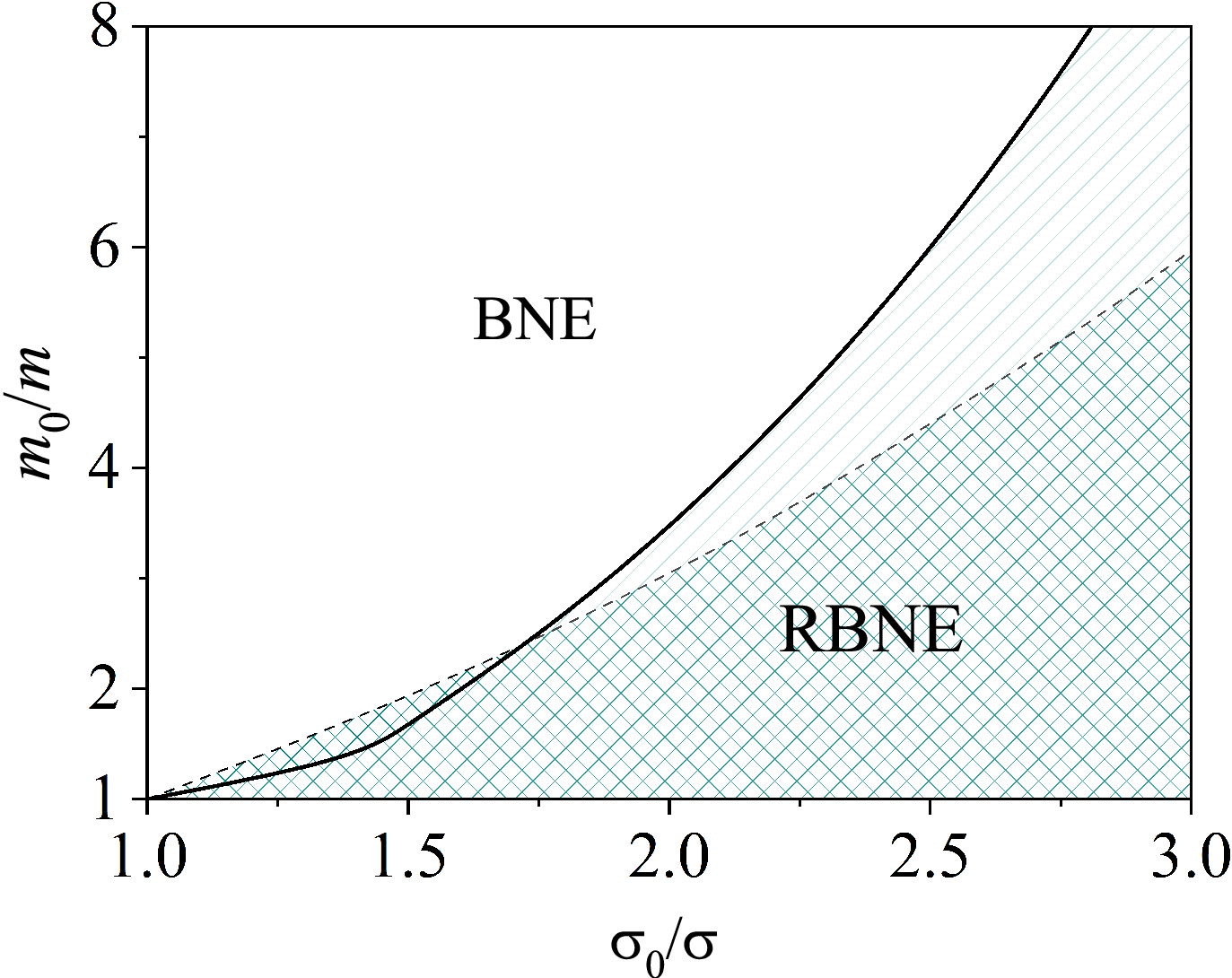}
\caption{(color online) Plot of the marginal segregation curve ($\Lambda=0$) for $d = 3$, $\alpha=\alpha_0=0.8$, $\phi=0.2$, $T^*_\text{ex}=0.1$, and $|g^*|\to 0$. The solid line corresponds to the segregation criterion for granular suspensions while the dashed line refers to the one derived for dry granular
mixtures. The points below the curve correspond to $\Lambda<0$ (RBNE), while the points above the curve correspond to $\Lambda>0$ (BNE).\label{fig_seg2}}
\end{figure}

The effects of the inelasticity in collisions on segregation are illustrated in Fig.\ \ref{fig_seg2} in the case of the common coefficient of restitution $\alpha=\alpha_0=0.8$. Notably, in contrast to the elastic case (depicted in the dashed line of Fig.\ \ref{fig_seg1}), we observe a pronounced enhancement of the RBNE for dry granular mixtures; this enhancement is more significant within the region where the intruders exhibit larger sizes. The inelastic nature of collisions leads to a consequent loss of kinetic energy and a decrease in temperature. Given the higher collision frequency in the hot plate, the dissipative nature of these collisions becomes more pronounced. This phenomenon bears similarity to the thermalizing influence exerted by the interstitial gas on particles located in the hotter zone. Consequently, when considering larger intruders, the BNE/RBNE diagram in a dry granular gas exhibits a closer resemblance to the corresponding diagram observed in a granular suspension. Interestingly, we find that the RBNE zone exhibits a reduction in size as the degree of inelasticity increases in the case of granular suspensions (solid line of Fig.\ \ref{fig_seg2}). This is attributed to the bath's ability to lower the temperature of the intruder particles near the hot plate, which in turn reduces their collision frequency. As a result, it becomes more challenging for the tracer particles to undergo collisions that result in rebound and subsequent return to the colder plate.


\subsubsection{Thermalized systems ($\partial_z T\to 0$)}

\begin{figure}
\includegraphics[width=0.7 \columnwidth,angle=0]{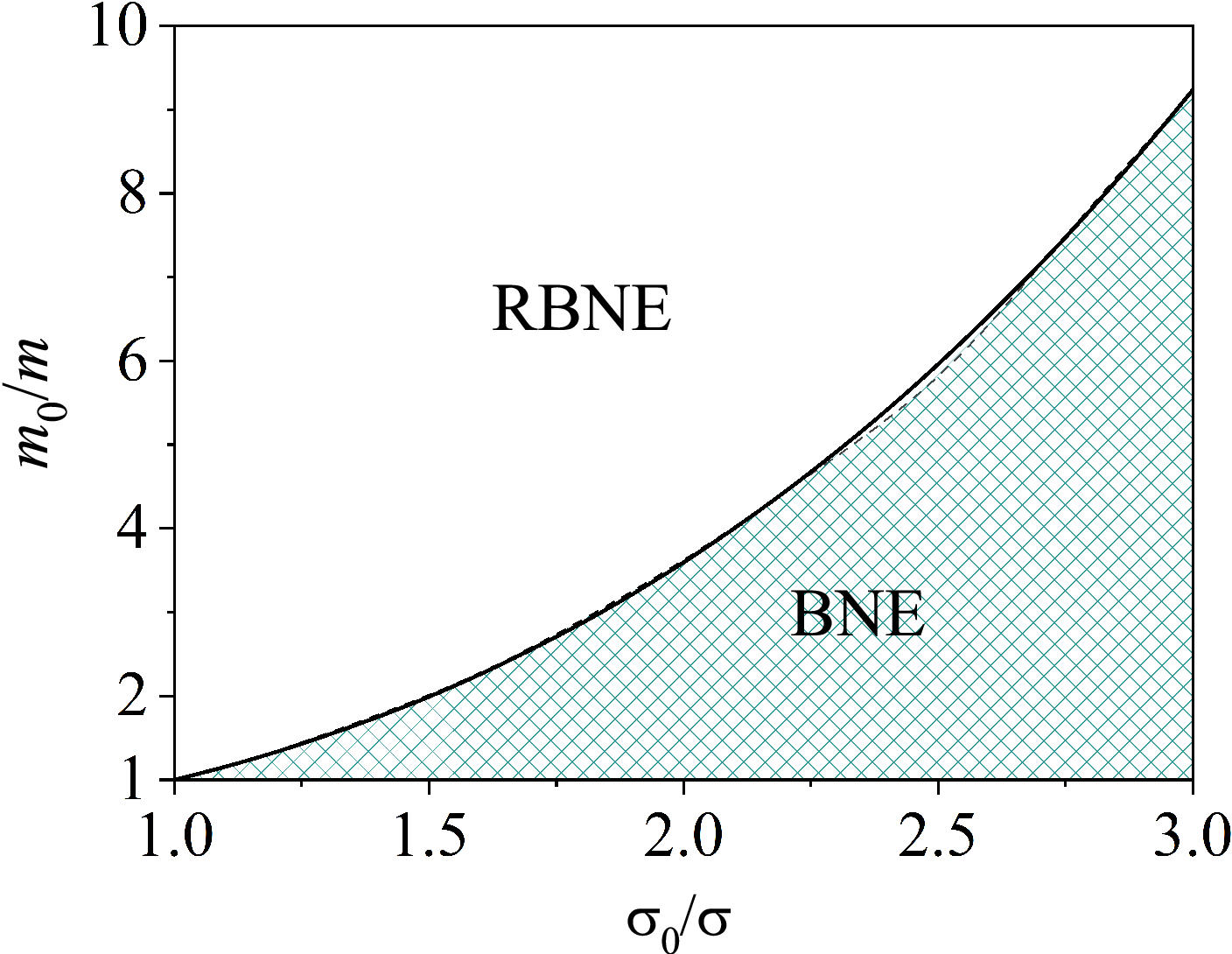}
\caption{(color online) Plot of the marginal segregation curve ($\Lambda=0$) for $d = 3$, $\alpha=\alpha_0=1$, $\phi=0.2$, $T^*_\text{ex}=0.1$, and $|g^*|\to \infty$. The solid line corresponds to the segregation
criterion for granular suspensions while the dashed line refers to the one derived for dry granular
mixtures. The points below the curve correspond to $\Lambda>0$ (BNE), while the points above the curve correspond to $\Lambda<0$ (RBNE).\label{fig_seg3}}
\end{figure}

\begin{figure}
\includegraphics[width=0.7 \columnwidth,angle=0]{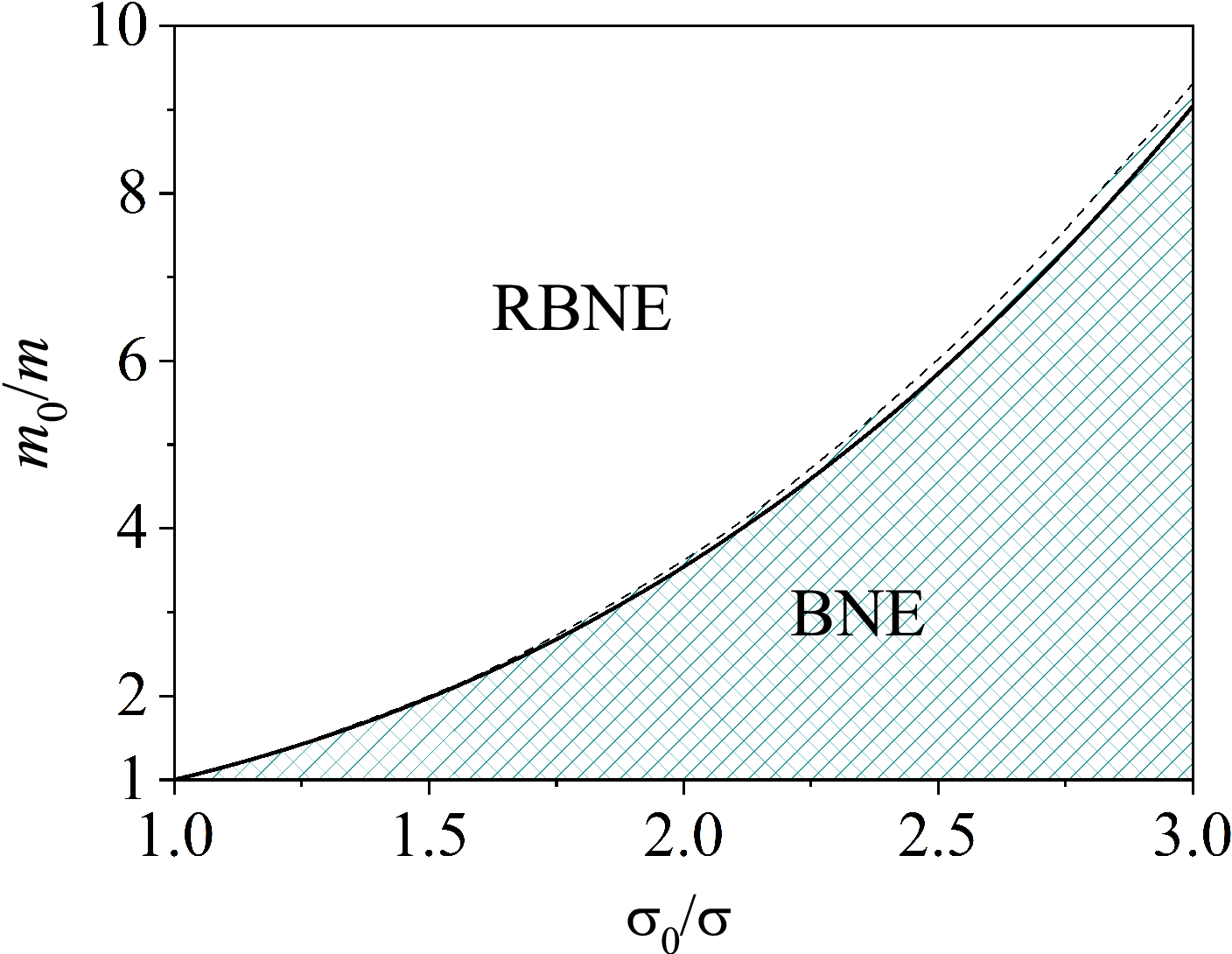}
\caption{(color online) Plot of the marginal segregation curve ($\Lambda=0$) for $d = 3$, $\alpha=\alpha_0=0.8$, $\phi=0.2$, $T^*_\text{ex}=0.1$, and $|g^*|\to \infty$. The solid line corresponds to the segregation
criterion for granular suspensions while the dashed line refers to the one derived for dry granular
mixtures. The points below the curve correspond to $\Lambda>0$ (BNE), while the points above the curve correspond to $\Lambda<0$ (RBNE).\label{fig_seg4}}
\end{figure}

We now consider the situation where the segregation dynamics is essentially driven by the gravitational force. In this case, $|g^*|\to \infty$ and hence, the temperature gradient can be neglected ($\partial_z T\to 0$). This condition can be achieved in experimental setups and numerical simulations involving shaken or sheared systems. \cite{HQL01, BEKR03, SBKR05, QDDWJZZ21, YAZMD23} Consequently, it is worthwhile to explore the influence of the interstitial gas on the segregation criterion in this particular case. Figures \ref{fig_seg3} and \ref{fig_seg4} show the marginal segregation curve for the same systems as depicted in Figs.\ \ref{fig_seg1} and \ref{fig_seg2}. As apparent from Figs.\ \ref{fig_seg3} and \ref{fig_seg4}, we observe that the impact of the gravitational force on segregation is much more important than that of the interstitial gas. In fact, both curves (with and without gas phase) practically collapse in a common curve. Furthermore, similar to granular mixtures driven by a stochastic external force (see Figs.\ 4 and 6 of Ref.\ \onlinecite{G09}), we see that the presence of gravity completely reverses the RBNE/BNE transition found in the phase diagrams in absence of gravity. Regarding the effect of inelasticity in collisions, our results clearly indicate that this effect is also negligible, even less significant than that observed in a driven granular mixtures. \cite{G09} 

Therefore, to provide an explanation for BNE/RBNE transition, we focus on the influence of gravity rather than dissipative dynamics or the force exerted by the interstitial gas. Evidently, when the intruder particles are heavier, they accumulate at the bottom of the container, leading to the observation of the RBNE effect. This explains the left region of Figs.\ \ref{fig_seg3} and \ref{fig_seg4}. Conversely, as the size ratio $\sigma_0/\sigma$ increases, the tracer particles suffer more collisions per time. The main consequence of these collisions is the existence of a ``buoyancy'' effect on the intruder, resulting from the pressure exerted by the sea of granular particles in the colder region. As a consequence, the intruder is lifted against gravity, and the BNE effect is observed.



\subsubsection{General case}

\begin{figure}
\includegraphics[width=0.7 \columnwidth,angle=0]{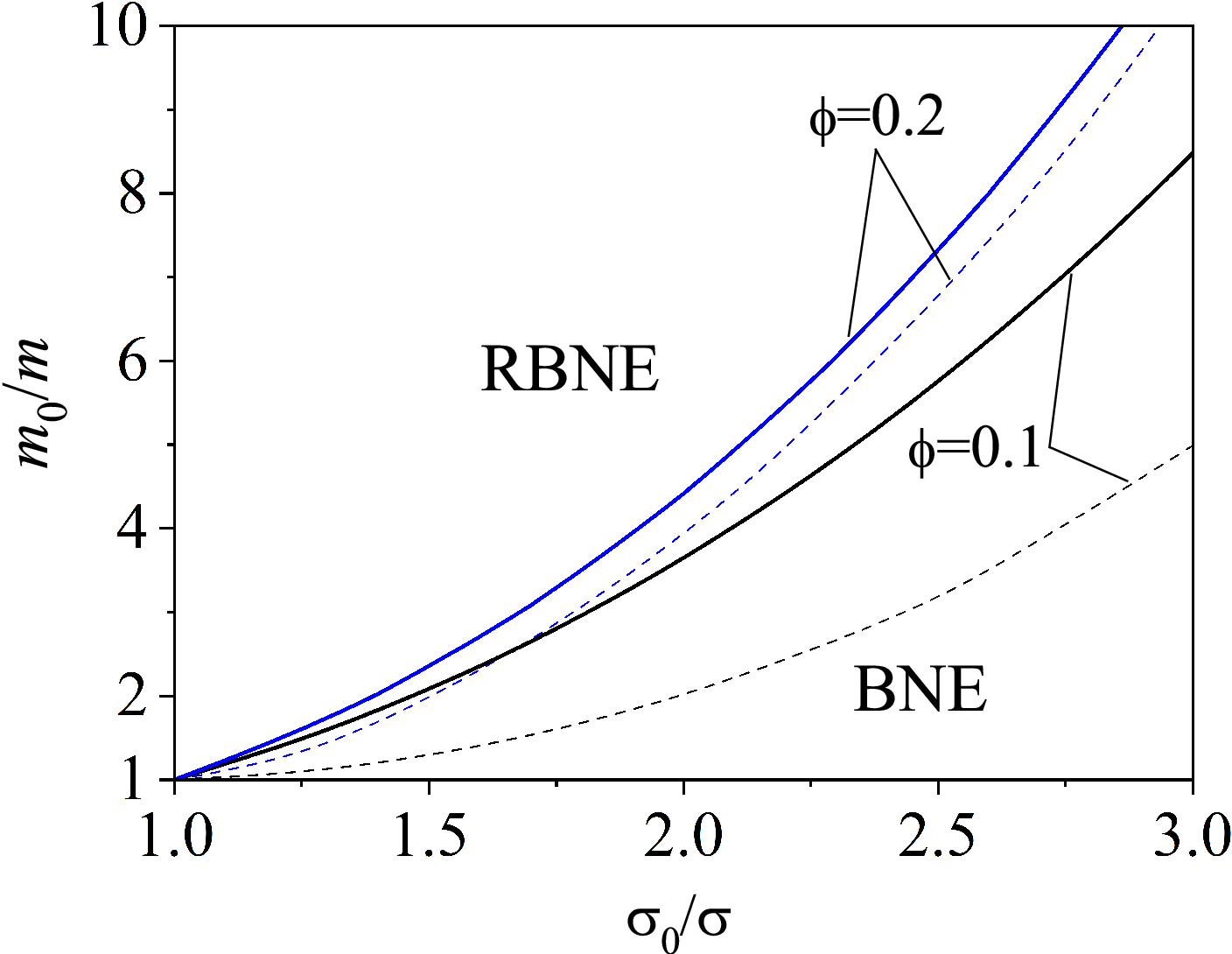}
\caption{(color online) Plot of the marginal segregation curve ($\Lambda=0$) for $d = 3$, $\alpha=\alpha_0=0.7$, $T^*_\text{ex}=0.1$, $|g^*|=1$, and two values of $\phi$. The solid line corresponds to the segregation
criterion for granular suspensions while the dashed line refers to the one derived for dry granular
mixtures. The points below the curve correspond to $\Lambda>0$ (BNE), while the points above the curve correspond to $\Lambda<0$ (RBNE).\label{fig_seg5}}
\end{figure}

Finally, we consider the general case where the effect of the temperature gradient is comparable to that of gravity. To illustrate it, Fig.\ \ref{fig_seg5} shows the marginal segregation curve for $|g^*|=1$, a (common) coefficient of restitution $\alpha=\alpha_0=0.7$, $T_\text{ex}^*=0.1$, and two values of the volume fraction $\phi$: $0.1$ and $0.2$. We observe that the effect of gravity in the segregation dynamics overcomes that of the temperature gradient, as seen from the reversion in the RBNE/BNE transition (compare Figs.\ \ref{fig_seg1} and \ref{fig_seg2} with Fig.\ \ref{fig_seg5}). This finding is consistent with the previous results obtained for dry granular mixtures.\cite{G09} We also see that the effect of the interstitial gas is to enhance the BNE effect, since the buoyancy increases with the diameter ratios $\sigma_0/\sigma$. With regards to density, an increase of shaking strength mimics the effect of decreasing the volume fraction $\phi$ in the BNE/RBNE transition. \cite{TAH03,ATH06} Figure \ref{fig_seg5} also illustrates that the BNE effect is intensified as density increases (or shaking strength is reduced). This behavior was observed in driven granular mixtures \cite{G09} and we see that is also applicable to both dry granular mixtures and suspensions.

\section{Discussion}
\label{sec9}

In this paper, the Enskog-Lorentz kinetic equation has been considered as the starting point to obtain the diffusion transport coefficients of a binary granular suspension where one of the species is present in tracer concentration. As in previous works, \cite{GKG20} 
the influence of the interstitial gas on grains has been accounted for via a drag force plus a stochastic Langevin term defined in terms of the (known) background temperature $T_\text{ex}$. For arbitrary concentration, the corresponding set of kinetic equations for the binary mixture were solved in a previous work \cite {GKG20} by means of the Chapman--Enskog method \cite{CC70} up to the first order in spatial gradients. From this solution, the Navier--Stokes transport coefficients of the mixture are given in terms of the solutions of a set of coupled linear integral equations. In particular, in the case of the mass flux, there are four relevant diffusion transport coefficients obeying integral equations that are usually solved by considering only the leading term in a Sonine polynomial expansion. Due to the technical difficulties embodied in the general problem, we have considered here the tracer limit for the sake of simplicity. In this limit, to solve the corresponding integral equations, we have retained in the Sonine polynomial expansion terms up to the second order and hence, we have explicitly determined the first and second Sonine approximations to the diffusion coefficients as functions of the parameter space of the system. 

It should be noted that the accuracy of the Sonine approximation may be questionable under the conditions of strong dissipation and/or quite disparate size or mass ratios. However, our results have shown
a relatively quite rapid convergence between both Sonine approximations in comparison with the dry (no gas phase) granular limit. In this context, we can conclude that the presence of the interstitial gas tends to enhance the convergence of the Sonine polynomial series expansion.

To assess the reliability of the Sonine approximations considered in this paper, theoretical results have been compared against DSMC simulations. First, a remarkable agreement is found between the theoretical predictions for the temperature ratio $T_0/T$ (which has been obtained by approximating the zeroth-order distributions by their Maxwellian forms) and the DSMC simulations. Additionally, by employing the Einstein formula for the mean square displacement of intruders moving in a granular suspension, the tracer diffusion coefficient $D_0$ has been computationally determined. In order to extend the study recently done in Ref.\ \onlinecite{GGABYG23} for the evaluation of $D_0$ when $\al=\al_0$, we have performed in this paper simulations when both coefficients of restitution are different ($\alpha \neq \alpha_0$). It is confirmed first that the (reduced) coefficient $D_0(\al_0)/D_0(1)$ [$D_0(1)$ being the tracer diffusion coefficient for elastic collisions] exhibits a non-monotonic dependence on $\alpha_0$. Furthermore, intentional selection of mixture parameters reveals stronger disparities between the results obtained from the second and first Sonine approximations for significantly small mass and size ratios. Nevertheless, these discrepancies remain smoother than those have been observed in the dry granular case. \cite{GV09,GV12} As expected from previous works, \cite{GGABYG23} although the theoretical predictions of the first Sonine approximation agree relatively well with simulations, 
we find that the results of the second Sonine approximation agree perfectly with the  DSMC simulations. Thus, the second Sonine approximation 
corrects the small discrepancies observed between the first Sonine solution and simulations, indicating an improved convergence of the solution.

The knowledge of the diffusion coefficients allow us to perform a linear stability analysis of the homogeneous steady state of the suspension. This analysis is relatively simple since the hydrodynamic equations for $n$, $\mathbf{U}$, and $T$ are decoupled from that for $n_0$. Moreover, given that a previous \cite{GGG19a} linear stability analysis had shown that the hydrodynamic equations obeying the fields $n$, $\mathbf{U}$, and $T$ are linearly stable, the hydrodynamic equation for $n_0$ becomes an autonomous differential equation.  We can at this point obtain the time evolution of  the Fourier transform to the (dimensionless) tracer density. Since both the first ($D^*_0[1]$) and second ($D^*_0[2]$) (reduced) Sonine approximations are always positive, any perturbation in the tracer density tends to vanish over sufficiently long times. As a consequence, the homogeneous steady state of the system (tracer plus granular suspension) has been shown to be \emph{linearly} stable.

As a second application, segregation induced by a thermal gradient and/or gravity has been also studied. The main target has been to assess the impact of the interstitial gas on the phase diagrams previously obtained in the absence of the gas phase. \cite{G11} In the absence of gravity, our results (see Figs.\ \ref{fig_seg1} and \ref{fig_seg2}) show that the effect of the gas phase on segregation is in general significant for large mass and/or diameter ratios. The main effect of the surrounding gas is to increase the size of the RBNE region (intruders attempt to accumulate near the hot plate) with respect to the one observed in the absence of the gas phase. On the other hand, in the situations where segregation is mainly driven by gravity (namely, when thermal gradient can be neglected) the phase diagrams with and without gas phase are practically identical (see Figs.\ \ref{fig_seg3} and \ref{fig_seg4}). In addition, as for dry granular mixtures, the influence of inelasticity of collisions on segregation is very tiny since the corresponding marginal segregation curve for elastic collisions collapses practically with the one obtained for inelastic collisions.           

As extensively discussed in this manuscript, the obtained results are contingent upon various approximations and assumptions. Initially, our calculations have been constrained to low Reynolds numbers, consistent with the framework of Stokes' law. Additionally, we have accounted for a similar influence on the dynamics of grains coming from gas-solid forces and collisional effects, as indicated by the Stokes number. Consequently, our study encompasses neither Brownian particles (lack of collisions) nor a dry granular gas (without an interstitial gas). Given these limitations, the key question arises regarding the existence of natural systems satisfying both conditions, which can be replicated in laboratory settings. In a previous work \cite{GGABYG23} where the authors of the present paper are involved, Section VII aims to address the above query   
within a simplified framework that incorporates several dimensionless parameters for monodisperse granular suspensions. These dimensionless parameters comprise the Reynolds number Re, Stokes number St, dimensionless Stokes friction coefficient $\gamma_\text{St}/\nu\equiv\gamma^*/R$, and the reduced background temperature $T^*_\text{ex}$. It is demonstrated that the resulting values used to describe a suspension of gold grains immersed in a hydrogen molecular gas fall within the previously discussed values for the relevant quantities in our model. Specifically, it becomes evident that the influence of collisions on diffusion coefficients is not negligible for many of the situations where the present suspension model applies. More details on this issue can be found in Section VII of Ref.\ \onlinecite{GGABYG23}.

In this work, we have used the Chapman--Enskog method \cite{CC70} conveniently adapted to dissipative dynamics as a reliable procedure to obtain the diffusion transport properties of intruders immersed in a granular suspension. 
Our calculations have been restricted to the Navier--Stokes domain (i.e. low Knudsen numbers). 
In the Navier-Stokes hydrodynamics regime, according to the Curie principle,\cite{GM84} shear stress cannot modify the mass flux due to the constraint imposed by linear transport coupling, which exclusively occurs between irreducible tensors sharing the same rank and parity. On the other hand, it is worth investigating tracer diffusion coefficients in a binary suspension where the excess component (granular gas) is strongly sheared. In this particular scenario, the state of uniform shear flow serves as the reference state for the Chapman-Enskog expansion. In this situation, due to the anisotropy induced by the shear flow, second-rank tensors are required to describe the diffusion process instead of the conventional scalar coefficients. In the case of dry granular mixtures, at given values of the coefficients of restitution, these tensors have been shown \cite{GT15} to exhibit a nonlinear dependence on the shear rate and, thus, on the inertial number $I$. \cite{JFP06} We want to study the shear-rate dependence of the diffusion coefficients when the dynamics of solid particles are influenced by the presence of an interstitial gas. Further investigation in this direction is planned for the near future.


The results reported in this paper have been derived for a suspension model where the effect of the background gas on the dynamics of grains has been accounted for via a fluid-solid force (coarse-grained description). It would be of interest to revisit the tracer diffusion problem by considering a coarse-grained model in which fluid-solid interactions are governed not only by the drag force (Stokes drag force) but also by the Archimedes force (gas pressure gradient). Furthermore, as an alternative perspective, a collisional model that explicitly accounts for the collisions between grains and particles of the surrounding molecular gas could be implemented. This sort of suspension model has been recently proposed \cite{GG22} and the results derived from this collisional model has been shown to reduce to those obtained from the coarse-grained approach \cite{GGG19a} when the grains are much heavier than the particles of the background gas. We expect a similar conclusion in the tracer diffusion problem analyzed in the present paper. 
Last but not least, it would also be desirable to carry on computer simulations to assess the reliability of the phase diagrams obtained here for thermal diffusion segregation of intruders. We plan to make some progresses along these lines in the near future.

\acknowledgments 
We acknowledge financial support from Grant PID2020-112936GB-I00 funded by MCIN/AEI/10.13039/501100011033, and from Grant IB20079 funded by Junta de Extremadura (Spain) and by ERDF A way of making Europe. The research of R.G.G. also has been supported by Plan Propio de Iniciación a la Investigación, Desarrollo Tecnológico e Innovación de la Universidad de Extremadura (ACCIÓN III).\\

\noindent\textbf{AUTHOR DECLARATIONS}\\
\\\textbf{Conflict of Interest}\\

The authors have no conflicts to disclose.\\

\noindent\textbf{Author Contributions}\\

\noindent\textbf{Rub\'en G\'omez Gonz\'alez:} Formal analysis (equal); Investigation
(equal); Software (equal); Writing – review \& editing (equal).\\

\noindent\textbf{Vicente Garz\'o:}  Conceptualization (equal); Investigation (equal);
Supervision (equal); Writing – original draft (equal).\\

\noindent\textbf{DATA AVAILABILITY}\\

The data that support the findings of this study are available from the corresponding author upon reasonable request.

\appendix
\section{The second Sonine coefficients $a$ and $c$ of the granular gas}
\label{appA}

The first-order velocity distribution function $f^{(1)}$ of the granular gas has the form \cite{GGG19a}
\beq
\label{a1}
f^{(1)}=\boldsymbol{\mathcal{A}}\cdot \nabla T+\boldsymbol{\mathcal{C}}\cdot \nabla n
+\boldsymbol{\mathcal{D}}:\nabla \mathbf{U}+E \nabla \cdot \mathbf{U}.
\eeq
To evaluate the diffusion transport coefficients of the tracer particles, only the quantities $\boldsymbol{\mathcal{A}}$ and $\boldsymbol{\mathcal{C}}$ are needed. Both quantities vanish in the first Sonine approximation. In the second Sonine approximation, $\boldsymbol{\mathcal{A}}$ and $\boldsymbol{\mathcal{C}}$ are given, respectively, by
\begin{equation}
\label{a2}
\boldsymbol{\mathcal{A}}({\bf V})\to -f_{\text{M}}({\bf V})a\; {\bf S}({\bf V}),\quad
\boldsymbol{\mathcal{C}}({\bf V})\to -f_{\text{M}}({\bf V})c\; {\bf S}({\bf V}),
\end{equation}
where the Maxwellian distribution $f_{\text{M}}({\bf V})$ is defined by Eq.\ \eqref{2.16} with the replacement $\mathbf{v}\to \mathbf{V}$ and $\mathbf{S}(\mathbf{V})$ is
\begin{equation}
\label{a3}
{\bf S}({\bf V})=\left(\frac{1}{2}mV^2-\frac{d+2}{2}T\right){\bf V}.
\end{equation}
The coefficients $a$ and $c$ are related to the kinetic contributions to the thermal conductivity $\kappa_k$ and the diffusive heat conductivity $\mu_k$. They are defined as
\beq
\label{a4}
\left\{a,c\right\}=\frac{2}{d+2}\frac{m}{nT^3}\left\{\kappa_{\text{k}}, \mu_{\text{k}}\right\}.
\eeq
Up to the second Sonine approximation, the coefficients $a$ and $c$ are \cite{GGG19a}
\beq
\label{a5}
a=\frac{1}{\nu T^2}\frac{1+3\frac{2^{d-3}}{d+2}\chi^{(0)}\phi\left(1+\alpha\right)^2
\left(2\alpha-1\right)}{\nu_{\kappa}^*+\gamma^*-\frac{3}{2}\zeta^*},
\eeq
\beqa
\label{a6}
c&=&\frac{1}{\nu n T}
\left(\nu_\kappa^*+3\gamma^*\right)^{-1}\Bigg\{a^*\Bigg[
\zeta^*\left(1+\phi\frac{\partial\ln\chi^{(0)}}{\partial\phi}\right)\nonumber\\
& & -2\left(\theta^{-1}-1\right)\gamma^* \phi \frac{\partial \ln R}{\partial \phi}\Bigg]+3\frac{2^{d-2}}{\left(d+2\right)}\chi^{(0)}\phi\left(1+\alpha\right)\nonumber\\
& &
\times
\left(1+\frac{1}{2}\phi\frac{\partial\ln\chi^{(0)}}{\partial\phi}\right)\al(\al-1)\Bigg\}.
\eeqa
In Eqs.\ \eqref{a5} and \eqref{a6}, $\chi^{(0)}$ is the pair correlation at zeroth order, $\gamma^*=\gamma/\nu=\lambda \theta^{-1/2}$, $a^*=\nu T^2 a$, and
\beqa
\label{a7}
\nu_\kappa^*&=&\frac{8}{\sqrt{2}(d+2)}\frac{\pi^{(d-1)/2}}{\Gamma\left(\frac{d}{2}\right)}
\frac{1+\alpha}{d}\chi^{(0)}\nonumber\\
& & \times\left[\frac{d-1}{2}+\frac{3}{16}(d+8)(1-\alpha)\right].
\eeqa
It must be recalled again that Eqs.\ \eqref{a5}--\eqref{a7} have been displayed neglecting the non-Gaussian corrections to the zeroth-order distribution function $f^{(0)}$. As said along the paper, the effect of these contributions on the Sonine coefficients $a$ and $c$ are significant only for quite extreme values of inelasticity. \cite{GGG19a}

\section{Sonine approximations to the diffusion transport coefficients}
\label{appB}

In this appendix we determine the first and second Sonine approximations to the diffusion transport coefficients $D_0$, $D$, $D^T$, and $D^U$. These coefficients are given in terms of the solutions of the set of integral equations \eqref{3.6}--\eqref{3.9}, where the inhomogeneous terms $\mathbf{A}_0$, $\mathbf{B_0}$, $\mathbf{C}_0$, and $\mathbf{E}_0$ are defined by Eqs.\ \eqref{3.11}--\eqref{3.14}, respectively. Up to the second Sonine approximation, the functions $\boldsymbol{\mathcal{A}}_{0}$, $\boldsymbol{\mathcal{B}}_{0}$, $\boldsymbol{\mathcal{C}}_{0}$, and $\boldsymbol{\mathcal{E}}_{0}$ are approximated by Eqs.\ \eqref{4.1}--\eqref{4.3.1}, respectively, while $\boldsymbol{\mathcal{A}}$ and $\boldsymbol{\mathcal{C}}$ are defined by Eq.\ \eqref{a2}. Let us evaluate each transport coefficient separately.

\begin{widetext}

\subsection{Thermal diffusion coefficient $D^T$}

We consider first the thermal diffusion coefficient $D^T$. To get it, we substitute $\boldsymbol{\mathcal{A}}$ and $\boldsymbol{\mathcal{A}}_0$ by their Sonine approximations and then, multiply the integral equation \eqref{3.6} by $m_0\mathbf{V}$ and integrate over velocity. After some algebra, one gets
\beq
\label{b1}
\Bigg(\nu_1-2\gamma \theta^{-1}-\frac{1}{2}\zeta^{(0)}+\gamma_0\Bigg)D^T+\frac{n_0T_0^{(0)2}}{\rho}\nu_2 a_0=Z_1,
\eeq
where
\beq
\label{b2}
Z_1=-\frac{n_0T_0^{(0)2}}{\rho}\nu_3 a+\frac{n_0T_0^{(0)}}{\rho}+\frac{n_0T}{\rho}\theta \Delta_{\theta,0}-\frac{p\rho_0}{\rho^2}
+\frac{m_0}{d\rho}\int d\mathbf{V}\; \mathbf{V}\cdot \boldsymbol{\mathcal{K}}_0\Bigg[T\frac{\partial f^{(0)}}{\partial T}\Bigg].
\eeq
Here, $\Delta_{\theta,0}=\partial \tau_0/\partial \theta$, $\tau_0=T_0^{(0)}/T$, and we have introduced the collision frequencies
\begin{equation}
\label{b3}
\nu_{1}=-\frac{1}{dn_0T_0^{(0)}}\int d{\bf v}\,m_0{\bf V}\cdot
J_{0}^{(0)}[f_{0,M}{\bf V},f^{(0)}],
\end{equation}
\begin{equation}
\label{b4}
\nu_{2}=-\frac{1}{dn_0T_0^{(0)2}} \int d{\bf v}\,m_0{\bf V}\cdot J_{0}^{(0)}[f_{0,M}{\bf
S}_0,f^{(0)}],
\end{equation}
\begin{equation}
\label{b5}
\nu_{3}=-\frac{1}{dn_0T_0^{(0)2}} \int d{\bf v}\,m_0{\bf V}\cdot
J_{0}^{(0)}[f_0^{(0)},f_{M}{\bf S}].
\end{equation}
In Eqs.\ \eqref{b1} and \eqref{b2}, $\zeta^{(0)}$ and $T_0^{(0)}$ are the zeroth-order contributions to the cooling rate $\zeta$ and the partial temperature $T_0$, respectively. In the Maxwellian approximation, the cooling rate $\zeta^{(0)}$ is defined by Eq.\ \eqref{2.17cooling} and $T_0^{(0)}$ is given by the physical solution to Eq.\ \eqref{2.29}. Upon obtaining \eqref{b2}, use has been made of the result
\beq
\label{b6}
\int d\mathbf{V} m_0 \mathbf{V}\cdot \mathbf{A}_0=-dn_0T_0^{(0)}-dn_0T \theta \Delta_{\theta,0}+\frac{d p}{\rho}\rho_0-
\int d{\bf v} m_0 \mathbf{V}\cdot \boldsymbol{\mathcal {K}}_{0}\Bigg[T\frac{\partial f^{(0)}}{\partial T}\Bigg].
\eeq

The collision integral of Eq.\ \eqref{b6} involving the operator $\boldsymbol{\mathcal {K}}_{0}$ can be computed when one replaces the distribution function $f^{(0)}$ by its Maxwellian form \eqref{2.26}. In this approximation, one has the identity
\beq
\label{b7}
T\frac{\partial f^{(0)}}{\partial T}=-\frac{1}{2}\frac{\partial}{\partial \mathbf{V}}\cdot\mathbf{V}f^{(0)}
\eeq
and hence, one obtains the result \cite{GHD07}
\beq
\label{b7b}
\int d{\bf v} m_0 \mathbf{V}\cdot \boldsymbol{\mathcal {K}}_{0}\Bigg[T\frac{\partial f^{(0)}}{\partial T}\Bigg]=2^{d-1} d n_0  \left(\frac{\overline{\sigma}}{\sigma}\right)^d \chi_{0}^{(0)}\phi M_0(1+\al_{0}).
\eeq
If only the first Sonine correction is retained in Eq.\ \eqref{b1} (which means $a_0=a=0$), then the (reduced) first Sonine approximation $D^{*T}[1]=(m\nu/x_0 T)D^T[1]$ to the thermal diffusion coefficient is
\beq
\label{b8}
D^{*T}[1]=-\Bigg(\nu_1^*-2\gamma^* \theta^{-1}-\frac{1}{2}\zeta^*+\gamma_0^*\Bigg)^{-1}
\Bigg[\frac{m_0}{m}p^*-\tau_0-\theta \Delta_{\theta,0}-2^{d-1}\left(\frac{\overline{\sigma}}{\sigma}\right)^d\chi_{0}^{(0)}M_0 \phi (1+\al_{0})\Bigg],
\eeq
where $\nu_i^*=\nu_i/\nu$ and $p^*=p/(nT)$ is given by Eq.\ \eqref{3.15}.

To close the evaluation of $D^T$ up to the second Sonine approximation, we multiply now Eq.\ \eqref{3.6} by $\mathbf{S}_0(\mathbf{V})$ and integrates over $\mathbf{v}$. After a tedious algebra, one achieves the result
\beq
\label{b9}
\Bigg(\nu_4+3\gamma_0-2\gamma \theta^{-1}-\frac{1}{2}\zeta^{(0)}\Bigg)a_0+\frac{\rho }{n_0T_0^{(0)2}}\Bigg[\nu_5+2\gamma_0\Bigg(1-\frac{1}{d+2}\frac{T_\text{ex}}{T_0^{(0)}}\Bigg)\Bigg]D^T
=Z_2,
\eeq
where
\beq
\label{b10}
Z_2=-\nu_6 a+\frac{1}{T_0^{(0)}}+\frac{2}{d(d+2)}\frac{m_0}{n_0 T_0^{(0)3}}\int d\mathbf{V}\; \mathbf{S}_0\cdot \boldsymbol{\mathcal{K}}_0\Bigg[T\frac{\partial f^{(0)}}{\partial T}\Bigg],
\eeq
\begin{equation}
\label{b11}
\nu_{4}=-\frac{2}{d(d+2)}\frac{m_0}{n_0 T_0^{(0)3}}\int d{\bf v}\,{\bf S}_0\cdot
J_{0}^{(0)}[f_{0,M}{\bf S}_0,f^{(0)}],
\end{equation}
\begin{equation}
\label{b12}
\nu_{5}=-\frac{2}{d(d+2)}\frac{m_0}{n_0 T_0^{(0)2}}\int d{\bf v}\,{\bf S}_0\cdot
J_{0}^{(0)}[f_{0,M}{\bf V},f^{(0)}],
\end{equation}
\begin{equation}
\label{b13}
\nu_{6}=-\frac{2}{d(d+2)}\frac{m_0}{n_0 T_0^{(0)3}}\int d{\bf v}\,{\bf S}_0\cdot
J_{0}^{(0)}[f_0^{(0)},f_{M}{\bf S}].
\end{equation}
As before, if one replaces $f_0^{(0)}$ by $f_{0,\text{M}}$, the collision integral involving the operator $\boldsymbol{\mathcal {K}}_{0}$ in Eq.\ \eqref{b10} is given by \cite{GHD07}
\begin{eqnarray}
\label{b14}
\int d{\bf v} \mathbf{S}_{0}({\bf V})\cdot \boldsymbol{{\cal K}}_{0}
\left[T\frac{\partial f^{(0)})}{\partial T}\right]&=&\frac{d}{4}n_0\frac{M_{0}T^2}{m}(1+\omega)^d \chi_0^{(0)}\phi (1+\alpha_0)\Bigg\{
\frac{M\tau_0}{M_0}\Big[(d+2)(M_{0}^2-1)\nonumber\\
& & +(2d-5-9\alpha_0)M_{0}M+ (d-1+3\alpha_0+6\alpha_0^2)M^2\Big]+6
M^2(1+\alpha_0)^2\Bigg\}.
\end{eqnarray}

The (reduced) thermal diffusion coefficient $D^{T*}$ is defined in Eq.\ \eqref{4.8}. Thus, in reduced units and by using matrix notation, Eqs.\ \eqref{b1} and \eqref{b9} can be rewritten as
\begin{equation}
\label{b15}
\left(
\begin{array}{cc}
\nu_1^*-2\gamma^* \theta^{-1}-\frac{1}{2}\zeta^*+\gamma_0^*&\tau_0^2\nu_{2}^*\\
\frac{\nu_5^*+2\gamma_0^*\Big(1-(\theta\tau_0)^{-1}\Big)}{\tau_0^2}&
\nu_4^*+3\gamma_0^*-2\gamma^* \theta^{-1}-\frac{1}{2}\zeta^{*}
\end{array}
\right) \left(
\begin{array}{c}
D^{T*}\\
a_0^* \end{array} \right)= \left(
\begin{array}{c}
X_1^*-\tau_0^2 \nu_{3}^* a^*\\
X_2^*-\nu_{6}^* a^*
\end{array}
\right).
\end{equation}
Here, $a_0^*=T^2 \nu a_0$, $a^*=T^2\nu a$, and
\begin{equation}
\label{b16}
X_1^*=\tau_0+\theta \Delta_{\theta,0}-\frac{p m_0}{\rho T}+\frac{1}{2} (1+\omega)^d
M_{0}\chi_0^{(0)}\phi (1+\alpha_0),
\end{equation}
\beq
\label{b17}
X_2^*=\frac{1}{\tau_0}+\frac{2}{d(d+2)} \frac{m_0 T}{n_0 T_0^{(0)3}}\int d\mathbf{V}\; \mathbf{S}_0\cdot \boldsymbol{\mathcal{K}}_0\Bigg[T\frac{\partial f^{(0)}}{\partial T}\Bigg].
\eeq

The second Sonine approximation $D^{T*}[2]$ to $D^{T*}$ can be obtained from Eq.\ \eqref{b15} and the result is
\beq
\label{b18}
D^{T*}[2]=\frac{(3\gamma_0^*+\nu_4^*-\frac{1}{2}\zeta^*-2\gamma^*\theta^{-1})(a^*\nu_3\tau_0^2-X_1^*)-
\nu_2^*\tau_0^2(a^*\nu_6^*-X_2^*)}{\nu_2^*(\nu_5^*+2\gamma_0^*(1-(\tau_0\theta)^{-1})-(\gamma_0^*+\nu_1^*
-\frac{1}{2}\zeta^*-2\gamma^*\theta^{-1})(3\gamma_0^*+\nu_4^*-\frac{1}{2}\zeta^*-2\gamma^*\theta^{-1})}.
\eeq
\end{widetext}
To obtain the dependence of $D^{T*}[1]$ and $D^{T*}[2]$ on the parameter space of the system, it still remains to determine the (reduced) collision frequencies $\nu_i^*$. Explicit expressions of them have been achieved in previous works \cite{GHD07,GV12,G19} when the distributions $f^{(0)}$ and $f_0^{(0)}$ are replaced by their Maxwellian approximations $f_{M}$ and $f_{0,M}$, respectively. These expressions will be provided in the Appendix \ref{appC} for the sake of completeness.

The determination of the first and second Sonine approximations of the diffusion coefficients $D_0$ and $D$ follows similar steps as those made for $D^T$. Below, only the final expressions will be provided.

\subsection{Tracer diffusion coefficient $D_0$}

\begin{widetext}
The (reduced) tracer diffusion coefficient $D_0^*$ is defined by Eq.\ \eqref{4.8}. It obeys the matrix equation
\begin{equation}
\label{b18.1}
\left(
\begin{array}{cc}
\nu_1^*+\gamma_0^*&\tau_0^2\nu_{2}^*\\
\frac{\nu_5^*+2\gamma_0^*\Big(1-(\theta\tau_0)^{-1}\Big)}{\tau_0^2}&
\nu_4^*+3\gamma_0^*
\end{array}
\right) \left(
\begin{array}{c}
D_0^{*}\\
b_0^* \end{array} \right)= \left(
\begin{array}{c}
\tau_0\\
0
\end{array}
\right),
\end{equation}
where $b_0^*=T \nu b_0$. Thus, the first $D_0^*[1]$ and second $D_0^*[2]$ Sonine approximations to $D_0^*$ read, respectively, as
\beq
\label{b19}
D_0^*[1]=\frac{\tau_0}{\nu_1^*+\gamma_0^*},
\eeq
\beq
\label{b20}
D_0^*[2]=\frac{(3\gamma_0^*+\nu_4^*)\tau_0}{(\gamma_0^*+\nu_1^*)(3\gamma_0^*+\nu_4^*)-\nu_2^*(\nu_5^*+
2\gamma_0^*(1-(\tau_0\theta)^{-1}))}.
\eeq

\subsection{Mutual diffusion coefficient $D$}

The (reduced) diffusion coefficient $D^*$ is defined in Eq.\ \eqref{4.9}. After some calculations, the first $D^*[1]$ and second $D^*[2]$ Sonine approximations to $D^*$ can be written, respectively, as
\beqa
\label{b21}
D^*[1]&=&(\nu_1^*+\gamma_0^*)^{-1}\Bigg\{\Bigg[\zeta^*\left(1+\phi\frac{\partial\ln\chi^{(0)}}{\partial\phi}\right)
-2\left(\theta^{-1}-1\right)\gamma^* \phi \frac{\partial \ln R}{\partial \phi}\Bigg]D^{T*}[1]+\phi \Delta_{\phi,0}-\frac{m_0}{m}\left(p^*+\phi\frac{\partial p^*}{\partial \phi}\right)\nonumber\\
& &  +
\frac{1}{2} M_{0}\phi
(1+\alpha_0) \left(\frac{1+\beta}{\beta}\right)\frac{\partial}{\partial \phi}
\left(\frac{\mu_0}{T}\right)_{T,n_0}\Bigg\},
\eeqa
\beq
\label{b22}
 D^*[2]=\frac{(3\gamma_0^*+\nu_4^*)(c^*\nu_3^*\tau_0^2-Y_1^*)-\nu^*_2\tau_0^2(c^*\nu_6^*-Y_2^*)}{\nu_2^*
 (\nu_5^*+2\gamma_0^*(1-(\tau_0\theta)^{-1}))-(\gamma_0^*+\nu_1^*)(3\gamma_0^*+\nu_4^*)}.
\eeq
In Eq.\ \eqref{b22}, the dimensionless quantities $Y_1^*$ and $Y_2^*$ are given by
\beqa
\label{b22.1} Y_1^*&=&\Bigg[\zeta^*\left(1+\phi\frac{\partial\ln\chi^{(0)}}{\partial\phi}\right)
-2\left(\theta^{-1}-1\right)\gamma^* \phi \frac{\partial \ln R}{\partial \phi}\Bigg]D^{T*}[2]\nonumber\\
& & 
+\phi \Delta_{\phi,0}-\frac{m_0}{m}\left(p^*+\phi\frac{\partial p^*}{\partial \phi}\right)+
\frac{1}{2} M_{0}\phi
(1+\alpha_0) \left(\frac{1+\beta}{\beta}\right)\frac{\partial}{\partial \phi}
\left(\frac{\mu_0}{T}\right)_{T,n_0},\nonumber\\
\eeqa
\begin{eqnarray}
\label{b22.2}
Y_2^*&=&\Bigg[\zeta^*\left(1+\phi\frac{\partial\ln\chi^{(0)}}{\partial\phi}\right)
-2\left(\theta^{-1}-1\right)\gamma^* \phi \frac{\partial \ln R}{\partial \phi}\Bigg]a_0^*+\tau_0^{-2}
\phi \Delta_{\phi,0}+\frac{1}{2(d+2)}\frac{M^2}{M_{0}}\phi (1+\alpha_0)\frac{\partial}{\partial \phi}
\left(\frac{\mu_0}{T}\right)_{T,n_0}\nonumber\\
& &\times 
 \Bigg\{ \Big[(d+8)M_{0}^2+(7+2d-9\alpha_0)M_{0}M+(2+d+3\alpha_0^2-
3\alpha_0)M^2\Big]\beta\nonumber\\
& & +3M^2(1+\alpha_0)^2\beta^3+
\Big[(d+2)M_{0}^2+(2d-5-9\alpha_0)M_{0}M+(d-1+3\alpha_0+6\alpha_0^2)
M^2\Big]\beta^2\nonumber\\
& & -(d+2)\beta(1+\beta)\Bigg\},
\end{eqnarray}
where 
\beq
\label{b22.3}
a_0^*=\frac{\tau_0^{-2}\left[\nu_5^*+2\gamma_0^*(1-(\tau_0\theta)^{-1})\right]\left(X_1^*-a^*\nu_3\tau_0^2\right)-
\left(\gamma_0^*+\nu_1^*-\frac{\zeta^*}{2}-2\gamma^*\theta^{-1}\right)\left(X_2^*-a^*\nu_6^*\right)}
{\nu_2^*\left[\nu_5^*+2\gamma_0^*(1-(\tau_0\theta)^{-1})\right]-(\gamma_0^*+\nu_1^*-\frac{1}{2}\zeta^*-2\gamma^*\theta^{-1})
(3\gamma_0^*+\nu_4^*-\frac{1}{2}\zeta^*-2\gamma^*\theta^{-1})}.
\eeq

\subsection{Velocity diffusion coefficient $D^U$}

The (reduced) velocity diffusion coefficient $D_U^*$ is defined by Eq.\ \eqref{4.9}. It obeys the matrix equation
\begin{equation}
\label{b23}
\left(
\begin{array}{cc}
\nu_1^*+\gamma_0^*&\tau_0^2\nu_{2}^*\\
\frac{\nu_5^*+2\gamma_0^*\Big(1-(\theta\tau_0)^{-1}\Big)}{\tau_0^2}&
\nu_4^*+3\gamma_0^*
\end{array}
\right) \left(
\begin{array}{c}
D_U^{*}\\
e_0^* \end{array} \right)= \left(
\begin{array}{c}
\gamma_0^*-\gamma^*\\
0
\end{array}
\right),
\end{equation}
where $e_0^*=(T^2/m_0) e_0$. The first $D_U^*[1]$ and second $D_U^*[2]$ Sonine approximations to $D_U^*$ are given, respectively, by
\beq
\label{b24}
D_U^*[1]=\frac{\gamma_0^*-\gamma^*}{\nu_1^*+\gamma_0^*},
\eeq
\beq
\label{b25}
D_U^*[2]=\frac{(3\gamma_0^*+\nu_4^*)(\gamma^*-\gamma_0^*)}{\nu_2^*(\nu_5^*+2\gamma_0^*(1-(\tau_0\theta)^{-1}))
-(\gamma_0^*+\nu_1^*)(3\gamma_0^*+\nu_4^*)}.
\eeq

\section{Collision frequencies}
\label{appC}

In this appendix we display the expressions of the collision frequencies $\nu_1^*$--$\nu_6^*$. They are given by \cite{GHD07,GV12}
\begin{equation}
\label{c1}
\nu_{1}^*=\frac{2\pi^{(d-1)/2}}{d\Gamma\left(\frac{d}{2}\right)}\left(\frac{\overline{\sigma}}{\sigma}
\right)^{d-1}\chi_0^{(0)}M(1+\alpha_0) \left(\frac{1+\beta}{\beta}\right)^{1/2},
\end{equation}
\begin{equation}
\label{c2}
\nu_{2}^*=\frac{\pi^{(d-1)/2}}
{d\Gamma\left(\frac{d}{2}\right)}\left(\frac{\overline{\sigma}}{\sigma}\right)^{d-1}\chi_0^{(0)}
M(1+\alpha_0)[\beta(1+\beta)]^{-1/2},
\end{equation}
\begin{equation}
\label{c3}
\nu_{3}^*=-\frac{\pi^{(d-1)/2}}
{d\Gamma\left(\frac{d}{2}\right)}\left(\frac{\overline{\sigma}}{\sigma}\right)^{d-1}\chi_0^{(0)}
\frac{M^2}{M_{0}}(1+\alpha_0) \beta^{5/2}(1+\beta)^{-1/2},
\end{equation}
\begin{equation}
\label{c4}
\nu_{4}^*=\frac{\pi^{(d-1)/2}}
{d(d+2)\Gamma\left(\frac{d}{2}\right)}\left(\frac{\overline{\sigma}}{\sigma}\right)^{d-1}
\chi_0^{(0)}M(1+\alpha_0)\left(\frac{\beta}{1+\beta}\right)^{3/2}
\left[A-(d+2)\frac{1+\beta}{\beta} B\right],
\end{equation}
\begin{equation}
\label{c5}
\nu_{5}^*=\frac{2\pi^{(d-1)/2}}
{d(d+2)\Gamma\left(\frac{d}{2}\right)}\left(\frac{\overline{\sigma}}{\sigma}\right)^{d-1}
\chi_0^{(0)}M(1+\alpha_0)\left(\frac{\beta}{1+\beta}\right)^{1/2}B,
\end{equation}
\begin{equation}
\label{c6}
\nu_{6}^*=-\frac{\pi^{(d-1)/2}}
{d(d+2)\Gamma\left(\frac{d}{2}\right)}\left(\frac{\overline{\sigma}}{\sigma}\right)^{d-1}
\chi_0^{(0)}\frac{M^2}{M_{0}}(1+\alpha_0)\left(\frac{\beta}{1+\beta}\right)^{3/2}
\left[C+(d+2)(1+\beta) D\right].
\end{equation}
In Eqs.\ \eqref{c1}--\eqref{c6}, we have introduced the quantities
\begin{eqnarray}
\label{c7}
A&=&2M^2\left(\frac{1+\beta}{\beta}\right)^{2}
\left(2\alpha_0^{2}-\frac{d+3}{2}\alpha_0+d+1\right)
\big[d+5+(d+2)\beta\big]-M(1+\beta) \Big\{\xi\beta^{-2}[(d+5)+(d+2)\beta]
\nonumber\\
& & \times
[(11+d)\alpha_0
-5d-7]-\beta^{-1}[20+d(15-7\alpha_0)+d^2(1-\alpha_0)-28\alpha_0] -(d+2)^2(1-\alpha_0)\Big\}
\nonumber\\
& & +3(d+3)\xi^2\beta^{-2}[d+5+(d+2)\beta]+ 2\xi\beta^{-1}[24+11d+d^2+(d+2)^2\beta]
\nonumber\\
& & +(d+2)\beta^{-1} [d+3+(d+8)\beta]-(d+2)(1+\beta)\beta^{-2}
[d+3+(d+2)\beta],\nonumber\\
\end{eqnarray}
\begin{eqnarray}
\label{c8}
B&=& (d+2)(1+2\xi)+M(1+\beta)\Big\{(d+2)(1-\alpha_0)
-[(11+d)\alpha_0-5d-7]\xi\beta^{-1}\Big\}+3(d+3)\xi^2\beta^{-1}\nonumber\\
& &+2M^2\left(2\alpha_0^{2}-\frac{d+3}{2}\alpha
_{0}+d+1\right)\beta^{-1}(1+\beta)^2- (d+2)\beta^{-1}(1+\beta),
\end{eqnarray}
\begin{eqnarray}
\label{c9}
C&=&
 2M^2(1+\beta)^2
\left(2\alpha_0^{2}-\frac{d+3}{2}\alpha_0+d+1\right)
\left[d+2+(d+5)\beta\right]-M(1+\beta) \Big\{\xi[d+2+(d+5)\beta][(11+d)\alpha_0
\nonumber\\
& &
-5d-7] +\beta[20+d(15-7\alpha_0)+d^2(1-\alpha_0)-28\alpha_0] +(d+2)^2(1-\alpha_0)\Big\}
+3(d+3)\xi^2[d+2+(d+5)\beta]
\nonumber\\
& & - 2\xi[(d+2)^2+(24+11d+d^2)\beta]
+(d+2)\beta [d+8+(d+3)\beta]-(d+2)(1+\beta)
[d+2+(d+3)\beta],
\end{eqnarray}
\begin{eqnarray}
\label{c10}
D&=& (d+2)(2\xi-\beta)+M(1+\beta)\Big\{(d+2)(1-\alpha_0)
+[(11+d)\alpha_0-5d-7]\lambda\Big\}-3(d+3)\xi^2\nonumber\\
& & -2M^2\left(2\alpha_0^{2}-\frac{d+3}{2}\alpha_{0}+d+1\right)
(1+\beta)^2+(d+2)(1+\beta).
\end{eqnarray}
In Eqs.\ \eqref{c7}--\eqref{c10}, $\xi=M_{0}(1-\tau_0^{-1})$.

\section{Derivatives of $\tau_0$ with respect to $\theta$ and $\phi$}
\label{appD}

It is quite apparent that to determine the transport coefficients $D^{T*}[2]$ and $D^*[2]$ one needs to evaluate the derivatives $\Delta_{\theta,0}=\partial \tau_0/\partial \theta$ and $\Delta_{\phi,0}=\partial \tau_0/\partial \phi$ in the steady state. To evaluate them, we start from Eq.\ (69) of Ref.\ \onlinecite{GKG20} in the tracer limit ($x_0\to 0$):
\begin{equation}
\label{d1}
\Omega^*\theta\frac{\partial\tau_0}{\partial\theta}=-\tau_0\Omega^*+\Omega_0^*,
\end{equation}
where 
\beq
\label{d2}
\Omega^*=2\gamma^*(\theta^{-1}-1)-\zeta^*,\quad 
\Omega_0^*=2\gamma_0^*(\theta^{-1}-\tau_0)-\tau_0\zeta_0^*,
\eeq
and $\zeta_0^*=\zeta_0/\nu$ is given by Eq.\ \eqref{2.27}. Since in the steady state, $\Omega^*=\Omega_0^*=0$, then according to Eq.\ \eqref{d1} the derivative $\partial\tau_0/\partial\theta$ becomes indeterminate. As in previous works, \cite{GKG20,KG13} the above problem can be solved by employing l'H\^opital's rule. Thus, we take first the derivative with respect to $\theta$ in both sides of Eq.\ \eqref{d1} and then take the steady-state condition ($\Omega^*=\Omega_0^*=0$). After some algebra, we achieve the following expression for $\Delta_{\theta,0}$:
\begin{equation}
\label{d4}
\Delta_{\theta,0}=\frac{\gamma^*_0
\theta^{-1}(3\theta^{-1}-\tau_0)-\gamma^*\tau_0\theta^{-1}(3\theta^{-1}-1)}{\gamma^*(3\theta^{-1}-1)-2\gamma_0^*-\zeta^*_0+\beta
\frac{\partial\zeta_0^*}{\partial\beta}}.
\end{equation}
Once the derivative is known, the derivative $\Delta_{\phi,0}\equiv \partial\tau_0/\partial\phi$ can be computed in a similar way. The expression is
\begin{equation}
\label{d5}
\Delta_{\phi,0}=\frac{\left[\zeta^*\frac{\partial\ln\chi^{(0)}}{\partial\phi}-2\frac{\partial\lambda}{\partial\phi}\theta^{-1/2}(\theta^{-1}-1)\right](\theta\Delta_{\theta,0}+\tau_0)+2\frac{\partial\lambda_0}{\partial\phi}\theta^{-1/2}(\theta^{-1}-\tau_0)-\tau_0\zeta_0^*\frac{\partial\ln \chi_0^{(0)}}{\partial\phi}}{2\gamma_0^*+\zeta^*_0-\beta\frac{\partial\zeta_0^*}{\partial\beta}}.
\end{equation}
Note that in the limit of mechanically equivalent particles, $\gamma^*=\gamma_0^*$, $\tau_0=1$ and hence, according to Eq.\ \eqref{d4} $\Delta_{\theta,0}=0$. As a result, since $\chi^{(0)}=\chi_0^{(0)}$ and $\zeta^*=\zeta^*_0$, Eq\ \eqref{d4} yields $\Delta_{\phi,0}=0$.

\end{widetext}

\bibliography{Brownian}

\begin{thebibliography}{85}%
\makeatletter
\providecommand \@ifxundefined [1]{%
 \@ifx{#1\undefined}
}%
\providecommand \@ifnum [1]{%
 \ifnum #1\expandafter \@firstoftwo
 \else \expandafter \@secondoftwo
 \fi
}%
\providecommand \@ifx [1]{%
 \ifx #1\expandafter \@firstoftwo
 \else \expandafter \@secondoftwo
 \fi
}%
\providecommand \natexlab [1]{#1}%
\providecommand \enquote  [1]{``#1''}%
\providecommand \bibnamefont  [1]{#1}%
\providecommand \bibfnamefont [1]{#1}%
\providecommand \citenamefont [1]{#1}%
\providecommand \href@noop [0]{\@secondoftwo}%
\providecommand \href [0]{\begingroup \@sanitize@url \@href}%
\providecommand \@href[1]{\@@startlink{#1}\@@href}%
\providecommand \@@href[1]{\endgroup#1\@@endlink}%
\providecommand \@sanitize@url [0]{\catcode `\\12\catcode `\$12\catcode
  `\&12\catcode `\#12\catcode `\^12\catcode `\_12\catcode `\%12\relax}%
\providecommand \@@startlink[1]{}%
\providecommand \@@endlink[0]{}%
\providecommand \url  [0]{\begingroup\@sanitize@url \@url }%
\providecommand \@url [1]{\endgroup\@href {#1}{\urlprefix }}%
\providecommand \urlprefix  [0]{URL }%
\providecommand \Eprint [0]{\href }%
\providecommand \doibase [0]{http://dx.doi.org/}%
\providecommand \selectlanguage [0]{\@gobble}%
\providecommand \bibinfo  [0]{\@secondoftwo}%
\providecommand \bibfield  [0]{\@secondoftwo}%
\providecommand \translation [1]{[#1]}%
\providecommand \BibitemOpen [0]{}%
\providecommand \bibitemStop [0]{}%
\providecommand \bibitemNoStop [0]{.\EOS\space}%
\providecommand \EOS [0]{\spacefactor3000\relax}%
\providecommand \BibitemShut  [1]{\csname bibitem#1\endcsname}%
\let\auto@bib@innerbib\@empty
\bibitem [{\citenamefont {Koch}(1990)}]{K90}%
  \BibitemOpen
  \bibfield  {author} {\bibinfo {author} {\bibfnamefont {D.~L.}\ \bibnamefont
  {Koch}},\ }\bibfield  {title} {\enquote {\bibinfo {title} {Kinetic theory for
  a monodisperse gas-solid suspension},}\ }\href@noop {} {\bibfield  {journal}
  {\bibinfo  {journal} {Phys. Fluids A}\ }\textbf {\bibinfo {volume} {2}},\
  \bibinfo {pages} {1711--1722} (\bibinfo {year} {1990})}\BibitemShut {NoStop}%
\bibitem [{\citenamefont {Gidaspow}(1994)}]{G94}%
  \BibitemOpen
  \bibfield  {author} {\bibinfo {author} {\bibfnamefont {D.}~\bibnamefont
  {Gidaspow}},\ }\href@noop {} {\emph {\bibinfo {title} {Multiphase Flow and
  Fluidization}}}\ (\bibinfo  {publisher} {Academic Press},\ \bibinfo {year}
  {1994})\BibitemShut {NoStop}%
\bibitem [{\citenamefont {Jackson}(2000)}]{J00}%
  \BibitemOpen
  \bibfield  {author} {\bibinfo {author} {\bibfnamefont {R.}~\bibnamefont
  {Jackson}},\ }\href@noop {} {\emph {\bibinfo {title} {The {D}ynamics of
  {F}luidized {P}articles}}}\ (\bibinfo  {publisher} {Cambridge University
  Press, New York},\ \bibinfo {year} {2000})\BibitemShut {NoStop}%
\bibitem [{\citenamefont {Koch}\ and\ \citenamefont {Hill}(2001)}]{KH01}%
  \BibitemOpen
  \bibfield  {author} {\bibinfo {author} {\bibfnamefont {D.~L.}\ \bibnamefont
  {Koch}}\ and\ \bibinfo {author} {\bibfnamefont {R.~J.}\ \bibnamefont
  {Hill}},\ }\bibfield  {title} {\enquote {\bibinfo {title} {Inertial effects
  in suspensions and porous-media flows},}\ }\href@noop {} {\bibfield
  {journal} {\bibinfo  {journal} {Annu. Rev. Fluid Mech.}\ }\textbf {\bibinfo
  {volume} {33}},\ \bibinfo {pages} {619--647} (\bibinfo {year}
  {2001})}\BibitemShut {NoStop}%
\bibitem [{\citenamefont {Zhao}\ and\ \citenamefont {Wang}(2021)}]{ZW21}%
  \BibitemOpen
  \bibfield  {author} {\bibinfo {author} {\bibfnamefont {B.}~\bibnamefont
  {Zhao}}\ and\ \bibinfo {author} {\bibfnamefont {J.}~\bibnamefont {Wang}},\
  }\bibfield  {title} {\enquote {\bibinfo {title} {Kinetic theory of
  polydisperse gas-solid flow: {N}avier-stokes transport coefficients},}\
  }\href@noop {} {\bibfield  {journal} {\bibinfo  {journal} {Phys. Fluids}\
  }\textbf {\bibinfo {volume} {33}},\ \bibinfo {pages} {103322} (\bibinfo
  {year} {2021})}\BibitemShut {NoStop}%
\bibitem [{\citenamefont {Chassagne}, \citenamefont {Bonamy},\ and\
  \citenamefont {Chauchat}(2023)}]{CBC23}%
  \BibitemOpen
  \bibfield  {author} {\bibinfo {author} {\bibfnamefont {R.}~\bibnamefont
  {Chassagne}}, \bibinfo {author} {\bibfnamefont {C.}~\bibnamefont {Bonamy}}, \
  and\ \bibinfo {author} {\bibfnamefont {J.}~\bibnamefont {Chauchat}},\
  }\bibfield  {title} {\enquote {\bibinfo {title} {A frictional-collisional
  model for bedload transport based on kinetic theory of granular flows:
  discrete and continuum approaches},}\ }\href@noop {} {\bibfield  {journal}
  {\bibinfo  {journal} {J. Fluid Mech.}\ }\textbf {\bibinfo {volume} {964}},\
  \bibinfo {pages} {A27} (\bibinfo {year} {2023})}\BibitemShut {NoStop}%
\bibitem [{\citenamefont {Garz\'o}\ \emph {et~al.}(2012)\citenamefont
  {Garz\'o}, \citenamefont {Tenneti}, \citenamefont {Subramaniam},\ and\
  \citenamefont {Hrenya}}]{GTSH12}%
  \BibitemOpen
  \bibfield  {author} {\bibinfo {author} {\bibfnamefont {V.}~\bibnamefont
  {Garz\'o}}, \bibinfo {author} {\bibfnamefont {S.}~\bibnamefont {Tenneti}},
  \bibinfo {author} {\bibfnamefont {S.}~\bibnamefont {Subramaniam}}, \ and\
  \bibinfo {author} {\bibfnamefont {C.~M.}\ \bibnamefont {Hrenya}},\ }\bibfield
   {title} {\enquote {\bibinfo {title} {Enskog kinetic theory for monodisperse
  gas-solid flows},}\ }\href@noop {} {\bibfield  {journal} {\bibinfo  {journal}
  {J. Fluid Mech.}\ }\textbf {\bibinfo {volume} {712}},\ \bibinfo {pages}
  {129--168} (\bibinfo {year} {2012})}\BibitemShut {NoStop}%
\bibitem [{\citenamefont {Tsao}\ and\ \citenamefont {Koch}(1995)}]{TK95}%
  \BibitemOpen
  \bibfield  {author} {\bibinfo {author} {\bibfnamefont {H.-K.}\ \bibnamefont
  {Tsao}}\ and\ \bibinfo {author} {\bibfnamefont {D.~L.}\ \bibnamefont
  {Koch}},\ }\bibfield  {title} {\enquote {\bibinfo {title} {Simple shear flows
  of dilute gas--solid suspensions},}\ }\href@noop {} {\bibfield  {journal}
  {\bibinfo  {journal} {J. Fluid Mech.}\ }\textbf {\bibinfo {volume} {296}},\
  \bibinfo {pages} {211--245} (\bibinfo {year} {1995})}\BibitemShut {NoStop}%
\bibitem [{\citenamefont {Sangani}\ \emph {et~al.}(1996)\citenamefont
  {Sangani}, \citenamefont {Mo}, \citenamefont {Tsao},\ and\ \citenamefont
  {Koch}}]{SMTK96}%
  \BibitemOpen
  \bibfield  {author} {\bibinfo {author} {\bibfnamefont {A.~S.}\ \bibnamefont
  {Sangani}}, \bibinfo {author} {\bibfnamefont {G.}~\bibnamefont {Mo}},
  \bibinfo {author} {\bibfnamefont {H.-K.}\ \bibnamefont {Tsao}}, \ and\
  \bibinfo {author} {\bibfnamefont {D.~L.}\ \bibnamefont {Koch}},\ }\bibfield
  {title} {\enquote {\bibinfo {title} {Simple shear flows of dense gas-solid
  suspensions at finite \textsc{S}tokes numbers},}\ }\href@noop {} {\bibfield
  {journal} {\bibinfo  {journal} {J. Fluid Mech.}\ }\textbf {\bibinfo {volume}
  {313}},\ \bibinfo {pages} {309--341} (\bibinfo {year} {1996})}\BibitemShut
  {NoStop}%
\bibitem [{\citenamefont {Wylie}\ \emph {et~al.}(2009)\citenamefont {Wylie},
  \citenamefont {Zhang}, \citenamefont {Li},\ and\ \citenamefont
  {Hengyi}}]{WZLH09}%
  \BibitemOpen
  \bibfield  {author} {\bibinfo {author} {\bibfnamefont {J.~J.}\ \bibnamefont
  {Wylie}}, \bibinfo {author} {\bibfnamefont {Q.}~\bibnamefont {Zhang}},
  \bibinfo {author} {\bibfnamefont {Y.}~\bibnamefont {Li}}, \ and\ \bibinfo
  {author} {\bibfnamefont {X.}~\bibnamefont {Hengyi}},\ }\bibfield  {title}
  {\enquote {\bibinfo {title} {Driven inelastic-particle systems with drag},}\
  }\href@noop {} {\bibfield  {journal} {\bibinfo  {journal} {Phys. Rev. E}\
  }\textbf {\bibinfo {volume} {79}},\ \bibinfo {pages} {031301} (\bibinfo
  {year} {2009})}\BibitemShut {NoStop}%
\bibitem [{\citenamefont {Parmentier}\ and\ \citenamefont
  {Simonin}(2012)}]{PS12}%
  \BibitemOpen
  \bibfield  {author} {\bibinfo {author} {\bibfnamefont {J.-F.}\ \bibnamefont
  {Parmentier}}\ and\ \bibinfo {author} {\bibfnamefont {O.}~\bibnamefont
  {Simonin}},\ }\bibfield  {title} {\enquote {\bibinfo {title} {Transition
  models from the quenched to ignited states for flows of inertial particles
  suspended in a simple sheared viscous fluid},}\ }\href@noop {} {\bibfield
  {journal} {\bibinfo  {journal} {J. Fluid Mech.}\ }\textbf {\bibinfo {volume}
  {711}},\ \bibinfo {pages} {147--160} (\bibinfo {year} {2012})}\BibitemShut
  {NoStop}%
\bibitem [{\citenamefont {Heussinger}(2013)}]{H13}%
  \BibitemOpen
  \bibfield  {author} {\bibinfo {author} {\bibfnamefont {C.}~\bibnamefont
  {Heussinger}},\ }\bibfield  {title} {\enquote {\bibinfo {title} {Shear
  thickening in granular suspensions: {I}nterparticle friction and dynamically
  correlated clusters},}\ }\href@noop {} {\bibfield  {journal} {\bibinfo
  {journal} {Phys. Rev. E}\ }\textbf {\bibinfo {volume} {88}},\ \bibinfo
  {pages} {050201 (R)} (\bibinfo {year} {2013})}\BibitemShut {NoStop}%
\bibitem [{\citenamefont {Wang}\ \emph {et~al.}(2014)\citenamefont {Wang},
  \citenamefont {Grob}, \citenamefont {Zippelius},\ and\ \citenamefont
  {Sperl}}]{WGZS14}%
  \BibitemOpen
  \bibfield  {author} {\bibinfo {author} {\bibfnamefont {T.}~\bibnamefont
  {Wang}}, \bibinfo {author} {\bibfnamefont {M.}~\bibnamefont {Grob}}, \bibinfo
  {author} {\bibfnamefont {A.}~\bibnamefont {Zippelius}}, \ and\ \bibinfo
  {author} {\bibfnamefont {M.}~\bibnamefont {Sperl}},\ }\bibfield  {title}
  {\enquote {\bibinfo {title} {Active microrheology of driven granular
  particles},}\ }\href@noop {} {\bibfield  {journal} {\bibinfo  {journal}
  {Phys. Rev. E}\ }\textbf {\bibinfo {volume} {89}},\ \bibinfo {pages}
  {{042}{209}} (\bibinfo {year} {2014})}\BibitemShut {NoStop}%
\bibitem [{\citenamefont {Saha}\ and\ \citenamefont {Alam}(2017)}]{SA17}%
  \BibitemOpen
  \bibfield  {author} {\bibinfo {author} {\bibfnamefont {S.}~\bibnamefont
  {Saha}}\ and\ \bibinfo {author} {\bibfnamefont {M.}~\bibnamefont {Alam}},\
  }\bibfield  {title} {\enquote {\bibinfo {title} {Revisiting ignited-quenched
  transition and the non-{N}ewtonian rheology of a sheared dilute gas-solid
  suspension},}\ }\href@noop {} {\bibfield  {journal} {\bibinfo  {journal} {J.
  Fluid Mech.}\ }\textbf {\bibinfo {volume} {833}},\ \bibinfo {pages}
  {206--246} (\bibinfo {year} {2017})}\BibitemShut {NoStop}%
\bibitem [{\citenamefont {Alam}, \citenamefont {Saha},\ and\ \citenamefont
  {Gupta}(2019)}]{ASG19}%
  \BibitemOpen
  \bibfield  {author} {\bibinfo {author} {\bibfnamefont {M.}~\bibnamefont
  {Alam}}, \bibinfo {author} {\bibfnamefont {S.}~\bibnamefont {Saha}}, \ and\
  \bibinfo {author} {\bibfnamefont {R.}~\bibnamefont {Gupta}},\ }\bibfield
  {title} {\enquote {\bibinfo {title} {Unified theory for a sheared gas-solid
  suspension: from rapid granular suspension to its small-{S}tokes-number
  limit},}\ }\href@noop {} {\bibfield  {journal} {\bibinfo  {journal} {J. Fluid
  Mech.}\ }\textbf {\bibinfo {volume} {870}},\ \bibinfo {pages} {1175--1193}
  (\bibinfo {year} {2019})}\BibitemShut {NoStop}%
\bibitem [{\citenamefont {Saha}\ and\ \citenamefont {Alam}(2020)}]{SA20}%
  \BibitemOpen
  \bibfield  {author} {\bibinfo {author} {\bibfnamefont {S.}~\bibnamefont
  {Saha}}\ and\ \bibinfo {author} {\bibfnamefont {M.}~\bibnamefont {Alam}},\
  }\bibfield  {title} {\enquote {\bibinfo {title} {Burnett-order constitutive
  relations, second moment anisotropy and co-existing states in sheared dense
  gas-solid suspensions},}\ }\href@noop {} {\bibfield  {journal} {\bibinfo
  {journal} {J. Fluid Mech.}\ }\textbf {\bibinfo {volume} {887}},\ \bibinfo
  {pages} {A9} (\bibinfo {year} {2020})}\BibitemShut {NoStop}%
\bibitem [{\citenamefont {G\'omez~Gonz\'alez}\ and\ \citenamefont
  {Garz\'o}(2022{\natexlab{a}})}]{GG22}%
  \BibitemOpen
  \bibfield  {author} {\bibinfo {author} {\bibfnamefont {R.}~\bibnamefont
  {G\'omez~Gonz\'alez}}\ and\ \bibinfo {author} {\bibfnamefont
  {V.}~\bibnamefont {Garz\'o}},\ }\bibfield  {title} {\enquote {\bibinfo
  {title} {Kinetic theory of granular particles immersed in a molecular gas},}\
  }\href@noop {} {\bibfield  {journal} {\bibinfo  {journal} {J. Fluid Mech.}\
  }\textbf {\bibinfo {volume} {943}},\ \bibinfo {pages} {A9} (\bibinfo {year}
  {2022}{\natexlab{a}})}\BibitemShut {NoStop}%
\bibitem [{\citenamefont {G\'omez~Gonz\'alez}, \citenamefont {Khalil},\ and\
  \citenamefont {Garz\'o}(2020)}]{GKG20}%
  \BibitemOpen
  \bibfield  {author} {\bibinfo {author} {\bibfnamefont {R.}~\bibnamefont
  {G\'omez~Gonz\'alez}}, \bibinfo {author} {\bibfnamefont {N.}~\bibnamefont
  {Khalil}}, \ and\ \bibinfo {author} {\bibfnamefont {V.}~\bibnamefont
  {Garz\'o}},\ }\bibfield  {title} {\enquote {\bibinfo {title} {Enskog kinetic
  theory for multicomponent granular suspensions},}\ }\href@noop {} {\bibfield
  {journal} {\bibinfo  {journal} {Phys. Rev. E}\ }\textbf {\bibinfo {volume}
  {101}},\ \bibinfo {pages} {012904} (\bibinfo {year} {2020})}\BibitemShut
  {NoStop}%
\bibitem [{\citenamefont {G\'omez~Gonz\'alez}\ and\ \citenamefont
  {Garz\'o}(2019)}]{GGG19a}%
  \BibitemOpen
  \bibfield  {author} {\bibinfo {author} {\bibfnamefont {R.}~\bibnamefont
  {G\'omez~Gonz\'alez}}\ and\ \bibinfo {author} {\bibfnamefont
  {V.}~\bibnamefont {Garz\'o}},\ }\bibfield  {title} {\enquote {\bibinfo
  {title} {Transport coefficients for granular suspensions at moderate
  densities},}\ }\href@noop {} {\bibfield  {journal} {\bibinfo  {journal} {J.
  Stat. Mech.}\ }\textbf {\bibinfo {volume} {093204}} (\bibinfo {year}
  {2019})}\BibitemShut {NoStop}%
\bibitem [{\citenamefont {Chapman}\ and\ \citenamefont {Cowling}(1970)}]{CC70}%
  \BibitemOpen
  \bibfield  {author} {\bibinfo {author} {\bibfnamefont {S.}~\bibnamefont
  {Chapman}}\ and\ \bibinfo {author} {\bibfnamefont {T.~G.}\ \bibnamefont
  {Cowling}},\ }\href@noop {} {\emph {\bibinfo {title} {The Mathematical Theory
  of Nonuniform Gases}}}\ (\bibinfo  {publisher} {Cambridge University Press,
  Cambridge},\ \bibinfo {year} {1970})\BibitemShut {NoStop}%
\bibitem [{\citenamefont {Sela}\ and\ \citenamefont {Goldhirsch}(1998)}]{SG98}%
  \BibitemOpen
  \bibfield  {author} {\bibinfo {author} {\bibfnamefont {N.}~\bibnamefont
  {Sela}}\ and\ \bibinfo {author} {\bibfnamefont {I.}~\bibnamefont
  {Goldhirsch}},\ }\bibfield  {title} {\enquote {\bibinfo {title} {Hydrodynamic
  equations for rapid flows of smooth inelastic spheres to \textsc{B}urnett
  order},}\ }\href@noop {} {\bibfield  {journal} {\bibinfo  {journal} {J. Fluid
  Mech.}\ }\textbf {\bibinfo {volume} {361}},\ \bibinfo {pages} {41--74}
  (\bibinfo {year} {1998})}\BibitemShut {NoStop}%
\bibitem [{\citenamefont {Serero}\ \emph {et~al.}(2006)\citenamefont {Serero},
  \citenamefont {Goldhirsch}, \citenamefont {Noskowicz},\ and\ \citenamefont
  {Tan}}]{SGNT06}%
  \BibitemOpen
  \bibfield  {author} {\bibinfo {author} {\bibfnamefont {D.}~\bibnamefont
  {Serero}}, \bibinfo {author} {\bibfnamefont {I.}~\bibnamefont {Goldhirsch}},
  \bibinfo {author} {\bibfnamefont {S.~H.}\ \bibnamefont {Noskowicz}}, \ and\
  \bibinfo {author} {\bibfnamefont {M.~L.}\ \bibnamefont {Tan}},\ }\bibfield
  {title} {\enquote {\bibinfo {title} {Hydrodynamics of granular gases and
  granular gas mixtures},}\ }\href@noop {} {\bibfield  {journal} {\bibinfo
  {journal} {J. Fluid Mech.}\ }\textbf {\bibinfo {volume} {554}},\ \bibinfo
  {pages} {237--258} (\bibinfo {year} {2006})}\BibitemShut {NoStop}%
\bibitem [{\citenamefont {Garz\'o}, \citenamefont {Dufty},\ and\ \citenamefont
  {Hrenya}(2007)}]{GDH07}%
  \BibitemOpen
  \bibfield  {author} {\bibinfo {author} {\bibfnamefont {V.}~\bibnamefont
  {Garz\'o}}, \bibinfo {author} {\bibfnamefont {J.~W.}\ \bibnamefont {Dufty}},
  \ and\ \bibinfo {author} {\bibfnamefont {C.~M.}\ \bibnamefont {Hrenya}},\
  }\bibfield  {title} {\enquote {\bibinfo {title} {Enskog theory for
  polydisperse granular mixtures. \textsc{I}. \textsc{N}avier--\textsc{S}tokes
  order transport},}\ }\href@noop {} {\bibfield  {journal} {\bibinfo  {journal}
  {Phys. Rev. E}\ }\textbf {\bibinfo {volume} {76}},\ \bibinfo {pages}
  {{031}{303}} (\bibinfo {year} {2007})}\BibitemShut {NoStop}%
\bibitem [{\citenamefont {Garz\'o}, \citenamefont {Hrenya},\ and\ \citenamefont
  {Dufty}(2007)}]{GHD07}%
  \BibitemOpen
  \bibfield  {author} {\bibinfo {author} {\bibfnamefont {V.}~\bibnamefont
  {Garz\'o}}, \bibinfo {author} {\bibfnamefont {C.~M.}\ \bibnamefont {Hrenya}},
  \ and\ \bibinfo {author} {\bibfnamefont {J.~W.}\ \bibnamefont {Dufty}},\
  }\bibfield  {title} {\enquote {\bibinfo {title} {Enskog theory for
  polydisperse granular mixtures. \textsc{II}. \textsc{S}onine polynomial
  approximation},}\ }\href@noop {} {\bibfield  {journal} {\bibinfo  {journal}
  {Phys. Rev. E}\ }\textbf {\bibinfo {volume} {76}},\ \bibinfo {pages}
  {{031}{304}} (\bibinfo {year} {2007})}\BibitemShut {NoStop}%
\bibitem [{\citenamefont {Garz\'o}(2021)}]{G21}%
  \BibitemOpen
  \bibfield  {author} {\bibinfo {author} {\bibfnamefont {V.}~\bibnamefont
  {Garz\'o}},\ }\bibfield  {title} {\enquote {\bibinfo {title} {Comment on
  “{K}inetic theory models for granular mixtures with unequal granular
  temperature: {H}ydrodynamic velocity” [{P}hys. {F}luids 33, 043321
  (2021)]},}\ }\href@noop {} {\bibfield  {journal} {\bibinfo  {journal} {Phys.
  Fluids}\ }\textbf {\bibinfo {volume} {33}},\ \bibinfo {pages} {089101}
  (\bibinfo {year} {2021})}\BibitemShut {NoStop}%
\bibitem [{\citenamefont {Garc\'ia~Chamorro}\ and\ \citenamefont
  {Garz\'o}(2023)}]{ChG23}%
  \BibitemOpen
  \bibfield  {author} {\bibinfo {author} {\bibfnamefont {M.}~\bibnamefont
  {Garc\'ia~Chamorro}}\ and\ \bibinfo {author} {\bibfnamefont {V.}~\bibnamefont
  {Garz\'o}},\ }\bibfield  {title} {\enquote {\bibinfo {title} {Assessment of
  kinetic theories for moderately dense granular binary mixtures: {S}hear
  viscosity coefficient},}\ }\href@noop {} {\bibfield  {journal} {\bibinfo
  {journal} {Phys. Fluids}\ }\textbf {\bibinfo {volume} {35}},\ \bibinfo
  {pages} {027121} (\bibinfo {year} {2023})}\BibitemShut {NoStop}%
\bibitem [{\citenamefont {Solsvik}\ and\ \citenamefont {Manger}(2021)}]{SM21}%
  \BibitemOpen
  \bibfield  {author} {\bibinfo {author} {\bibfnamefont {J.}~\bibnamefont
  {Solsvik}}\ and\ \bibinfo {author} {\bibfnamefont {E.}~\bibnamefont
  {Manger}},\ }\bibfield  {title} {\enquote {\bibinfo {title} {Kinetic theory
  models for granular mixtures with unequal granular temperature:
  {H}ydrodynamic velocity},}\ }\href@noop {} {\bibfield  {journal} {\bibinfo
  {journal} {Phys. Fluids}\ }\textbf {\bibinfo {volume} {33}},\ \bibinfo
  {pages} {043321} (\bibinfo {year} {2021})}\BibitemShut {NoStop}%
\bibitem [{\citenamefont {Garz\'o}\ and\ \citenamefont
  {Montanero}(2004)}]{GM04}%
  \BibitemOpen
  \bibfield  {author} {\bibinfo {author} {\bibfnamefont {V.}~\bibnamefont
  {Garz\'o}}\ and\ \bibinfo {author} {\bibfnamefont {J.~M.}\ \bibnamefont
  {Montanero}},\ }\bibfield  {title} {\enquote {\bibinfo {title} {Diffusion of
  impurities in a granular gas},}\ }\href@noop {} {\bibfield  {journal}
  {\bibinfo  {journal} {Phys. Rev. E}\ }\textbf {\bibinfo {volume} {69}},\
  \bibinfo {pages} {{021}{301}} (\bibinfo {year} {2004})}\BibitemShut {NoStop}%
\bibitem [{\citenamefont {Garz\'o}\ and\ \citenamefont
  {Montanero}(2007)}]{GM07}%
  \BibitemOpen
  \bibfield  {author} {\bibinfo {author} {\bibfnamefont {V.}~\bibnamefont
  {Garz\'o}}\ and\ \bibinfo {author} {\bibfnamefont {J.~M.}\ \bibnamefont
  {Montanero}},\ }\bibfield  {title} {\enquote {\bibinfo {title}
  {\textsc{N}avier--\textsc{S}tokes transport coefficients of $d$-dimensional
  granular binary mixtures at low-density},}\ }\href@noop {} {\bibfield
  {journal} {\bibinfo  {journal} {J. Stat. Phys.}\ }\textbf {\bibinfo {volume}
  {129}},\ \bibinfo {pages} {27--58} (\bibinfo {year} {2007})}\BibitemShut
  {NoStop}%
\bibitem [{\citenamefont {Garz\'o}\ and\ \citenamefont
  {Vega~Reyes}(2009)}]{GV09}%
  \BibitemOpen
  \bibfield  {author} {\bibinfo {author} {\bibfnamefont {V.}~\bibnamefont
  {Garz\'o}}\ and\ \bibinfo {author} {\bibfnamefont {F.}~\bibnamefont
  {Vega~Reyes}},\ }\bibfield  {title} {\enquote {\bibinfo {title} {Mass
  transport of impurities in a moderately dense granular gas},}\ }\href@noop {}
  {\bibfield  {journal} {\bibinfo  {journal} {Phys. Rev. E}\ }\textbf {\bibinfo
  {volume} {79}},\ \bibinfo {pages} {{041}{303}} (\bibinfo {year}
  {2009})}\BibitemShut {NoStop}%
\bibitem [{\citenamefont {Garz\'o}\ and\ \citenamefont
  {Vega~Reyes}(2012)}]{GV12}%
  \BibitemOpen
  \bibfield  {author} {\bibinfo {author} {\bibfnamefont {V.}~\bibnamefont
  {Garz\'o}}\ and\ \bibinfo {author} {\bibfnamefont {F.}~\bibnamefont
  {Vega~Reyes}},\ }\bibfield  {title} {\enquote {\bibinfo {title} {Segregation
  of an intruder in a heated granular gas},}\ }\href@noop {} {\bibfield
  {journal} {\bibinfo  {journal} {Phys. Rev. E}\ }\textbf {\bibinfo {volume}
  {85}},\ \bibinfo {pages} {{021}{308}} (\bibinfo {year} {2012})}\BibitemShut
  {NoStop}%
\bibitem [{\citenamefont {G\'omez~Gonz\'alez}\ \emph
  {et~al.}(2023)\citenamefont {G\'omez~Gonz\'alez}, \citenamefont {Abad},
  \citenamefont {Bravo~Yuste},\ and\ \citenamefont {Garz\'o}}]{GGABYG23}%
  \BibitemOpen
  \bibfield  {author} {\bibinfo {author} {\bibfnamefont {R.}~\bibnamefont
  {G\'omez~Gonz\'alez}}, \bibinfo {author} {\bibfnamefont {E.}~\bibnamefont
  {Abad}}, \bibinfo {author} {\bibfnamefont {S.}~\bibnamefont {Bravo~Yuste}}, \
  and\ \bibinfo {author} {\bibfnamefont {V.}~\bibnamefont {Garz\'o}},\
  }\bibfield  {title} {\enquote {\bibinfo {title} {Diffusion of intruders in
  granular suspensions: Enskog theory and random walk interpretation},}\
  }\href@noop {} {\bibfield  {journal} {\bibinfo  {journal} {Phys. Rev. E}\
  }\textbf {\bibinfo {volume} {108}},\ \bibinfo {pages} {024903} (\bibinfo
  {year} {2023})}\BibitemShut {NoStop}%
\bibitem [{\citenamefont {Bird}(1994)}]{B94}%
  \BibitemOpen
  \bibfield  {author} {\bibinfo {author} {\bibfnamefont {G.~A.}\ \bibnamefont
  {Bird}},\ }\href@noop {} {\emph {\bibinfo {title} {Molecular Gas Dynamics and
  the Direct Simulation Monte Carlo of Gas Flows}}}\ (\bibinfo  {publisher}
  {Clarendon, Oxford},\ \bibinfo {year} {1994})\BibitemShut {NoStop}%
\bibitem [{\citenamefont {G\'omez~Gonz\'alez}\ and\ \citenamefont
  {Garz\'o}(2022{\natexlab{b}})}]{GG22a}%
  \BibitemOpen
  \bibfield  {author} {\bibinfo {author} {\bibfnamefont {R.}~\bibnamefont
  {G\'omez~Gonz\'alez}}\ and\ \bibinfo {author} {\bibfnamefont
  {V.}~\bibnamefont {Garz\'o}},\ }\bibfield  {title} {\enquote {\bibinfo
  {title} {Kinetic theory of binary granular suspensions at low density.
  {T}hermal diffusion segregation},}\ }in\ \href@noop {} {\emph {\bibinfo
  {booktitle} {Nonequilibrium {T}hermodynamics and {F}luctuation
  {K}inetics}}},\ \bibinfo {editor} {edited by\ \bibinfo {editor}
  {\bibfnamefont {L.}~\bibnamefont {Brenig}}, \bibinfo {editor} {\bibfnamefont
  {N.}~\bibnamefont {Brilliantov}}, \ and\ \bibinfo {editor} {\bibfnamefont
  {M.}~\bibnamefont {Tlidi}}}\ (\bibinfo  {publisher} {Fundamental {T}heories
  of {P}hysics 208, Springer},\ \bibinfo {year} {2022})\BibitemShut {NoStop}%
\bibitem [{\citenamefont {Williams}\ and\ \citenamefont
  {MacKintosh}(1996)}]{WM96}%
  \BibitemOpen
  \bibfield  {author} {\bibinfo {author} {\bibfnamefont {D.~R.~M.}\
  \bibnamefont {Williams}}\ and\ \bibinfo {author} {\bibfnamefont {F.~C.}\
  \bibnamefont {MacKintosh}},\ }\bibfield  {title} {\enquote {\bibinfo {title}
  {Driven granular media in one dimension: {C}orrelations and equation of
  state},}\ }\href@noop {} {\bibfield  {journal} {\bibinfo  {journal} {Phys.
  Rev. E}\ }\textbf {\bibinfo {volume} {54}},\ \bibinfo {pages} {R9--R12}
  (\bibinfo {year} {1996})}\BibitemShut {NoStop}%
\bibitem [{\citenamefont {van Noije}\ and\ \citenamefont {Ernst}(1998)}]{NE98}%
  \BibitemOpen
  \bibfield  {author} {\bibinfo {author} {\bibfnamefont {T.~P.~C.}\
  \bibnamefont {van Noije}}\ and\ \bibinfo {author} {\bibfnamefont {M.~H.}\
  \bibnamefont {Ernst}},\ }\bibfield  {title} {\enquote {\bibinfo {title}
  {Velocity distributions in homogeneous granular fluids: the free and heated
  case},}\ }\href@noop {} {\bibfield  {journal} {\bibinfo  {journal} {Granular
  Matter}\ }\textbf {\bibinfo {volume} {1}},\ \bibinfo {pages} {57--64}
  (\bibinfo {year} {1998})}\BibitemShut {NoStop}%
\bibitem [{\citenamefont {Lutsko}, \citenamefont {Brey},\ and\ \citenamefont
  {Dufty}(2002)}]{LBD02}%
  \BibitemOpen
  \bibfield  {author} {\bibinfo {author} {\bibfnamefont {J.~F.}\ \bibnamefont
  {Lutsko}}, \bibinfo {author} {\bibfnamefont {J.~J.}\ \bibnamefont {Brey}}, \
  and\ \bibinfo {author} {\bibfnamefont {J.~W.}\ \bibnamefont {Dufty}},\
  }\bibfield  {title} {\enquote {\bibinfo {title} {Diffusion in a granular
  fluid. \textsc{II}. \textsc{S}imulation},}\ }\href@noop {} {\bibfield
  {journal} {\bibinfo  {journal} {Phys. Rev. E}\ }\textbf {\bibinfo {volume}
  {65}},\ \bibinfo {pages} {{051}{304}} (\bibinfo {year} {2002})}\BibitemShut
  {NoStop}%
\bibitem [{\citenamefont {Dahl}\ \emph {et~al.}(2002)\citenamefont {Dahl},
  \citenamefont {Hrenya}, \citenamefont {Garz\'o},\ and\ \citenamefont
  {Dufty}}]{DHGD02}%
  \BibitemOpen
  \bibfield  {author} {\bibinfo {author} {\bibfnamefont {S.~R.}\ \bibnamefont
  {Dahl}}, \bibinfo {author} {\bibfnamefont {C.~M.}\ \bibnamefont {Hrenya}},
  \bibinfo {author} {\bibfnamefont {V.}~\bibnamefont {Garz\'o}}, \ and\
  \bibinfo {author} {\bibfnamefont {J.~W.}\ \bibnamefont {Dufty}},\ }\bibfield
  {title} {\enquote {\bibinfo {title} {Kinetic temperatures for a granular
  mixture},}\ }\href@noop {} {\bibfield  {journal} {\bibinfo  {journal} {Phys.
  Rev. E}\ }\textbf {\bibinfo {volume} {66}},\ \bibinfo {pages} {{041}{301}}
  (\bibinfo {year} {2002})}\BibitemShut {NoStop}%
\bibitem [{\citenamefont {Mitrano}\ \emph {et~al.}(2011)\citenamefont
  {Mitrano}, \citenamefont {Dhal}, \citenamefont {Cromer}, \citenamefont
  {Pacella},\ and\ \citenamefont {Hrenya}}]{MDCPH11}%
  \BibitemOpen
  \bibfield  {author} {\bibinfo {author} {\bibfnamefont {P.~P.}\ \bibnamefont
  {Mitrano}}, \bibinfo {author} {\bibfnamefont {S.~R.}\ \bibnamefont {Dhal}},
  \bibinfo {author} {\bibfnamefont {D.~J.}\ \bibnamefont {Cromer}}, \bibinfo
  {author} {\bibfnamefont {M.~S.}\ \bibnamefont {Pacella}}, \ and\ \bibinfo
  {author} {\bibfnamefont {C.~M.}\ \bibnamefont {Hrenya}},\ }\bibfield  {title}
  {\enquote {\bibinfo {title} {Instabilities in the homogeneous cooling of a
  granular gas: {A} quantitative assessment of kinetic-theory predictions},}\
  }\href@noop {} {\bibfield  {journal} {\bibinfo  {journal} {Phys. Fluids}\
  }\textbf {\bibinfo {volume} {23}},\ \bibinfo {pages} {{093}{303}} (\bibinfo
  {year} {2011})}\BibitemShut {NoStop}%
\bibitem [{\citenamefont {Mitrano}, \citenamefont {Garz\'o},\ and\
  \citenamefont {Hrenya}(2014)}]{MGH14}%
  \BibitemOpen
  \bibfield  {author} {\bibinfo {author} {\bibfnamefont {P.~P.}\ \bibnamefont
  {Mitrano}}, \bibinfo {author} {\bibfnamefont {V.}~\bibnamefont {Garz\'o}}, \
  and\ \bibinfo {author} {\bibfnamefont {C.~M.}\ \bibnamefont {Hrenya}},\
  }\bibfield  {title} {\enquote {\bibinfo {title} {Instabilities in granular
  binary mixtures at moderate densities},}\ }\href@noop {} {\bibfield
  {journal} {\bibinfo  {journal} {Phys. Rev. E}\ }\textbf {\bibinfo {volume}
  {89}},\ \bibinfo {pages} {{020}{201}(R)} (\bibinfo {year}
  {2014})}\BibitemShut {NoStop}%
\bibitem [{\citenamefont {Garz\'o}(2019)}]{G19}%
  \BibitemOpen
  \bibfield  {author} {\bibinfo {author} {\bibfnamefont {V.}~\bibnamefont
  {Garz\'o}},\ }\href@noop {} {\emph {\bibinfo {title} {Granular Gaseous
  Flows}}}\ (\bibinfo  {publisher} {Springer Nature, Cham},\ \bibinfo {year}
  {2019})\BibitemShut {NoStop}%
\bibitem [{\citenamefont {Van~der {H}oef}, \citenamefont {Beetstra},\ and\
  \citenamefont {Kuipers}(2005)}]{HBK05}%
  \BibitemOpen
  \bibfield  {author} {\bibinfo {author} {\bibfnamefont {M.~A.}\ \bibnamefont
  {Van~der {H}oef}}, \bibinfo {author} {\bibfnamefont {R.}~\bibnamefont
  {Beetstra}}, \ and\ \bibinfo {author} {\bibfnamefont {J.~A.~M.}\ \bibnamefont
  {Kuipers}},\ }\bibfield  {title} {\enquote {\bibinfo {title}
  {Lattice-{B}oltzmann simulations of low-{R}eynolds number flow past mon- and
  bidisperse arrays of spheres: resutls for the permeability drag force},}\
  }\href@noop {} {\bibfield  {journal} {\bibinfo  {journal} {J. Fluid Mech.}\
  }\textbf {\bibinfo {volume} {528}},\ \bibinfo {pages} {233--254} (\bibinfo
  {year} {2005})}\BibitemShut {NoStop}%
\bibitem [{\citenamefont {Beetstra}, \citenamefont {Van~der {H}oef},\ and\
  \citenamefont {Kuipers}(2007)}]{BHK07}%
  \BibitemOpen
  \bibfield  {author} {\bibinfo {author} {\bibfnamefont {R.}~\bibnamefont
  {Beetstra}}, \bibinfo {author} {\bibfnamefont {M.~A.}\ \bibnamefont {Van~der
  {H}oef}}, \ and\ \bibinfo {author} {\bibfnamefont {J.~A.~M.}\ \bibnamefont
  {Kuipers}},\ }\bibfield  {title} {\enquote {\bibinfo {title} {Erratum: {D}rag
  force of intermediate {R}eynolds number flow past mon- and bidisperse arrays
  of spheres [{A}{I}{C}h{E} {J}. 53, 489-591 (2007)]},}\ }\href@noop {}
  {\bibfield  {journal} {\bibinfo  {journal} {AIChE J.}\ }\textbf {\bibinfo
  {volume} {53}},\ \bibinfo {pages} {3020} (\bibinfo {year}
  {2007})}\BibitemShut {NoStop}%
\bibitem [{\citenamefont {Yin}\ and\ \citenamefont {Sundaresan}(2009)}]{YS09b}%
  \BibitemOpen
  \bibfield  {author} {\bibinfo {author} {\bibfnamefont {X.}~\bibnamefont
  {Yin}}\ and\ \bibinfo {author} {\bibfnamefont {S.}~\bibnamefont
  {Sundaresan}},\ }\bibfield  {title} {\enquote {\bibinfo {title}
  {Fluid-particle drag in low-{R}eynolds-number polydisperse gas-solid
  suspensions},}\ }\href@noop {} {\bibfield  {journal} {\bibinfo  {journal}
  {AIChE}\ }\textbf {\bibinfo {volume} {55}},\ \bibinfo {pages} {1352--1368}
  (\bibinfo {year} {2009})}\BibitemShut {NoStop}%
\bibitem [{\citenamefont {Koch}\ and\ \citenamefont {Sangani}(1999)}]{KS99}%
  \BibitemOpen
  \bibfield  {author} {\bibinfo {author} {\bibfnamefont {D.~L.}\ \bibnamefont
  {Koch}}\ and\ \bibinfo {author} {\bibfnamefont {A.~S.}\ \bibnamefont
  {Sangani}},\ }\bibfield  {title} {\enquote {\bibinfo {title} {Particle
  pressure and marginal stability limits for a homogeneous monodiperse
  gas-fluidized bed: Kinetic theory and numerical simulations},}\ }\href@noop
  {} {\bibfield  {journal} {\bibinfo  {journal} {J. Fluid Mech.}\ }\textbf
  {\bibinfo {volume} {400}},\ \bibinfo {pages} {229} (\bibinfo {year}
  {1999})}\BibitemShut {NoStop}%
\bibitem [{\citenamefont {van Kampen}(2007)}]{K07}%
  \BibitemOpen
  \bibfield  {author} {\bibinfo {author} {\bibfnamefont {N.~G.}\ \bibnamefont
  {van Kampen}},\ }\href@noop {} {\emph {\bibinfo {title} {Stochastic Processes
  in Physics and Chemistry}}}\ (\bibinfo  {publisher} {North Holland,
  Amsterdam},\ \bibinfo {year} {2007})\BibitemShut {NoStop}%
\bibitem [{\citenamefont {G\'omez~Gonz\'alez}\ and\ \citenamefont
  {Garz\'o}(2021)}]{GGG21}%
  \BibitemOpen
  \bibfield  {author} {\bibinfo {author} {\bibfnamefont {R.}~\bibnamefont
  {G\'omez~Gonz\'alez}}\ and\ \bibinfo {author} {\bibfnamefont
  {V.}~\bibnamefont {Garz\'o}},\ }\bibfield  {title} {\enquote {\bibinfo
  {title} {Time-dependent homogeneous states of binary granular suspensions},}\
  }\href@noop {} {\bibfield  {journal} {\bibinfo  {journal} {Phys. Fluids}\
  }\textbf {\bibinfo {volume} {33}},\ \bibinfo {pages} {093315} (\bibinfo
  {year} {2021})}\BibitemShut {NoStop}%
\bibitem [{Note1()}]{Note1}%
  \BibitemOpen
  \bibinfo {note} {Note that there is a mistake in Eq.\ (35) for the
  coefficient $D^*$. Equation \protect \eqref {4.14.2} corrects this
  mistake.}\BibitemShut {Stop}%
\bibitem [{\citenamefont {Mason}(1954)}]{M54}%
  \BibitemOpen
  \bibfield  {author} {\bibinfo {author} {\bibfnamefont {E.~A.}\ \bibnamefont
  {Mason}},\ }\bibfield  {title} {\enquote {\bibinfo {title} {Transport
  properties of gases obeying a modified {B}uckingham potential},}\ }\href@noop
  {} {\bibfield  {journal} {\bibinfo  {journal} {J. Chem. Phys.}\ }\textbf
  {\bibinfo {volume} {22}},\ \bibinfo {pages} {169--192} (\bibinfo {year}
  {1954})}\BibitemShut {NoStop}%
\bibitem [{\citenamefont {L\'opez~de Haro}\ and\ \citenamefont
  {Cohen}(1984)}]{LC84}%
  \BibitemOpen
  \bibfield  {author} {\bibinfo {author} {\bibfnamefont {M.}~\bibnamefont
  {L\'opez~de Haro}}\ and\ \bibinfo {author} {\bibfnamefont {E.~G.~D.}\
  \bibnamefont {Cohen}},\ }\bibfield  {title} {\enquote {\bibinfo {title} {The
  {E}nskog theory for multicomponent mixtures. {III}. {T}ransport properties of
  dense binary mixtures with one tracer component},}\ }\href@noop {} {\bibfield
   {journal} {\bibinfo  {journal} {J. Chem. Phys.}\ }\textbf {\bibinfo {volume}
  {80}},\ \bibinfo {pages} {408--415} (\bibinfo {year} {1984})}\BibitemShut
  {NoStop}%
\bibitem [{\citenamefont {Carnahan}\ and\ \citenamefont
  {Starling}(1969)}]{CS69}%
  \BibitemOpen
  \bibfield  {author} {\bibinfo {author} {\bibfnamefont {N.~F.}\ \bibnamefont
  {Carnahan}}\ and\ \bibinfo {author} {\bibfnamefont {K.~E.}\ \bibnamefont
  {Starling}},\ }\bibfield  {title} {\enquote {\bibinfo {title} {Equation of
  state for nonattracting rigid spheres},}\ }\href@noop {} {\bibfield
  {journal} {\bibinfo  {journal} {J. Chem. Phys.}\ }\textbf {\bibinfo {volume}
  {51}},\ \bibinfo {pages} {635--636} (\bibinfo {year} {1969})}\BibitemShut
  {NoStop}%
\bibitem [{\citenamefont {Boublik}(1970)}]{B70}%
  \BibitemOpen
  \bibfield  {author} {\bibinfo {author} {\bibfnamefont {T.}~\bibnamefont
  {Boublik}},\ }\bibfield  {title} {\enquote {\bibinfo {title} {Hard-sphere
  equation of state},}\ }\href@noop {} {\bibfield  {journal} {\bibinfo
  {journal} {J. Chem. Phys.}\ }\textbf {\bibinfo {volume} {53}},\ \bibinfo
  {pages} {471} (\bibinfo {year} {1970})}\BibitemShut {NoStop}%
\bibitem [{\citenamefont {Reed}\ and\ \citenamefont {Gubbins}(1973)}]{RG73}%
  \BibitemOpen
  \bibfield  {author} {\bibinfo {author} {\bibfnamefont {T.~M.}\ \bibnamefont
  {Reed}}\ and\ \bibinfo {author} {\bibfnamefont {K.~E.}\ \bibnamefont
  {Gubbins}},\ }\href@noop {} {\emph {\bibinfo {title} {Applied Statistical
  Mechanics}}}\ (\bibinfo  {publisher} {MacGraw-Hill, New York},\ \bibinfo
  {year} {1973})\BibitemShut {NoStop}%
\bibitem [{\citenamefont {McLennan}(1989)}]{M89}%
  \BibitemOpen
  \bibfield  {author} {\bibinfo {author} {\bibfnamefont {J.~A.}\ \bibnamefont
  {McLennan}},\ }\href@noop {} {\emph {\bibinfo {title} {Introduction to
  Nonequilibrium Statistical Mechanics}}}\ (\bibinfo  {publisher}
  {Prentic--Hall, New Yersey},\ \bibinfo {year} {1989})\BibitemShut {NoStop}%
\bibitem [{\citenamefont {Montanero}\ and\ \citenamefont
  {Garz\'o}(2002)}]{MG02}%
  \BibitemOpen
  \bibfield  {author} {\bibinfo {author} {\bibfnamefont {J.~M.}\ \bibnamefont
  {Montanero}}\ and\ \bibinfo {author} {\bibfnamefont {V.}~\bibnamefont
  {Garz\'o}},\ }\bibfield  {title} {\enquote {\bibinfo {title} {Monte {C}arlo
  simulation of the homogeneous cooling state for a granular mixture},}\
  }\href@noop {} {\bibfield  {journal} {\bibinfo  {journal} {Granular Matter}\
  }\textbf {\bibinfo {volume} {4}},\ \bibinfo {pages} {17--24} (\bibinfo {year}
  {2002})}\BibitemShut {NoStop}%
\bibitem [{\citenamefont {Khalil}\ and\ \citenamefont {Garz\'o}(2014)}]{KG14}%
  \BibitemOpen
  \bibfield  {author} {\bibinfo {author} {\bibfnamefont {N.}~\bibnamefont
  {Khalil}}\ and\ \bibinfo {author} {\bibfnamefont {V.}~\bibnamefont
  {Garz\'o}},\ }\bibfield  {title} {\enquote {\bibinfo {title} {Homogeneous
  states in driven granular mixtures: {E}nskog kinetic theory versus molecular
  dynamics simulations},}\ }\href@noop {} {\bibfield  {journal} {\bibinfo
  {journal} {J. Chem. Phys.}\ }\textbf {\bibinfo {volume} {140}},\ \bibinfo
  {pages} {{164}{901}} (\bibinfo {year} {2014})}\BibitemShut {NoStop}%
\bibitem [{\citenamefont {Rosato}\ \emph {et~al.}(1987)\citenamefont {Rosato},
  \citenamefont {Strandburg}, \citenamefont {Prinz},\ and\ \citenamefont
  {Swendsen}}]{RSPS87}%
  \BibitemOpen
  \bibfield  {author} {\bibinfo {author} {\bibfnamefont {A.}~\bibnamefont
  {Rosato}}, \bibinfo {author} {\bibfnamefont {K.~J.}\ \bibnamefont
  {Strandburg}}, \bibinfo {author} {\bibfnamefont {F.}~\bibnamefont {Prinz}}, \
  and\ \bibinfo {author} {\bibfnamefont {R.~H.}\ \bibnamefont {Swendsen}},\
  }\bibfield  {title} {\enquote {\bibinfo {title} {Why the {B}razil nuts are on
  top: {S}ize segregation of particulate matter by shaking},}\ }\href@noop {}
  {\bibfield  {journal} {\bibinfo  {journal} {Phys. Rev. Lett.}\ }\textbf
  {\bibinfo {volume} {58}},\ \bibinfo {pages} {1038--1040} (\bibinfo {year}
  {1987})}\BibitemShut {NoStop}%
\bibitem [{\citenamefont {Duran}, \citenamefont {Rajchenbach},\ and\
  \citenamefont {Cl\'ement}(1993)}]{DRC93}%
  \BibitemOpen
  \bibfield  {author} {\bibinfo {author} {\bibfnamefont {J.}~\bibnamefont
  {Duran}}, \bibinfo {author} {\bibfnamefont {J.}~\bibnamefont {Rajchenbach}},
  \ and\ \bibinfo {author} {\bibfnamefont {E.}~\bibnamefont {Cl\'ement}},\
  }\bibfield  {title} {\enquote {\bibinfo {title} {Arching effect model for
  particle size segregation},}\ }\href@noop {} {\bibfield  {journal} {\bibinfo
  {journal} {Phys. Rev. Lett.}\ }\textbf {\bibinfo {volume} {70}},\ \bibinfo
  {pages} {2431--2434} (\bibinfo {year} {1993})}\BibitemShut {NoStop}%
\bibitem [{\citenamefont {Cooke}\ \emph {et~al.}(1996)\citenamefont {Cooke},
  \citenamefont {Warr}, \citenamefont {Huntley},\ and\ \citenamefont
  {Ball}}]{CWHB96}%
  \BibitemOpen
  \bibfield  {author} {\bibinfo {author} {\bibfnamefont {W.}~\bibnamefont
  {Cooke}}, \bibinfo {author} {\bibfnamefont {S.}~\bibnamefont {Warr}},
  \bibinfo {author} {\bibfnamefont {J.~M.}\ \bibnamefont {Huntley}}, \ and\
  \bibinfo {author} {\bibfnamefont {R.~C.}\ \bibnamefont {Ball}},\ }\bibfield
  {title} {\enquote {\bibinfo {title} {Particle size segregation in a
  two-dimensional bed undergoing vertical vibration},}\ }\href@noop {}
  {\bibfield  {journal} {\bibinfo  {journal} {Phys. Rev. E}\ }\textbf {\bibinfo
  {volume} {53}},\ \bibinfo {pages} {2812} (\bibinfo {year}
  {1996})}\BibitemShut {NoStop}%
\bibitem [{\citenamefont {Shinbrot}\ and\ \citenamefont {Muzzio}(1998)}]{SM98}%
  \BibitemOpen
  \bibfield  {author} {\bibinfo {author} {\bibfnamefont {T.}~\bibnamefont
  {Shinbrot}}\ and\ \bibinfo {author} {\bibfnamefont {F.~J.}\ \bibnamefont
  {Muzzio}},\ }\bibfield  {title} {\enquote {\bibinfo {title} {Reverse buoyancy
  in shaken granular beds},}\ }\href@noop {} {\bibfield  {journal} {\bibinfo
  {journal} {Phys. Rev. Lett.}\ }\textbf {\bibinfo {volume} {81}},\ \bibinfo
  {pages} {4365--4368} (\bibinfo {year} {1998})}\BibitemShut {NoStop}%
\bibitem [{\citenamefont {Hong}, \citenamefont {Quinn},\ and\ \citenamefont
  {Luding}(2001)}]{HQL01}%
  \BibitemOpen
  \bibfield  {author} {\bibinfo {author} {\bibfnamefont {D.~C.}\ \bibnamefont
  {Hong}}, \bibinfo {author} {\bibfnamefont {P.~V.}\ \bibnamefont {Quinn}}, \
  and\ \bibinfo {author} {\bibfnamefont {S.}~\bibnamefont {Luding}},\
  }\bibfield  {title} {\enquote {\bibinfo {title} {Reverse {B}razil nut
  problem: {C}ompetition between percolation and condensation},}\ }\href@noop
  {} {\bibfield  {journal} {\bibinfo  {journal} {Phys. Rev. Lett.}\ }\textbf
  {\bibinfo {volume} {86}},\ \bibinfo {pages} {3423--3426} (\bibinfo {year}
  {2001})}\BibitemShut {NoStop}%
\bibitem [{\citenamefont {Arnarson}\ and\ \citenamefont
  {Willits}(1998)}]{AW98}%
  \BibitemOpen
  \bibfield  {author} {\bibinfo {author} {\bibfnamefont {B.}~\bibnamefont
  {Arnarson}}\ and\ \bibinfo {author} {\bibfnamefont {J.~T.}\ \bibnamefont
  {Willits}},\ }\bibfield  {title} {\enquote {\bibinfo {title} {Thermal
  diffusion in binary mixtures of smooth, nearly elastic spheres with and
  without gravity},}\ }\href@noop {} {\bibfield  {journal} {\bibinfo  {journal}
  {Phys. Fluids}\ }\textbf {\bibinfo {volume} {10}},\ \bibinfo {pages}
  {1324--1328} (\bibinfo {year} {1998})}\BibitemShut {NoStop}%
\bibitem [{\citenamefont {Jenkins}\ and\ \citenamefont {Yoon}(2002)}]{JY02}%
  \BibitemOpen
  \bibfield  {author} {\bibinfo {author} {\bibfnamefont {J.~T.}\ \bibnamefont
  {Jenkins}}\ and\ \bibinfo {author} {\bibfnamefont {D.~K.}\ \bibnamefont
  {Yoon}},\ }\bibfield  {title} {\enquote {\bibinfo {title} {Segregation in
  binary mixtures under gravity},}\ }\href@noop {} {\bibfield  {journal}
  {\bibinfo  {journal} {Phys. Rev. Lett.}\ }\textbf {\bibinfo {volume} {88}},\
  \bibinfo {pages} {{194}{301}} (\bibinfo {year} {2002})}\BibitemShut {NoStop}%
\bibitem [{\citenamefont {Trujillo}, \citenamefont {Alam},\ and\ \citenamefont
  {Herrmann}(2003)}]{TAH03}%
  \BibitemOpen
  \bibfield  {author} {\bibinfo {author} {\bibfnamefont {L.}~\bibnamefont
  {Trujillo}}, \bibinfo {author} {\bibfnamefont {M.}~\bibnamefont {Alam}}, \
  and\ \bibinfo {author} {\bibfnamefont {H.~J.}\ \bibnamefont {Herrmann}},\
  }\bibfield  {title} {\enquote {\bibinfo {title} {Segregation in a fluidized
  binary granular mixture: {C}ompetition between buoyancy and geometric
  forces},}\ }\href@noop {} {\bibfield  {journal} {\bibinfo  {journal}
  {Europhys. Lett.}\ }\textbf {\bibinfo {volume} {64}},\ \bibinfo {pages}
  {190--196} (\bibinfo {year} {2003})}\BibitemShut {NoStop}%
\bibitem [{\citenamefont {Alam}, \citenamefont {Trujillo},\ and\ \citenamefont
  {Herrmann}(2006)}]{ATH06}%
  \BibitemOpen
  \bibfield  {author} {\bibinfo {author} {\bibfnamefont {M.}~\bibnamefont
  {Alam}}, \bibinfo {author} {\bibfnamefont {L.}~\bibnamefont {Trujillo}}, \
  and\ \bibinfo {author} {\bibfnamefont {H.~J.}\ \bibnamefont {Herrmann}},\
  }\bibfield  {title} {\enquote {\bibinfo {title} {Hydrodynamic theory for
  reverse {B}razil nut segregation and the non-monotonic ascension dynamics},}\
  }\href@noop {} {\bibfield  {journal} {\bibinfo  {journal} {J. Stat. Phys.}\
  }\textbf {\bibinfo {volume} {124}},\ \bibinfo {pages} {587--623} (\bibinfo
  {year} {2006})}\BibitemShut {NoStop}%
\bibitem [{\citenamefont {Yoon}\ and\ \citenamefont {Jenkins}(2006)}]{YJ06}%
  \BibitemOpen
  \bibfield  {author} {\bibinfo {author} {\bibfnamefont {D.~K.}\ \bibnamefont
  {Yoon}}\ and\ \bibinfo {author} {\bibfnamefont {J.~T.}\ \bibnamefont
  {Jenkins}},\ }\bibfield  {title} {\enquote {\bibinfo {title} {The influence
  of different species' granular temperatures on segregation in a binary
  mixture of dissipative grains},}\ }\href@noop {} {\bibfield  {journal}
  {\bibinfo  {journal} {Phys. Fluids}\ }\textbf {\bibinfo {volume} {18}},\
  \bibinfo {pages} {{073}{303}} (\bibinfo {year} {2006})}\BibitemShut {NoStop}%
\bibitem [{\citenamefont {Garz\'o}(2006)}]{G06}%
  \BibitemOpen
  \bibfield  {author} {\bibinfo {author} {\bibfnamefont {V.}~\bibnamefont
  {Garz\'o}},\ }\bibfield  {title} {\enquote {\bibinfo {title} {Segregation in
  granular binary mixtures: {T}hermal diffusion},}\ }\href@noop {} {\bibfield
  {journal} {\bibinfo  {journal} {Europhys. Lett.}\ }\textbf {\bibinfo {volume}
  {75}},\ \bibinfo {pages} {521--527} (\bibinfo {year} {2006})}\BibitemShut
  {NoStop}%
\bibitem [{\citenamefont {Garz\'o}(2008)}]{G08a}%
  \BibitemOpen
  \bibfield  {author} {\bibinfo {author} {\bibfnamefont {V.}~\bibnamefont
  {Garz\'o}},\ }\bibfield  {title} {\enquote {\bibinfo {title} {Brazil-nut
  effect versus reverse {B}razil-nut effect in a moderately granular dense
  gas},}\ }\href@noop {} {\bibfield  {journal} {\bibinfo  {journal} {Phys. Rev.
  E}\ }\textbf {\bibinfo {volume} {78}},\ \bibinfo {pages} {{020}{301} (R)}
  (\bibinfo {year} {2008})}\BibitemShut {NoStop}%
\bibitem [{\citenamefont {Garz\'o}(2009)}]{G09}%
  \BibitemOpen
  \bibfield  {author} {\bibinfo {author} {\bibfnamefont {V.}~\bibnamefont
  {Garz\'o}},\ }\bibfield  {title} {\enquote {\bibinfo {title} {Segregation by
  thermal diffusion in moderately dense granular mixtures},}\ }\href@noop {}
  {\bibfield  {journal} {\bibinfo  {journal} {Eur. Phys. J. E}\ }\textbf
  {\bibinfo {volume} {29}},\ \bibinfo {pages} {261--274} (\bibinfo {year}
  {2009})}\BibitemShut {NoStop}%
\bibitem [{\citenamefont {Galvin}, \citenamefont {Dahl},\ and\ \citenamefont
  {Hrenya}(2005)}]{GDH05}%
  \BibitemOpen
  \bibfield  {author} {\bibinfo {author} {\bibfnamefont {J.~E.}\ \bibnamefont
  {Galvin}}, \bibinfo {author} {\bibfnamefont {S.~R.}\ \bibnamefont {Dahl}}, \
  and\ \bibinfo {author} {\bibfnamefont {C.~M.}\ \bibnamefont {Hrenya}},\
  }\bibfield  {title} {\enquote {\bibinfo {title} {On the role of
  non-equipartition in the dynamics of rapidly flowing granular mixtures},}\
  }\href@noop {} {\bibfield  {journal} {\bibinfo  {journal} {J. Fluid Mech.}\
  }\textbf {\bibinfo {volume} {528}},\ \bibinfo {pages} {207--232} (\bibinfo
  {year} {2005})}\BibitemShut {NoStop}%
\bibitem [{\citenamefont {Brey}, \citenamefont {Ruiz-Montero},\ and\
  \citenamefont {Moreno}(2005)}]{BRM05}%
  \BibitemOpen
  \bibfield  {author} {\bibinfo {author} {\bibfnamefont {J.~J.}\ \bibnamefont
  {Brey}}, \bibinfo {author} {\bibfnamefont {M.~J.}\ \bibnamefont
  {Ruiz-Montero}}, \ and\ \bibinfo {author} {\bibfnamefont {F.}~\bibnamefont
  {Moreno}},\ }\bibfield  {title} {\enquote {\bibinfo {title} {Energy partition
  and segregation for an intruder in a vibrated granular system under
  gravity},}\ }\href@noop {} {\bibfield  {journal} {\bibinfo  {journal} {Phys.
  Rev. Lett.}\ }\textbf {\bibinfo {volume} {95}},\ \bibinfo {pages}
  {{098}{001}} (\bibinfo {year} {2005})}\BibitemShut {NoStop}%
\bibitem [{\citenamefont {Brey}, \citenamefont {Ruiz-Montero},\ and\
  \citenamefont {Moreno}(2006)}]{BRM06}%
  \BibitemOpen
  \bibfield  {author} {\bibinfo {author} {\bibfnamefont {J.~J.}\ \bibnamefont
  {Brey}}, \bibinfo {author} {\bibfnamefont {M.~J.}\ \bibnamefont
  {Ruiz-Montero}}, \ and\ \bibinfo {author} {\bibfnamefont {F.}~\bibnamefont
  {Moreno}},\ }\bibfield  {title} {\enquote {\bibinfo {title} {Hydrodynamic
  profiles for an impurity in an open vibrated granular gas},}\ }\href@noop {}
  {\bibfield  {journal} {\bibinfo  {journal} {Phys. Rev. E}\ }\textbf {\bibinfo
  {volume} {73}},\ \bibinfo {pages} {{031}{301}} (\bibinfo {year}
  {2006})}\BibitemShut {NoStop}%
\bibitem [{\citenamefont {Brito}\ \emph {et~al.}(2008)\citenamefont {Brito},
  \citenamefont {Enr\'{\i}quez}, \citenamefont {Godoy},\ and\ \citenamefont
  {Soto}}]{BEGS08}%
  \BibitemOpen
  \bibfield  {author} {\bibinfo {author} {\bibfnamefont {R.}~\bibnamefont
  {Brito}}, \bibinfo {author} {\bibfnamefont {H.}~\bibnamefont
  {Enr\'{\i}quez}}, \bibinfo {author} {\bibfnamefont {S.}~\bibnamefont
  {Godoy}}, \ and\ \bibinfo {author} {\bibfnamefont {R.}~\bibnamefont {Soto}},\
  }\bibfield  {title} {\enquote {\bibinfo {title} {Segregation induced by
  inelasticity in a vibrofluidized granular mixture},}\ }\href@noop {}
  {\bibfield  {journal} {\bibinfo  {journal} {Phys. Rev. E}\ }\textbf {\bibinfo
  {volume} {77}},\ \bibinfo {pages} {{061}{301}} (\bibinfo {year}
  {2008})}\BibitemShut {NoStop}%
\bibitem [{\citenamefont {Brito}\ and\ \citenamefont {Soto}(2009)}]{BS09}%
  \BibitemOpen
  \bibfield  {author} {\bibinfo {author} {\bibfnamefont {R.}~\bibnamefont
  {Brito}}\ and\ \bibinfo {author} {\bibfnamefont {R.}~\bibnamefont {Soto}},\
  }\bibfield  {title} {\enquote {\bibinfo {title} {Competition of {B}razil nut
  effect, buoyancy, and inelasticity induced segregation in a granular
  mixture},}\ }\href@noop {} {\bibfield  {journal} {\bibinfo  {journal} {Eur.
  Phys. J. Special Topics}\ }\textbf {\bibinfo {volume} {179}},\ \bibinfo
  {pages} {207--219} (\bibinfo {year} {2009})}\BibitemShut {NoStop}%
\bibitem [{\citenamefont {Kincaid}, \citenamefont {Cohen},\ and\ \citenamefont
  {L\'opez~de Haro}(1987)}]{KCL87}%
  \BibitemOpen
  \bibfield  {author} {\bibinfo {author} {\bibfnamefont {J.~M.}\ \bibnamefont
  {Kincaid}}, \bibinfo {author} {\bibfnamefont {E.~G.~D.}\ \bibnamefont
  {Cohen}}, \ and\ \bibinfo {author} {\bibfnamefont {M.}~\bibnamefont
  {L\'opez~de Haro}},\ }\bibfield  {title} {\enquote {\bibinfo {title} {The
  {E}nskog theory for multicomponent mixtures. iv. {T}hermal diffusion},}\
  }\href@noop {} {\bibfield  {journal} {\bibinfo  {journal} {J. Chem. Phys.}\
  }\textbf {\bibinfo {volume} {86}},\ \bibinfo {pages} {963--975} (\bibinfo
  {year} {1987})}\BibitemShut {NoStop}%
\bibitem [{\citenamefont {Wildman}, \citenamefont {Huntley},\ and\
  \citenamefont {Parker}(2001)}]{WHP01}%
  \BibitemOpen
  \bibfield  {author} {\bibinfo {author} {\bibfnamefont {R.~D.}\ \bibnamefont
  {Wildman}}, \bibinfo {author} {\bibfnamefont {J.~M.}\ \bibnamefont
  {Huntley}}, \ and\ \bibinfo {author} {\bibfnamefont {D.~J.}\ \bibnamefont
  {Parker}},\ }\bibfield  {title} {\enquote {\bibinfo {title} {Granular
  temperature profiles in three-dimensional vibrofluidized granular beds},}\
  }\href@noop {} {\bibfield  {journal} {\bibinfo  {journal} {Phys. Rev. E}\
  }\textbf {\bibinfo {volume} {63}},\ \bibinfo {pages} {{061}{311}} (\bibinfo
  {year} {2001})}\BibitemShut {NoStop}%
\bibitem [{\citenamefont {Breu}\ \emph {et~al.}(2003)\citenamefont {Breu},
  \citenamefont {Ensner}, \citenamefont {Kruelle},\ and\ \citenamefont
  {Rehberg}}]{BEKR03}%
  \BibitemOpen
  \bibfield  {author} {\bibinfo {author} {\bibfnamefont {A.~P.~J.}\
  \bibnamefont {Breu}}, \bibinfo {author} {\bibfnamefont {H.~M.}\ \bibnamefont
  {Ensner}}, \bibinfo {author} {\bibfnamefont {C.~A.}\ \bibnamefont {Kruelle}},
  \ and\ \bibinfo {author} {\bibfnamefont {I.}~\bibnamefont {Rehberg}},\
  }\bibfield  {title} {\enquote {\bibinfo {title} {Reversing the {B}razil-nut
  effect: {C}ompetition between percolation and condensation},}\ }\href@noop {}
  {\bibfield  {journal} {\bibinfo  {journal} {Phys. Rev. Lett.}\ }\textbf
  {\bibinfo {volume} {90}},\ \bibinfo {pages} {{014}{302}} (\bibinfo {year}
  {2003})}\BibitemShut {NoStop}%
\bibitem [{\citenamefont {Schautz}\ \emph {et~al.}(2005)\citenamefont
  {Schautz}, \citenamefont {Brito}, \citenamefont {Kruelle},\ and\
  \citenamefont {Rehberg}}]{SBKR05}%
  \BibitemOpen
  \bibfield  {author} {\bibinfo {author} {\bibfnamefont {T.}~\bibnamefont
  {Schautz}}, \bibinfo {author} {\bibfnamefont {R.}~\bibnamefont {Brito}},
  \bibinfo {author} {\bibfnamefont {C.~A.}\ \bibnamefont {Kruelle}}, \ and\
  \bibinfo {author} {\bibfnamefont {I.}~\bibnamefont {Rehberg}},\ }\bibfield
  {title} {\enquote {\bibinfo {title} {A horizontal {B}razil-nut effect and its
  reverse},}\ }\href@noop {} {\bibfield  {journal} {\bibinfo  {journal} {Phys.
  Rev. Lett.}\ }\textbf {\bibinfo {volume} {95}},\ \bibinfo {pages}
  {{028}{001}} (\bibinfo {year} {2005})}\BibitemShut {NoStop}%
\bibitem [{\citenamefont {Qiao}\ \emph {et~al.}(2021)\citenamefont {Qiao},
  \citenamefont {Duan}, \citenamefont {Dong}, \citenamefont {Wang},
  \citenamefont {Jiang}, \citenamefont {Zhu},\ and\ \citenamefont
  {Zhao}}]{QDDWJZZ21}%
  \BibitemOpen
  \bibfield  {author} {\bibinfo {author} {\bibfnamefont {J.}~\bibnamefont
  {Qiao}}, \bibinfo {author} {\bibfnamefont {C.}~\bibnamefont {Duan}}, \bibinfo
  {author} {\bibfnamefont {K.}~\bibnamefont {Dong}}, \bibinfo {author}
  {\bibfnamefont {W.}~\bibnamefont {Wang}}, \bibinfo {author} {\bibfnamefont
  {H.}~\bibnamefont {Jiang}}, \bibinfo {author} {\bibfnamefont
  {H.}~\bibnamefont {Zhu}}, \ and\ \bibinfo {author} {\bibfnamefont
  {Y.}~\bibnamefont {Zhao}},\ }\bibfield  {title} {\enquote {\bibinfo {title}
  {Dem study of segregation degree and velocity of binary granular mixtures
  subject to vibration},}\ }\href@noop {} {\bibfield  {journal} {\bibinfo
  {journal} {Powder Technol.}\ }\textbf {\bibinfo {volume} {382}},\ \bibinfo
  {pages} {107--117} (\bibinfo {year} {2021})}\BibitemShut {NoStop}%
\bibitem [{\citenamefont {Shao}\ \emph {et~al.}(2023)\citenamefont {Shao},
  \citenamefont {Li}, \citenamefont {Wang}, \citenamefont {Sun},\ and\
  \citenamefont {Huang}}]{YAZMD23}%
  \BibitemOpen
  \bibfield  {author} {\bibinfo {author} {\bibfnamefont {Y.}~\bibnamefont
  {Shao}}, \bibinfo {author} {\bibfnamefont {A.}~\bibnamefont {Li}}, \bibinfo
  {author} {\bibfnamefont {Z.}~\bibnamefont {Wang}}, \bibinfo {author}
  {\bibfnamefont {M.}~\bibnamefont {Sun}}, \ and\ \bibinfo {author}
  {\bibfnamefont {D.}~\bibnamefont {Huang}},\ }\bibfield  {title} {\enquote
  {\bibinfo {title} {Resonance-induced acceleration of the {RBNE}-{BNE}
  segregation inversion},}\ }\href@noop {} {\bibfield  {journal} {\bibinfo
  {journal} {AIChE J.}\ }\textbf {\bibinfo {volume} {e18101}} (\bibinfo {year}
  {2023})}\BibitemShut {NoStop}%
\bibitem [{\citenamefont {Garz\'o}(2011)}]{G11}%
  \BibitemOpen
  \bibfield  {author} {\bibinfo {author} {\bibfnamefont {V.}~\bibnamefont
  {Garz\'o}},\ }\bibfield  {title} {\enquote {\bibinfo {title} {Thermal
  diffusion segregation in granular binary mixtures described by the {E}nskog
  equation},}\ }\href@noop {} {\bibfield  {journal} {\bibinfo  {journal} {New
  J. Phys.}\ }\textbf {\bibinfo {volume} {13}},\ \bibinfo {pages} {{055}{020}}
  (\bibinfo {year} {2011})}\BibitemShut {NoStop}%
\bibitem [{\citenamefont {de~Groot}\ and\ \citenamefont {Mazur}(1984)}]{GM84}%
  \BibitemOpen
  \bibfield  {author} {\bibinfo {author} {\bibfnamefont {S.~R.}\ \bibnamefont
  {de~Groot}}\ and\ \bibinfo {author} {\bibfnamefont {P.}~\bibnamefont
  {Mazur}},\ }\href@noop {} {\emph {\bibinfo {title} {Nonequilibrium
  Thermodynamics}}}\ (\bibinfo  {publisher} {Dover, New York},\ \bibinfo {year}
  {1984})\BibitemShut {NoStop}%
\bibitem [{\citenamefont {Garz\'o}\ and\ \citenamefont {Trizac}(2015)}]{GT15}%
  \BibitemOpen
  \bibfield  {author} {\bibinfo {author} {\bibfnamefont {V.}~\bibnamefont
  {Garz\'o}}\ and\ \bibinfo {author} {\bibfnamefont {E.}~\bibnamefont
  {Trizac}},\ }\bibfield  {title} {\enquote {\bibinfo {title} {Generalized
  transport coefficients for inelastic {M}axwell mixtures under shear flow},}\
  }\href@noop {} {\bibfield  {journal} {\bibinfo  {journal} {Phys. Rev. E}\
  }\textbf {\bibinfo {volume} {92}},\ \bibinfo {pages} {{052}{202}} (\bibinfo
  {year} {2015})}\BibitemShut {NoStop}%
\bibitem [{\citenamefont {Jop}, \citenamefont {Forterre},\ and\ \citenamefont
  {Pouliquen}(2006)}]{JFP06}%
  \BibitemOpen
  \bibfield  {author} {\bibinfo {author} {\bibfnamefont {P.}~\bibnamefont
  {Jop}}, \bibinfo {author} {\bibfnamefont {Y.}~\bibnamefont {Forterre}}, \
  and\ \bibinfo {author} {\bibfnamefont {O.}~\bibnamefont {Pouliquen}},\
  }\bibfield  {title} {\enquote {\bibinfo {title} {A constitutive law for dense
  granular flows},}\ }\href@noop {} {\bibfield  {journal} {\bibinfo  {journal}
  {Nature}\ }\textbf {\bibinfo {volume} {441}},\ \bibinfo {pages} {727--730}
  (\bibinfo {year} {2006})}\BibitemShut {NoStop}%
\bibitem [{\citenamefont {Khalil}\ and\ \citenamefont {Garz\'o}(2013)}]{KG13}%
  \BibitemOpen
  \bibfield  {author} {\bibinfo {author} {\bibfnamefont {N.}~\bibnamefont
  {Khalil}}\ and\ \bibinfo {author} {\bibfnamefont {V.}~\bibnamefont
  {Garz\'o}},\ }\bibfield  {title} {\enquote {\bibinfo {title} {Transport
  coefficients for driven granular mixtures at low-density},}\ }\href@noop {}
  {\bibfield  {journal} {\bibinfo  {journal} {Phys. Rev. E}\ }\textbf {\bibinfo
  {volume} {88}},\ \bibinfo {pages} {052201} (\bibinfo {year}
  {2013})}\BibitemShut {NoStop}%
\end{thebibliography}%

\end{document}